\documentclass[a4paper,11pt]{article}
\pdfoutput=1
\usepackage{jcappub}
\usepackage{aas_macros}
\usepackage{amsmath}
\usepackage{accents}
\usepackage{calc}
\usepackage{subcaption}
\usepackage[normalem]{ulem}
\usepackage[dvipsnames]{xcolor}
\usepackage{rotating}
\usepackage{tablefootnote}
\usepackage{enumitem} 
\usepackage{orcidlink}
\usepackage{xspace}
\usepackage{nicefrac}
\usepackage{fontawesome5}
\usepackage{hyperref}
\usepackage{booktabs}
\usepackage{xcolor}
\usepackage{graphicx}
\usepackage{tikz}
\usepackage{caption}
\usepackage{geometry}
\usepackage{diagbox}
\usepackage{multirow}
\usepackage{makecell}
\usepackage[export]{adjustbox}
\usepackage{listings}

\makeatletter
\newcommand{\github}[1]{%
   \href{#1}{\faGithubSquare}%
}
\makeatother

\newcommand{\codename}{\textsc{Disco-Dj}\xspace}

\newcommand{\dd}{\ensuremath{\text{d}}}
\newcommand{\vecb}[1]{\ensuremath{\boldsymbol{#1}}}
\newcommand{\bnabla}{\ensuremath{\boldsymbol{\nabla}}}
\newcommand{\mPsi}{{\mathit \Psi}}
\newcommand{\overcirc}[1]{%
  \smash{\overset{\raisebox{-100.0ex}{$\scriptstyle *$}}{#1}}%
}

\title{\boldmath DISCO-DJ II: a differentiable particle-mesh code for cosmology}

\author[a]{Florian List\orcidlink{0000-0002-3741-179X}}
\author[a,b]{\!Oliver Hahn\orcidlink{0000-0001-9440-1152}}
\author[a,b]{\!Thomas Fl\"{o}ss\orcidlink{0000-0002-8245-780X}}
\author[a]{\!Lukas Winkler\orcidlink{0000-0002-6792-6743}}
\affiliation{University of Vienna}
\affiliation[a]{\it Department of Astrophysics, Türkenschanzstraße 17, 1180 Vienna, Austria}
\affiliation[b]{\it Department of Mathematics, Oskar-Morgenstern-Platz 1, 1090 Vienna, Austria}
\emailAdd{florian.list@univie.ac.at}

\abstract{
The mildly non-linear regime of cosmic structure formation holds much of the information that upcoming large-scale structure surveys aim to exploit, making fast and accurate predictions on these scales essential.
We present the $N$-body module of \codename{} (\textbf{DI}fferentiable \textbf{S}imulations for \textbf{CO}smology -- \textbf{D}one with \textbf{J}\textsc{ax}), designed to deliver high-fidelity, GPU-accelerated, and differentiable particle-mesh simulations tailored for cosmological inference.
Theory-informed time integrators such as the recently introduced \textsc{BullFrog} method allow for accurate predictions already with few time steps (e.g.\ $6$~steps for per-cent-level accuracy in terms of the present-day power spectrum at $k \approx 0.2 \, h / \mathrm{Mpc}$ using $N = 512^3$ particles, which takes just a few seconds). 
To control discreteness effects and achieve high accuracy, the code incorporates a suite of advanced techniques, for example a custom non-uniform FFT implementation for force evaluation.
Both forward- and reverse-mode differentiation are supported, with memory requirements independent of the number of time steps; in the reverse case, this is achieved through an adjoint formulation.
We extensively study the effect of various numerical parameters on the accuracy. As an application of \codename, we perform field-level inference by recovering $\sigma_8$ and the initial conditions from a noisy \textsc{Gadget} matter density field. Coupled with our recently introduced Einstein--Boltzmann solver, the \codename ecosystem provides a self-consistent, fully differentiable pipeline for modelling the large-scale structure of the universe.
\href{https://github.com/cosmo-sims/DISCO-DJ}{\faGithubSquare}
}

\begin{document}
\maketitle
\section{Introduction}
\label{sec:intro}
How much cosmological information is encoded in the large-scale structure (LSS) of the Universe? The present-day matter distribution is shaped by both the initial conditions of the Universe and its expansion history, making it a powerful probe of cosmology and fundamental physics. Recent analyses of galaxy clustering and weak lensing have placed increasingly tight constraints on parameters such as the total matter density $\Omega_m$, the amplitude of primordial fluctuations $\sigma_8$, and the dark energy equation-of-state parameter $w$ \cite{PhysRevD.105.023520, adame2024desi, adame2025desi}. These results have been largely obtained using summary statistics such as the two-point correlation function or its Fourier-space analogue, the power spectrum, and more recently the bispectrum.

While these statistics efficiently capture information in the near-Gaussian regime, they are suboptimal on small, non-linear scales where non-Gaussian features of the matter field carry significant information. Crucially, the raw, uncompressed observational data still contains far more information than what is accessible via low-order $n$-point statistics alone. To fully exploit the constraining power of the LSS, especially in light of upcoming fourth-generation surveys such as \textsc{Euclid}~\cite{Euclid} and the Vera C.~Rubin Observatory’s \textsc{LSST}~\cite{LSST}, it is imperative to develop more powerful analysis methods.

This has motivated a growing interest in \textit{field-level} and \textit{machine learning-based} approaches \cite{Nguyen2024HowLevel, 10.1093/mnrasl/slab081, stadler2023cosmology, zhou2024accurate, stopyra2024towards, kostic2023consistency, doeser2024bayesian, jasche2019physical, Bayer:2023rmj, simon2025benchmarking, porqueres2023field, lemos2024field, rossello2025differentiable, andrews2023bayesian, horowitz2025differentiable, modi2021cosmicrim, 2025ApJ...989..207B, Floss:2023ylq, Bottema:2025vww, 2025MNRAS.540..716M, 2022JCAP...08..003T, Giri:2022nzt, Giri:2023mpg, Ho:2024whi, Chen:2024exy} -- with many works focusing particularly on initial condition reconstruction (e.g.\ \cite{weinberg1992, Gramann1993, croft1997reconstruction, Frisch2002, 2008MNRAS.389..497K, Jasche2013, 2017PhRvD..96b3505S, Feng_2018, 2023arXiv230313056J, List:2023jwo, Shallue2023, 2024MNRAS.527L.173L, savchenko2024mean}) -- which aim to extract the maximum amount of cosmological information by modelling the full matter or galaxy density field. 
Rather than compressing the data into summary statistics, field-level inference aims to recover both the global cosmological parameters and the full realisation of the initial conditions that gave rise to the observed structures in the Universe. Key to this approach is a \textit{forward model} of structure formation, which evolves the initial density field -- typically represented as a grid of Fourier- or real-space amplitudes -- into a late-time field. These forward models encode gravitational dynamics, cosmic expansion, and potentially baryonic effects or observational systematics, allowing for direct comparisons between simulations and observed fields.

The resulting inference problem is extraordinarily high-dimensional: even at modest resolutions of $N = 256^3$ to $512^3$ particles or grid cells, the number of latent variables describing the initial conditions reaches $\mathcal{O}(10^8)$, far exceeding the handful of global parameters like $\Omega_m$ or $\sigma_8$. This joint inference of initial conditions and cosmological parameters therefore pushes the limits of traditional sampling techniques and demands both accurate models and efficient computational strategies.

Traditional Markov Chain Monte Carlo (MCMC) methods scale poorly and suffer from slow convergence, especially when the posterior distribution is highly degenerate or curved. Incorporating gradient information, as done in Hamiltonian Monte Carlo (HMC; \cite{DUANE1987216, neal2011mcmc}) or related techniques such as microcanonical sampling \cite{robnik2023microcanonical, robnik2025metropolis}, significantly improves sampling efficiency by enabling more informed proposals that respect the local geometry of the posterior.
However, computing derivatives through cosmological forward models is non-trivial. Finite-difference approaches are conceptually simple, but become computationally prohibitive in high-dimensional spaces and are often dominated by numerical noise. Analytic differentiation, while more accurate, requires painstaking derivations and re-implementation of model internals, which is particularly cumbersome for non-linear simulations. This has historically limited the feasibility of derivative-based inference pipelines for complex structure formation models (although see, e.g., \cite{jasche2019physical}).

In recent years, the rise of automatic differentiation (autodiff) frameworks -- driven by advances in machine learning -- has opened new possibilities for differentiable simulations. Autodiff-enabled simulation tools allow for the computation of exact gradients w.r.t.\ initial conditions and cosmological parameters, enabling end-to-end differentiable forward models. These models can then be used for gradient-based optimisation, Fisher matrix forecasting, and fully differentiable HMC sampling.
In cosmology, several such forward models have recently been introduced \cite{li2022pmwd, modi2021flowpm, 2025MNRAS.tmp..937R, zhou2024accurate, horowitz2025differentiable, rossello2025differentiable, zack_li_2023_10065126}.
In Ref.~\cite{Hahn2023DISCO-DJCosmology}, we presented a differentiable Einstein--Boltzmann solver as part of our \codename framework, confirmed excellent agreement with the industry standard codes \textsc{CAMB} \cite{2011ascl.soft02026L} and \textsc{CLASS} \cite{lesgourgues2011cosmic}, and demonstrated its usefulness -- for example to forecast cosmological parameter constraints with the \textsc{Euclid} survey.

\begin{figure}
  \centering
  \begin{tikzpicture}    
    \node[anchor=south west, inner sep=0] (fig) at (0,0)
      {\includegraphics[width=\linewidth]{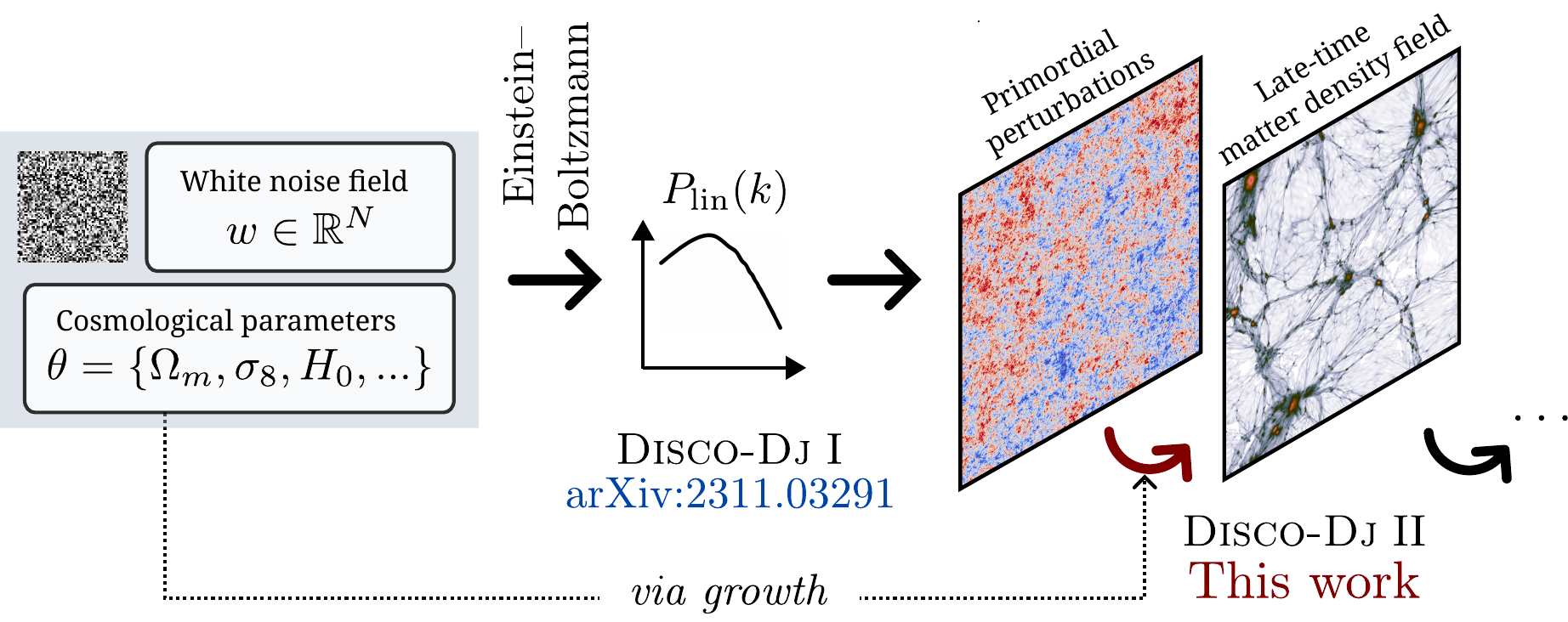}};
    \begin{scope}[shift={(fig.south west)}, x={(fig.south east)}, y={(fig.north west)}]
  \node[anchor=south west, inner sep=0pt] at (0.35, 0.17)
    {\href{https://arxiv.org/abs/2311.03291}{\phantom{\rule{0.23\linewidth}{0.03\linewidth}}}};      
    \end{scope}
  \end{tikzpicture}
  \caption{High-level structure of \codename. Given a white noise realisation $w \in \mathbb{R}^N$ of the initial density fluctuations -- whose dimensionality $N$ typically equals the number of $N$-body particles -- and a set of cosmological parameters $\theta$, \codename enables forward modelling the non-linear gravitational collapse in an \textbf{end-to-end autodifferentiable} and \textbf{GPU-accelerated} manner using \textsc{Jax} \cite{jax2018github}. The linear power spectrum is computed by solving the (linearised) Einstein--Boltzmann equations as presented in Ref.~\cite{Hahn2023DISCO-DJCosmology} (\textsc{Disco-Dj~I}). The non-linear structure formation module is introduced in this work (\textsc{Disco-Dj~II}), consisting of perturbative models and a fast $N$-body PM code with theory-informed time integrators. Extensions of \codename regarding bias modelling etc. are currently in preparation.
  }
  \label{fig:sketch}
\end{figure}

In this work, we present the non-linear structure formation model of \codename, centred around a GPU-accelerated particle-mesh (PM) $N$-body simulation code featuring theory-informed time integrators and an array of discreteness suppression techniques, complemented with an arbitrary-order implementation of Lagrangian perturbation theory (LPT). Combined with our Einstein--Boltzmann solver, \codename provides an end-to-end differentiable pipeline for making fast and accurate predictions of the large-scale structure, see Fig.~\ref{fig:sketch}. While primarily designed for gradient-based inference, \codename is also a powerful stand-alone tool, e.g.\ for generating large suites of training data for emulators and machine learning pipelines for applications such as likelihood-free inference, generative models, etc.

The structure of this paper is as follows. In Sec.~\ref{sec:implementation}, we describe the mathematical background and the numerical methods implemented in \codename. In Sec.~\ref{sec:validation}, we validate our implementation, study temporal and spatial convergence, and demonstrate its computational performance. Furthermore, we systematically study the impact of different numerical techniques and parameters on the accuracy of the predictions. The insights gained from this analysis can be expected to carry over to other cosmological (PM) codes.
Section~\ref{sec:inference} presents a proof-of-concept application where the differentiability of \codename is leveraged to perform cosmological field-level inference. We conclude this work and discuss some directions for future development in Sec.~\ref{sec:conclusions}.

{\textit{Note on notation:} For clarity, we will use bold vector notation only for vectorial quantities in configuration- and $k$-space, not for scalar quantities represented on a discrete set (such as the density contrast $\delta$ when evaluated on a discrete grid, etc.). }

\section{Methods and implementation}
\label{sec:implementation}

\subsection{Overview}
\begin{figure}
    \centering
    \includegraphics[width=1.0\linewidth]{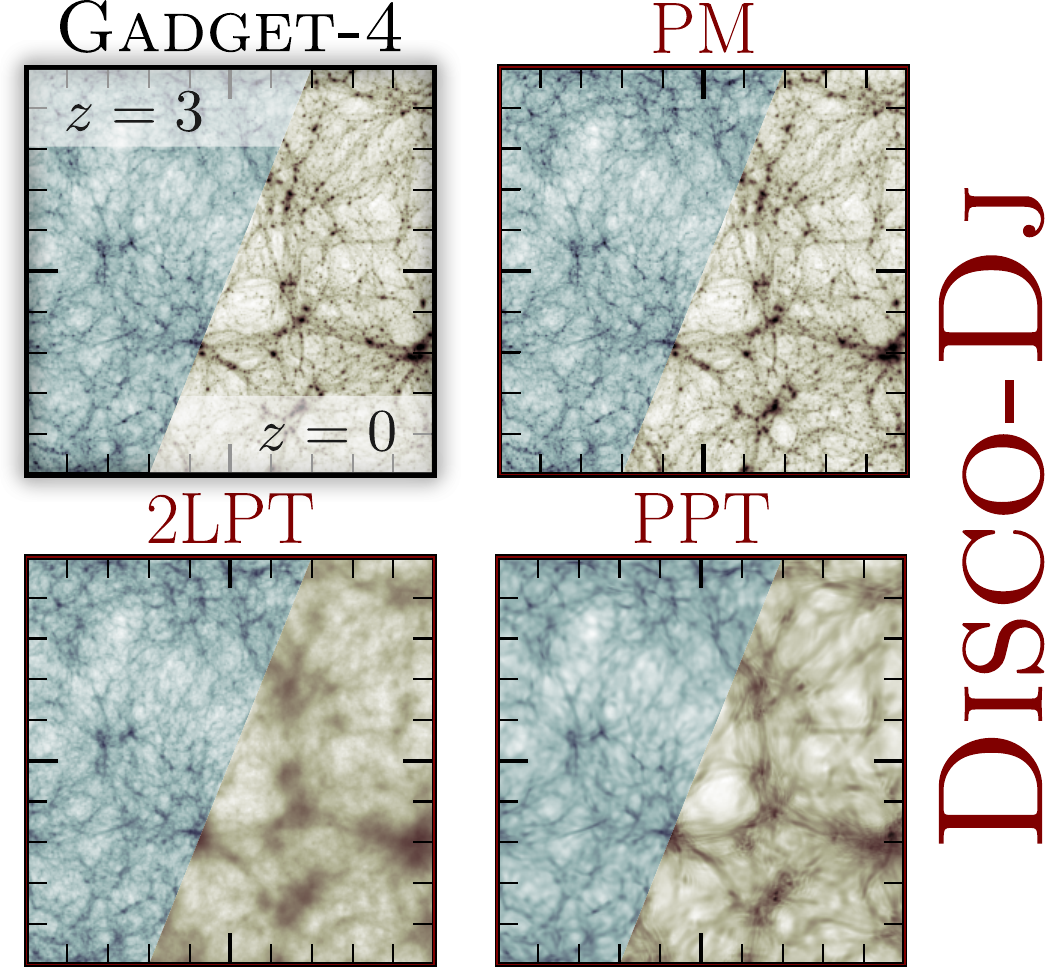}
    \caption{Simulated matter density slices through a box of comoving side length $L = 100 \, \mathrm{Mpc} / h$ at redshifts $z = 3$ (left half) and $z = 0$ (right half), computed with \textsc{Gadget-4} (upper left) and with various methods implemented in \codename. The slices are averaged over $25 \, \mathrm{Mpc} / h$ along the third dimension. Specifically, we consider a particle-mesh (PM) simulation with 10 \textsc{BullFrog} time steps (upper right), second-order Lagrangian perturbation theory (2LPT, lower left), and propagator perturbation theory (PPT, lower right). The number of particles / fluid elements is $N = 1024^3$ for \textsc{Gadget-4} and $N = 512^3$ for all methods in \codename. The colour mapping is logarithmic w.r.t.\ density, and the colour bar limits are individual for each redshift, but shared across the four panels. At $z = 3$, the predictions by the perturbative methods in the bottom row still agree fairly well with the \textsc{Gadget-4} reference. In contrast, at $z = 0$ the filamentary structures with these methods are smeared out, and interference patterns are visible in the PPT panel. Judging by eye, our PM simulation reproduces the reference very accurately, although small differences can be spotted by eye, e.g.\ the ellipticity of the cluster slightly below the vertical centre near the right edge. For quantitative results, see Sec.~\ref{sec:validation}.}
    \label{fig:methods}
\end{figure}

All forward models implemented in \codename are concerned with computing an approximate solution to the cosmological Vlasov--Poisson system,
\label{eq:Vlasov-Poisson}
\begin{align}
    \frac{\dd f}{\dd t} = \partial_t f + \frac{\vecb{p}}{a^2}\cdot\bnabla_{\vecb{x}} f + \frac{\vecb{g}}{a} \cdot\bnabla_{\vecb{p}} f = 0, \qquad\textrm{where}\qquad \vecb{g}:= -\bnabla_{\vecb{x}} \varphi,
\end{align}
see e.g.\ \cite{2022LRCA....8....1A, Rampf2021CosmologicalProblems} for recent reviews. Here, $f = f(\vecb{x}, \vecb{p}, t)$ is the phase-space distribution function, where the coordinates $\vecb{x}$ are comoving in a universe expanding with a scale factor $a(t)$. In practice, particles in a cube $\vecb{x} \in [0, L)^3$ are considered for a given box size $L$, equipped with a flat torus topology (i.e.\ with periodic boundary conditions, see \cite{2025arXiv250820606A} for a possible way to accommodate curved geometries). 

The gravitational acceleration is the negative gradient of the gravitational potential $\varphi$, which is related to the density contrast $\delta$ via Poisson's equation
\begin{equation}
    \Delta_{\vecb{x}} \varphi = \frac{3 \Omega_m H_0^2}{2} \delta,
    \label{eq:poisson}
\end{equation}
where $\delta$ itself is defined by marginalising the phase-space distribution $f$ over the momentum space
\begin{equation}
    1 + \delta = \int f \, \dd^3 p, \quad \mathrm{normalised \ as} \quad \int \delta \, \dd^3 x = 0.
\end{equation}
Moreover, $\Omega_m$ is the present-day matter density parameter, and $H_0$ is the Hubble constant, i.e.\ today's value of the Hubble parameter $H(a)$.

\subsection{Lagrangian perturbation theory}
\label{sec:lpt}
While the Vlasov equation describes the full phase-space evolution of matter, computing its moments over velocity yields fluid equations, where the zeroth and first moments define the mass density and momentum density, respectively. The evolution equation for the $i$th moment depends on the $(i+1)$th moment, and one therefore obtains an infinite hierarchy of equations, known as the Vlasov (or Boltzmann) hierarchy. In the cold limit, however, where there is a single velocity associated with each position, the velocity dispersion tensor and all higher-order cumulants vanish, terminating the Vlasov hierarchy already after the first moment (prior to shell-crossing, which gives rise to higher-order moments, see e.g.\ \cite{2009PhRvD..80d3504P, 2015MNRAS.454.3920H}). The resulting two fluid equations (plus the Poisson equation) are amenable to perturbative treatment. For instance, standard (Eulerian) perturbation theory \cite{Bernardeau2002Large-scaleTheory} expands the Eulerian density and velocity divergence as power laws and recursively solves the resulting recursion relations. This approach is valid as long as the density perturbations remain small, i.e.\ $\delta \ll 1$. 

Lagrangian perturbation theory (LPT; e.g.\ \cite{Zeldovich:1970, Buchert:1993, Bouchet:1995}), on the other hand, provides a solution to the fluid equations by expressing them in Lagrangian coordinates $\vecb{q}$ and perturbatively expanding the displacement $\vecb{\mPsi}(\vecb{q}, t) = \vecb{x}(\vecb{q}, t) - \vecb{q}$, i.e.\ the vector field pointing from initial (Lagrangian) positions $\vecb{q}$ to the Eulerian position $\vecb{x}(\vecb{q}, t)$ of the associated fluid element at time $t$. The continuity equation then implies that the density contrast can be reconstructed from the Jacobian determinant of this mapping:
\begin{equation}
    1 + \delta(\vecb{x}, t) = \left(\det \bnabla_{\vecb{q}} \otimes \vecb{x}(\vecb{q}, t)\right)^{-1}, \qquad \text{prior to shell-crossing.}
\end{equation}
At first order, LPT yields the well-known Zeldovich approximation \cite{Zeldovich:1970}, while second-order LPT (2LPT) captures quadratic corrections to the particle trajectories, etc. The validity of the LPT series expansion breaks down after shell-crossing, that is, when particle trajectories cross for the first time.

Apart from serving as a stand-alone analytical tool for predicting structure formation on mildly non-linear scales -- potentially applied to UV-filtered fields and supplemented by counterterms on smaller scales (e.g.\ \cite{Schmidt2020AnStructure}) -- 2LPT is nowadays the prevalent method for generating initial conditions for N-body simulations. Recently, however, it has been shown that using 3LPT allows initialising cosmological simulations at later times, thereby reducing discreteness effects in $N$-body simulations that are most critical at early times \cite{Michaux:2021, 2025arXiv250606200A, 2025MNRAS.542..735M}.
An alternative approach is to start cosmological simulations directly at $a = 0$ by means of LPT-informed time integrators (see below) and explicitly suppressing discreteness effects, see Ref.~\cite{List2024StartingBang} for more details.

In \codename, we implement the LPT recursion relations at arbitrary order in the form as derived by Ref.~\cite{Rampf2020}. Our default implementation uses the `$D^n$ approximation' (also known as `Einstein--de Sitter (EdS) approximation') of the growth, where the $n$th order growth function is approximated as proportional to the respective power of the linear growth factor, i.e.\ $\propto D^n$, ignoring higher-order terms in $\Omega_\Lambda$. 
In addition, we implemented a version of $n$LPT for $n \leq 3$ with the exact $\Lambda$CDM growth functions. In that case, the multiple terms that arise at each order cannot be lumped together, as their growth is no longer independent. This means that being able to evaluate the LPT fields at different times requires the storage of three separate fields already at 3LPT, with the number of terms at each order $n$ asymptotically growing quadratically for $n \to \infty$, which quickly becomes infeasible in terms of memory. In any case, the effect of higher-order corrections is completely negligible at early times and for a realistic $\Lambda$CDM cosmology, the deviation of the full growth from the EdS approximation is $< 1\%$ at second order and $< 2\%$ for all three third-order growth functions even at $z = 0$, see Appendix~\ref{sec:growth}.

Our default implementation of LPT thus computes the time-Taylor series of the growing-mode solution
\begin{equation}
    \vecb{\psi}(\vecb{q}, D)=\sum_{n=1}^{n_{\mathrm{max}}}
     \vecb{\psi}^{(n)}(\vecb{q}) D^n,
\label{eq:LPT}
\end{equation}
up to a given order $n_{\mathrm{max}}$, where the linear growth factor $D$ is used as the time variable, and the purely spatial terms $\vecb{\psi}^{(n)}$ absorb factors coming from the $D^n$ approximation of the growth, e.g.\ $-3/7$ for the 2LPT term, as the second-order growth function is given by $-(3/7) D^2 + O(\Omega_\Lambda)$. Then, the displacement field is split into a longitudinal part $L =\bnabla_{\vecb{q}} \cdot \vecb{\psi}$ and a transverse part $\vecb{T} = \bnabla_{\vecb{q}} \times \vecb{\psi}$ via the Helmholtz decomposition as
\begin{equation}
\label{eq:helmholtz}
    \vecb{\psi} = \Delta^{-1}_{\vecb{q}} (\bnabla_{\vecb{q}} L - \bnabla_{\vecb{q}} \times \vecb{T}).
\end{equation}
At first order $n = 1$, this yields $L^{(1)} = -\Delta_{\vecb{q}} \varphi_{\mathrm{ini}}$ and $T^{(1)} \equiv \vecb{0}$, where $\varphi_{\mathrm{ini}}$ is the initial potential, and hence $\vecb{\psi} = -D \bnabla_{\vecb{q}} \varphi_{\mathrm{ini}}$, i.e.\ the Zeldovich approximation.
At higher orders $n > 1$, recursion relations arise \cite{Zheligovsky2014Time-analyticityFlow, Rampf2020}, where the longitudinal component is
\begin{subequations}
\label{eq:LPT_L_and_T}
\begin{equation}
    L^{(n)} = \sum_{0<s<n} \frac{\frac{3-n}{2}-s^2-(n-s)^2}{(n+\frac 3 2)(n-1)} \mu_{2,L}^{(s, n-s)} + \!\!\!\! \sum_{n_1+n_2+n_3=n} \frac{\frac{3-n}{2}-n_1^2-n_2^2-n_3^2}{(n+\frac 3 2)(n-1)} \mu_{3,L}^{\left(n_1, n_2, n_3\right)},
\label{eq:LPT_L}
\end{equation}
and the transversal component
\begin{equation}
\boldsymbol{T}^{(n)} = \frac{1}{2} \sum_{0<s<n} \frac{n-2 s}{n} \boldsymbol{\mu}_{2,T}^{(s,n-s)}.
\label{eq:LPT_T}
\end{equation}
\end{subequations}
The spatial kernels in these expressions are defined as
\begin{subequations}
\label{eq:LPT_kernels}
\begin{align}
    \mu_{2,L}^{(n_1,n_2)} &= \frac{1}{2} \left(\psi_{i,i}^{(n_1)} \psi_{j,j}^{(n_2)} - \psi_{i,j}^{(n_1)} \psi_{j,i}^{(n_2)} \right), \\
    \mu_{3,L}^{(n_1,n_2,n_3)} &= \frac{1}{6} \epsilon_{ikl} \epsilon_{jmn} \psi_{i,j}^{(n_1)} \psi_{k,m}^{(n_2)} \psi_{l,n}^{(n_3)}, \\    
    (\mu_{2,T}^{(n_1,n_2)})_i &= \epsilon_{ijk} \psi_{l,j}^{(n_1)} \psi_{l,k}^{(n_2)},
\end{align}
\end{subequations}
where $\epsilon_{ijk}$ is the Levi--Civita symbol, Einstein's sum convention is adopted, and $\psi_{i,j}$ denotes the derivative of the $i$th component of $\vecb{\psi}$ w.r.t.\ the $j$th coordinate of the Lagrangian position vector $\vecb{q}$. Note again that we only consider growing modes.

Our specific implementation closely follows the algorithms given in Ref.~\cite[App.~A]{Rampf2020}, where \texttt{for}-loops over the indices are mostly replaced by \texttt{jax.lax.scan} operations. This is because \textsc{Python} \texttt{for}-loops in \textsc{Jax} are unrolled by the XLA compiler, leading to slow compilation, whereas \texttt{jax.lax.scan} lowers to a single compiled loop. Gradients and the inverse Laplacian in Eq.~\eqref{eq:helmholtz} are applied in Fourier space using the exact kernels $\mathrm{i} \vecb{k}$ and $-k^2$, respectively, where $\vecb{k}$ is the wave vector and $k = |\vecb{k}|$ is the wave number. In order to avoid convolutions in Fourier space, we perform the products of multiple terms arising in the spatial kernels in Eq.~\eqref{eq:LPT_kernels} in real space, applying a padding and subsequent cropping according to Orszag's 3/2 rule in order to de-alias the fields \cite{orszag1971elimination}. Note, however, that we truncate the fields to the desired resolution at each order and do not carry along the increasingly higher Fourier modes as done in the `no-mode-left-behind' approach by Ref.~\cite{Schmidt2020AnStructure}, as we noticed only a very small effect in our experiments, which would come at the cost of a huge increase in memory scaling as $(3/2)^n$ in terms of the LPT order $n$. We emphasise again that the factors in Eqs.~\eqref{eq:LPT_L_and_T} connect the $n$th-order growth factor to $D^n$, allowing us to lump together all spatial kernels at each order. The output of \codename's LPT routine is given by the $n$ displacement fields $\{\vecb{\psi}^{(s)}\}_{s=1}^n$, which can then be used to compute the LPT displacement field at arbitrary times by evaluating the sum in Eq.~\eqref{eq:LPT}, truncated at an arbitrary order $1 \leq n \leq n_{\mathrm{max}}$. Our 3LPT implementation with full $\Lambda$CDM growth is very similar; however, the three individual 3LPT contributions are stored separately, and each of them is multiplied by the associated third-order growth factor at any given time. 

Figure~\ref{fig:methods} illustrates the density field arising from 2LPT in comparison with an accurate $N$-body simulation performed with the \textsc{Gadget-4} simulation code \cite{Springel2020}, at redshifts $z = 3$ and $z = 0$. Visually, the 2LPT matches the reference well at $z = 3$, whereas the filamentary structures are significantly puffier at $z = 0$; unsurprisingly, as this is far beyond the perturbative regime for the selected box size of $L = 100 \, \mathrm{Mpc} / h$.
Results from the other methods presented herein -- namely a semiclassical approach (see the next section) and a PM simulation -- are also shown.

\subsection{Propagator perturbation theory}
\label{sec:ppt}
An alternate perturbative approach to solving the Vlasov--Poisson system is given by Propagator perturbation theory (PPT, \cite{Uhlemann2019SemiclassicalStructure}), which employs a semiclassical (i.e.\ quantum-physics-inspired) description of structure formation. PPT converts the initial gravitational potential to a wave function $\Psi$ according to
\begin{equation}
    \Psi^{\mathrm{ini}}(\vecb{q}) = \exp\left(-\mathrm{i} \frac{\varphi_{\mathrm{ini}}(\vecb{q})}{\hbar}\right),
\end{equation}
where $\hbar > 0$ is a small number that effectively acts as a softening scale, which should be chosen as $\hbar > |\Delta \phi| / \pi$ due to the Nyquist--Shannon sampling theorem, with $\Delta \phi$ being the difference of the gravitational potential between neighbouring evaluation points of the computation grid \cite{Porqueres2020AData}.

One can then define a transition amplitude $K$, which propagates $\Psi^{\mathrm{ini}}$ to an arbitrary growth-factor time $D$ as
\begin{equation}
    \Psi(\vecb{x}, D) = \int K(\vecb{x}, \vecb{q}, D) \, \Psi^{\mathrm{ini}}(\vecb{q}) \, \mathrm{d}^3 q.
\end{equation}
The simplest choice for the transition amplitude is given by the counterpart of the Zeldovich approximation in LPT, i.e.\ motion with constant velocity in terms of growth-factor time $D$. This propagator corresponds to the transition amplitude 
\begin{equation}
    K(\vecb{x}, \vecb{q}, D) = (2 \pi \mathrm{i} \hbar D)^{-\nicefrac{3}{2}} \exp\left(\frac{\mathrm{i}}{\hbar} S(\vecb{x}, \vecb{q}, D) \right).
\end{equation}
Here, $S$ denotes the classical action from which the transition amplitude is derived via the Dirac--Feynman trick, given by
\begin{equation}
    S(\vecb{x}, \vecb{q}, D) = \frac{1}{2} \frac{\vecb{x} - \vecb{q}}{D}.
\end{equation}
For beyond-leading-order extensions and the application to two distinct fluids (such as cold dark matter and baryons), we refer the reader to Refs.~\cite{2021MNRAS.503..406R, 2021MNRAS.503..426H}.
Finally, the density contrast $\delta$ and momentum density $\vecb{j} = (1 + \delta) \frac{\mathrm{d}\vecb{x}}{\mathrm{d}D}$ follow from the wave function $\Psi$ as 
\begin{subequations}
\begin{align}
    1 + \delta &= \Psi \Bar{\Psi}, \\
    \vecb{j} &= \frac{\mathrm{i} \hbar}{2} \left(\Psi \bnabla_{\vecb{x}} \bar{\Psi} - \bar{\Psi} \bnabla_{\vecb{x}} \Psi \right).
\end{align}
\end{subequations}

Note that in contrast to LPT, which yields a displacement for any \textit{Lagrangian} position, PPT allows evaluating the wave function at arbitrary \textit{Eulerian} positions. Thus, the (momentum) density field can be directly computed on a Eulerian grid, without the need to interpolate displaced particles onto a grid as in LPT. Extensions of the propagator to redshift-space distortions and the construction of wave functions that model other variables such as the optical depth, as done in Ref.~\cite{Porqueres2020AData} are straightforward to implement.

\subsection{Time stepping}
\label{sec:timestepping}
We now turn towards the ingredients for PM simulations. A key component is the time integration scheme used for the temporal discretisation of the Vlasov--Poisson system \eqref{eq:Vlasov-Poisson}. To make a connection with the previous sections, we note that it is possible in fact to design integrators that exactly trace perturbative trajectories up to a certain order, while higher-order truncation errors decrease as the step size is reduced \cite{Feng:2016, list2024perturbation, Rampf_2025}. A single PM time step would then compute, e.g., 2LPT trajectories (up to higher-order terms, and with the caveat that discretisation errors arise due to the Eulerian PM-based force computation, compared to the Lagrangian evaluation of quantities in LPT).

\codename supports different drift-kick-drift leapfrog/Verlet steppers of the form

\begin{subequations}
\begin{align}
    \vecb{X}_i^{n+\nicefrac{1}{2}} &= \vecb{X}_i^{n} + \tau_1 \vecb{V}^{n}_i, \label{eq:drift_1}\\
    \vecb{V}_i^{n+1} &= \alpha \vecb{V}_i^{n} + \beta \vecb{A}(\vecb{X}_i^{n+\nicefrac{1}{2}}), \label{eq:kick} \\
    \vecb{X}_i^{n+1} &= \vecb{X}_i^{n+\nicefrac{1}{2}} + \tau_2 \vecb{V}^{n+1}_i. \label{eq:drift_2}
\end{align}
\label{eq:dkd}
\end{subequations}
Here, $\vecb{X}_i^n$ denotes the position of particle $i$ at the $n$th time step, $\vecb{A}(\vecb{X}_i^{n+\nicefrac{1}{2}})$ is the acceleration evaluated in the middle of the full time step (i.e.\ after the first drift), $\tau = \tau_1 + \tau_2$ is the time step size, and $\alpha$ and $\beta$ are coefficients that may depend on both the current time and time step. For the velocity variable $\vecb{V}_i^n$ (and the associated time step variable $\tau$), different choices are supported, such as 
\begin{itemize}
    \item the canonical momentum variable $\vecb{V} = a^2 \dot{\vecb{X}}$ with $\tau = \int_{a_n}^{a_{n+1}} (H(a) a^3)^{-1} \, \dd a$ and
    \item the growth-time velocity $\vecb{V} = \dd \vecb{X} / \dd D = \dot{\vecb{X}} / (a H(a) D'(a))$ with $\tau = D_{n+1} - D_n$,
\end{itemize}
where the overdot denotes a derivative w.r.t.\ cosmic time. 

When canonical momentum is chosen as the velocity variable, the kick coefficients correspond to the standard symplectic leapfrog integrator in drift-kick-drift form, i.e.\ $\alpha \equiv 1$ and $\beta = \int_{a_n}^{a_{n+1}} (H(a) a^2)^{-1} \, \dd a$. The symplecticity of this integrator makes it well suited for simulating bound systems that have decoupled from the Hubble flow, where the resulting conservation of phase-space volume is highly desirable in order to prevent a secular in- or decrease of energy (see e.g. the comparison with a non-symplectic Runge--Kutta integrator in \cite[Fig.~4]{Springel:2005}). On cosmological scales, however, defining the velocity in terms of growth factor is more natural and, in practice, highly beneficial: in fact, a single drift off an unperturbed lattice at time zero with $D$-time velocity $\vecb{V} = \dd \vecb{X} / \dd D = - \bnabla_{\vecb{q}} \varphi_{\mathrm{ini}}$ is equivalent to the Zeldovich approximation, which well approximates structure formation on large scales. In particular -- unlike for the canonical momentum -- the quantities in the leapfrog scheme \eqref{eq:dkd} remain bounded for $a \to 0$.

Ref.~\cite{list2024perturbation} used this idea to introduce a class of `Zeldovich consistent' integrators, which lead to $N$-body trajectories that are by construction consistent with the Zeldovich trajectory after every step -- as long as the Zeldovich approximation is valid (i.e.\ particle trajectories have not yet crossed). This property is ensured by using the $D$-time velocity $\vecb{V} = \dd \vecb{X} / \dd D$, and by further requiring that $\beta = (1 - \alpha) D_{n+\nicefrac{1}{2}}$. One possible choice of $\alpha$ and $\beta$ satisfying this relation is the \textsc{FastPM} integrator \cite{Feng:2016}, which uses
\begin{equation}
    \alpha = \zeta(a(D_n)) / \zeta(a(D_{n+1})), \qquad \beta = (1 - \alpha) D_{n+\nicefrac{1}{2}},
\end{equation}
where $\zeta(a) = H(a) a^3 D'(a)$. As shown in Ref.~\cite{list2024perturbation}, this is the only choice that defines a \textit{symplectic} Zeldovich consistent integrator. 

Recently, Ref.~\cite{Rampf_2025} built onto the analysis of Ref.~\cite{list2024perturbation} and derived the \textsc{BullFrog} integrator, which corresponds to the unique choice of $\alpha$ and $\beta$ that aligns the $N$-body trajectory with 2LPT after each time step in its regime of validity:
\begin{equation}
    \alpha = \frac{E'(D_{n+1}) - \xi_{n+\nicefrac{1}{2}}}{E'(D_n) - \xi_{n+\nicefrac{1}{2}}}, \qquad \beta = (1 - \alpha) D_{n+\nicefrac{1}{2}},
\end{equation}
where 
\begin{equation}
    \xi = D_{n+\nicefrac{1}{2}}^{-1} \left(E_n + E'(D_n) \frac{D_{n+1} - D_n}{2}\right) - D_{n+\nicefrac{1}{2}},
\end{equation}
and $E = -(3/7) D^2 - (3/1001) \Lambda D^5 + O(\Lambda^2 D^8)$ with $\Lambda := \Omega_\Lambda / \Omega_m$ is the second-order growth function (see Appendix \ref{sec:growth} for the defining ordinary differential equation).
\footnote{For completeness, let us remark that the growth factor used by \textsc{BullFrog} is \textit{not} normalised according to $D(a = 1) = 1$, but rather agrees asymptotically with the Einstein--de Sitter growth factor $D(a) \asymp a$ at early times. Although \codename otherwise uses the normalised growth for consistency with the common conventions in cosmology, we do not make this distinction explicit here in order not to overload notation.}
In view of its superior performance, \textsc{BullFrog} is the default integrator in \textsc{DiscoDJ}, yielding accurate predictions already with very few steps.

Another important choice is that of the time step spacing. \codename supports uniform steps w.r.t.\ scale factor $a$, logarithmic scale factor $\log a$, superconformal time $\mathfrak{t}$ (defined via $\dd \mathfrak{t} = a^{-2} \dd t$), and growth-factor time $D$. The midpoint is centred in terms of the specified variable. For instance, when using \textsc{BullFrog} with uniform steps in $D$, one has $\tau_1 = D_{n+\nicefrac{1}{2}} - D_{n}$, where $D_{n} = D_0 + n \Delta D$ and $D_{n+\nicefrac{1}{2}} = D_0 + (n + 1/2) \Delta D$, whereas for uniform steps in $a$, one has $\tau_1 = D(a_{n+\nicefrac{1}{2}}) - D(a_n)$, where $a_n = a_0 + n \Delta a$ and $a_{n+\nicefrac{1}{2}} = a_0 + (n + 1/2) \Delta a$. Alternatively, a list of scale factors can be explicitly provided, enabling the computation of derivatives w.r.t.\ the evaluation times used by the time integrator, which could be used e.g.\ to find an optimal time step distribution via gradient-based optimisation. 

For all steppers, initial conditions for $n = 0$ are, by default, determined with LPT. Alternatively, it is also possible to perform a single (discreteness-suppressed) \textsc{BullFrog} step from time zero to the desired starting time, where the `pre-initialisation' at time zero uses $\vecb{X} = \vecb{q}$ and $\vecb{V} = \mathrm{d}\vecb{X}/\mathrm{d}D = - \bnabla_{\vecb{q}} \varphi_{\mathrm{ini}}$, see Ref.~\cite{List2024StartingBang}. (Note that the \textsc{PowerFrog} stepper used in that reference performs the same initialisation step from time zero as \textsc{BullFrog} -- apart from higher-order terms in $\Omega_\Lambda$, which the latter includes, but which are negligible at early times.)

\subsection{Force evaluation: particle-mesh and non-uniform FFT}
\label{sec:force}

\begin{table}[h]
\centering
\small
\begin{tabular}{@{}l >{$}l<{$} >{$}c<{$} >{$}c<{$}@{}}
\toprule
& \textbf{Real: } W(x) & \textbf{Fourier: } \hat{W}(k) & \textbf{Discrete support} \, [\text{cells}] \\
\midrule
\textbf{CIC} 
& \begin{cases}
1 - |x|, & |x| \leq 1 \\
0, & \text{else}
\end{cases}
& \mathrm{sinc}^2\!\left( \tfrac{k}{2} \right) 
& 2^d \overset{3\text{D}}{=} 8 \\
\textbf{TSC} 
& \begin{cases}
\frac{3}{4} - x^2, & |x| \leq \tfrac{1}{2} \\
\tfrac{1}{2}\!\left(\tfrac{3}{2} - |x|\right)^2, & \tfrac{1}{2} < |x| \leq \tfrac{3}{2} \\
0, & \text{else}
\end{cases}
& \mathrm{sinc}^3\!\left( \tfrac{k}{2} \right) 
& 3^d \overset{3\text{D}}{=} 27 \\
\textbf{PCS} 
& \begin{cases}
\tfrac{1}{6}(4 - 6x^2 + 3|x|^3), & |x| \leq 1 \\
\tfrac{1}{6}(2 - |x|)^3, & 1 < |x| \leq 2 \\
0, & \text{else}
\end{cases}
& \mathrm{sinc}^4\!\left( \tfrac{k}{2} \right) 
& 4^d \overset{3\text{D}}{=} 64 \\
\bottomrule
\end{tabular}
\caption{1D interpolation kernels in real and Fourier space: CIC (linear), TSC (quadratic), PCS (cubic spline). Here, $\mathrm{sinc}(k) = \sin(\pi k) / (\pi k)$. The 3D kernels are products over the three 1D kernels for each axis. The spatial variable $x$ is in units of grid cells here, and the wave number $k$ in units of the Nyquist wave number $k_{\mathrm{Nyquist}}$.}
\label{tab:interpolation_kernels}
\end{table}

\renewcommand{\arraystretch}{1.2}
\setlength{\tabcolsep}{4pt}

\begin{table}[h]
\centering
\small
\begin{tabular}{@{}l l >{$}c<{$} >{$}c<{$}@{}}
\toprule
& \textbf{Order} & \textbf{Real} & \textbf{Fourier} \\
\midrule
\multirow{4}{*}{\rotatebox[origin=c]{90}{\textbf{Gradient}}} 
& Exact & \partial_x & \mathrm{i}k \\
& 2     & \tfrac{1}{2}[{-1}\ 0\ 1] & \tfrac{\mathrm{i}}{\pi} \sin(\pi k) \\
& 4     & \hspace{0.15cm} \tfrac{1}{12}[1\ {-8}\ 0\ 8\ {-1}] & \tfrac{\mathrm{i}}{6\pi}[-\sin(2\pi k) + 8\sin(\pi k)] \\
& 6     & \hspace{-0.13cm} \tfrac{1}{60}[{-1}\ 9\ {-45}\ 0\ 45\ {-9}\ 1] & \tfrac{\mathrm{i}}{30\pi}[\sin(3\pi k) - 9\sin(2\pi k) + 45\sin(\pi k)] \\
\midrule
\multirow{4}{*}{\rotatebox[origin=c]{90}{\textbf{Laplacian}}} 
& Exact & \partial_x^2 & -k^2 \\
& 2     & \hspace{0.17cm} [{1}\ {-2}\ 1] & -\tfrac{2}{\pi^2}[\cos(\pi k) - 1] \\
& 4     & \tfrac{1}{12}[{-1}\ 16\ {-30}\ 16\ {-1}] & -\tfrac{1}{6\pi^2}[\cos(2\pi k) - 16\cos(\pi k) + 15] \\
& 6     & \tfrac{1}{180}[2\ {-27}\ 270\ {-490}\ 270\ {-27}\ 2] & -\tfrac{1}{90\pi^2}[-2\cos(3\pi k) + 27\cos(2\pi k) - 270\cos(\pi k) + 245] \\
\bottomrule
\end{tabular}
\caption{Finite-difference stencils and their Fourier transforms for first- and second derivative operators of various orders. \codename computes gradients and Laplacians in Fourier space, where the above listed operators are available. The wave number $k$ is in units of the Nyquist wave number $k_{\mathrm{Nyquist}}$ here, and the 3D kernels are given by the product of the three 1D kernels along each axis.}
\label{tab:derivative_kernels}
\end{table}

\renewcommand{\arraystretch}{1.6}
\setlength{\tabcolsep}{4pt}

\begin{figure}
    \centering
    \includegraphics[width=1\linewidth]{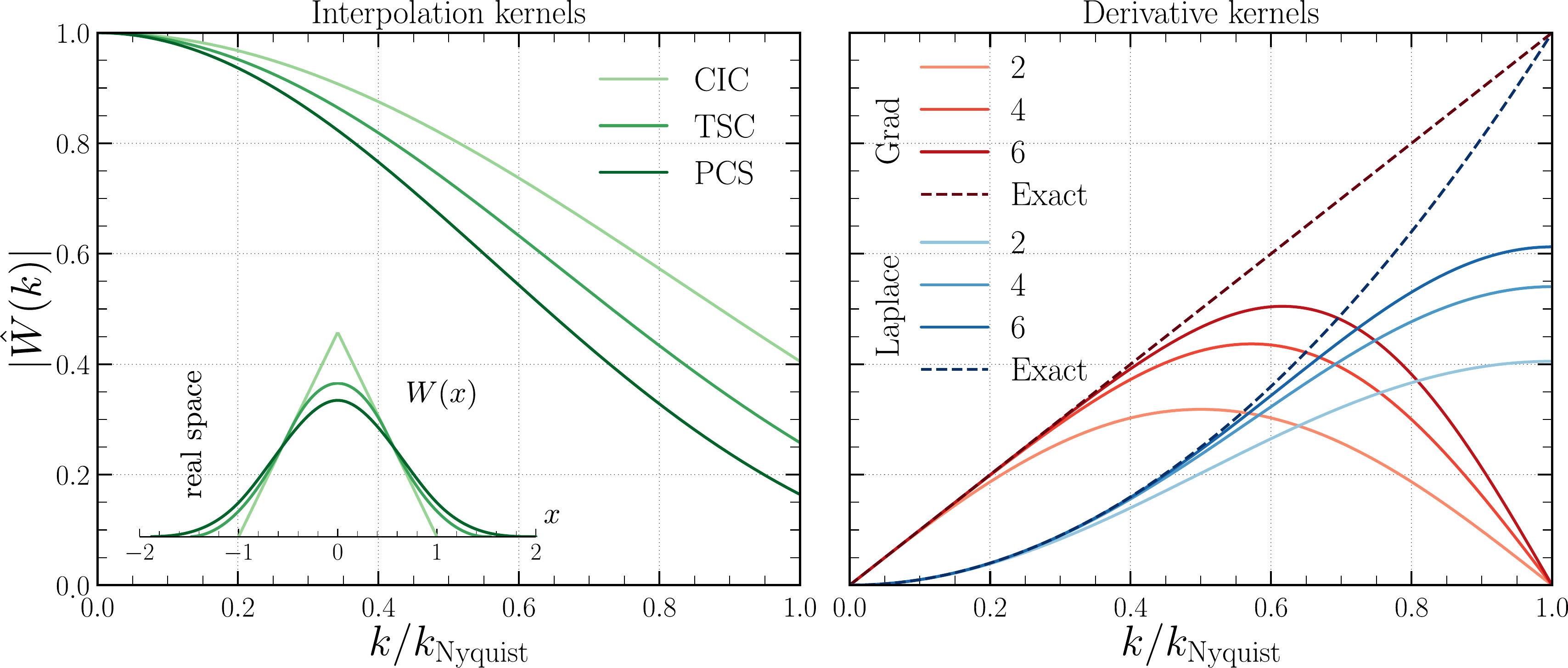}
    \caption{Fourier-transformed kernels used for interpolation (left) and derivatives (right), where the wave number $k$ is given in units of the Nyquist wave number $k_{\mathrm{Nyquist}}$. As for the interpolation kernels, a larger support (such as for PCS) suppresses the power more strongly. The inset plot illustrates the kernels in real space, where $x$ has units of grid cells, and the $y$-axis is such that the areas under the curves integrate to unity. The finite-difference derivative kernels closely match the exact derivatives on large scales, but fall off as $k \to k_{\mathrm{Nyquist}}$, with lower order implying an earlier drop.}
    \label{fig:kernels_fourier}
\end{figure}

Force evaluation is the computationally most expensive part of $N$-body simulations.
\codename supports two force computation schemes, (1) the particle-mesh (PM) method, and (2) force computation based on the non-uniform Fast Fourier Transform (NUFFT), using a custom \textsc{Jax}-based NUFFT implementation. 

\paragraph{Fourier density with the PM method}
For the PM method, the particle masses are interpolated onto a regular mesh via a localised mass assignment kernel in order to obtain a particle-based density estimate. Specifically, cloud-in-cell (CIC), triangular-shaped cloud (TSC), and piecewise cubic spline (PCS) interpolation are available, see Table~\ref{tab:interpolation_kernels} for the definitions of the kernels in real and Fourier space, as well as their support, which directly affects memory and runtime. In particular, note that the first derivative of the commonly employed CIC kernel is piecewise constant, and the second derivative vanishes. Thus, switching to higher-order kernels (or NUFFT, see below) may be required when \codename is used as a forward model within schemes that rely on additional smoothness, e.g.\ (quasi-)Newton methods. The density contrast on the mesh is then transformed to Fourier space via a real FFT.

\paragraph{Fourier density with the NUFFT method}
The NUFFT-based method also computes the Fourier-space density contrast from the particle positions; however, it does so in a more principled way by extending the very definition of the discrete Fourier transform (DFT) to non-uniform data points. This makes it a natural choice in the context of $N$-body simulations, where the particles are arbitrarily distributed within the simulation box. Specifically, the non-uniform analogue of the forward discrete Fourier transform (known as adjoint NDFT) is given by
\begin{equation}
    \hat{\delta}_{\vecb{k}} = \sum_{i=1}^N M_{i} e^{-\frac{2\pi\mathrm{i}}{L} \vecb{k} \cdot \vecb{X}_{i}},
    \label{eq:NUDFTII}
\end{equation}
where $\hat{\delta}_{\vecb{k}}$ is the discrete Fourier density contrast on the reciprocal lattice with dimensions $N_d \times N_d \times N_d$ (i.e.\ $\vecb{k} \in \{0, \ldots, N_d - 1\}^3$) with $N_d^3 = N$, and $M_{i}$ is the mass of the $i$th particle. 
In our implementation, all particles have the same mass, so we can set $M_{i} \equiv 1$. Note that for uniformly spaced particles on a grid, the above definition reduces to the standard forward DFT. Thus, at least conceptually, the NUDFT provides a direct mapping from particle positions to the Fourier-space density contrast without requiring any intermediate grid or local mass assignment kernel. This stands in contrast to the standard PM approach, where the interpolation step inevitably introduces aliasing and a low-pass filter.

In practice, however, efficient NUDFT implementations still rely on interpolating particle contributions onto a regular mesh to enjoy the $O(N \log N)$ runtime complexity of the standard FFT (per dimension), which requires uniformly spaced points in order for the divide-and-conquer technique underlying the FFT to be applicable. The key difference is that this interpolation is performed in a principled way: NUFFT employs smooth, compactly supported window functions whose spectral leakage can be rigorously controlled. This allows one to bound the approximation error in Fourier space w.r.t.\ the true DFT given in Eq.~\eqref{eq:NUDFTII}, up to a user-specified tolerance. In our implementation, we adopt the `exponential of semicircle' kernel \cite{barnett2019parallel, barnett2021aliasing}, which has favourable convergence properties when increasing the size of its support. The regular mesh size per dimension $N_d$ is typically chosen slightly larger than $N^{\nicefrac{1}{3}}$ to suppress aliasing. By default -- and throughout this work -- we use the `low-upsampling' value of $N_d = (5/4) N^{\nicefrac{1}{3}}$ suggested by Ref.~\cite{barnett2019parallel}. After the FFT is performed, only the Fourier modes below the particle Nyquist mode are retained. As for the support of the NUFFT interpolation kernel, \codename provides the option to either specify the desired accuracy of the Fourier density (and the kernel size is then determined accordingly), or to directly specify the kernel size. We choose a kernel size of $5^3$ here because we empirically found that the improvement from larger kernels is small.

\paragraph{Deconvolution of the interpolation kernel}
Since interpolation from the particle positions onto the grid suppresses power, see Fig.~\ref{fig:kernels_fourier}, it is common practice to deconvolve the resulting Fourier-space overdensity via division by the squared Fourier kernel (`squared' because of the interpolation back to the particles, see below). While our NUFFT implementation always performs a deconvolution, this choice is left to the user in PM simulations. In the PM case, it turns out that applying a deconvolution places an upper limit on the PM resolution in relation to the number of particles to not blow up small-scale noise. Therefore, whether a larger PM grid used without deconvolutions or a smaller PM grid with deconvolutions yields better results depends on the specific scenario, see e.g.\ Fig.~\ref{fig:mass_assignment_study_small_large} in the appendix.

\paragraph{Computing the forces}
With the Fourier density contrast at hand, the gravitational potential is computed by solving the Poisson equation \eqref{eq:poisson} with periodic boundary conditions in Fourier space by means of a real FFT, and the gravitational force follows as the negative potential gradient. When using localised mass assignment, \codename implements the exact Fourier transform of the Laplace operator ($-k^2$) and gradient operator ($\mathrm{i} \vecb{k}$), as well as Fourier kernels corresponding to finite difference approximations at various orders, see Table~\ref{tab:derivative_kernels}. Results with different kernels are presented in Appendix~\ref{sec:effect_of_derivative_kernel}. The finite-difference kernels act as low-pass filters, see Fig.~\ref{fig:kernels_fourier} for the amplitude of the different derivatives in Fourier space, and thus de-alias implicitly to some extent. When NUFFT is used to obtain the Fourier density, the exact spectral gradient kernel $\mathrm{i} \vecb{k}$ is the default in \codename. Alternatively, in this case, we also support gradient computation via applying the gradient operator to the interpolation kernel (exponential of a semicircle) -- similarly to the way in which gradients are computed in smoothed-particle hydrodynamics. In practice, we notice little difference between these two options, as the interpolation kernel is anyway chosen large enough to ensure spectral accuracy.

Once the gravitational force $\hat{\vecb{g}}$ has been computed in Fourier space, its three components are interpolated back to the particles, either with the same localised CIC/TSC/PCS kernel as for the density computation, or using NUFFT as
\begin{equation}
    \vecb{g}_{i} = \vecb{A}(\vecb{X}_i) = \sum_{\vecb{k} \in \{0, \ldots, N_d-1\}^3} \hat{\vecb{g}}_{\vecb{k}} e^{\frac{2 \pi \mathrm{i}}{L} \vecb{k} \cdot \vecb{X}_{i}},
\end{equation}
where $\vecb{g}_{i}$ is the acceleration that acts on particle $i$. 
To reduce the memory footprint, we provide the option to specify the chunk size for the vectorisation of the interpolation operations. When setting this parameter to a value strictly smaller than the total number of particles, the code loops over chunks, thereby reducing the memory load at the cost of a slight increase in runtime.

\subsection{Automatic differentiation}
\label{sec:autodiff}
\codename supports both forward and reverse-mode autodiff. Forward-mode differentiation is most efficient whenever a high-dimensional output quantity is differentiated w.r.t.\ a low-dimensional input, as the derivatives are propagated from the inputs to the outputs, building one column of the Jacobian at a time. This makes it suitable when computing e.g.\ the derivative of a density field w.r.t.\ one or few cosmological parameters. 

Reverse-mode autodiff, on the other hand, is suited for functions with a high-dimensional input and a low-dimensional output. It computes one row of the Jacobian at a time, performing a backward pass (`backpropagation') from the outputs to the inputs and building a single row of the Jacobian at a time. In the context of deep learning, this situation is typical when optimising a scalar loss function w.r.t.\ millions of trainable parameters via gradient descent. In \codename, reverse-mode differentiation is suitable, e.g., for minimising scalar loss functions such as likelihoods, for computing gradients in the context of HMC, or more generally whenever gradients w.r.t.\ the high-dimensional initial white noise field are required. 

To ensure that the gradient computation of the forces -- the computationally most expensive step -- is efficient and also allows for chunking the particles in order to reduce the memory footprint just as in the forward pass, we implemented custom Jacobian-vector product (JVP) and vector-Jacobian product (VJP) operations for the interpolation onto (\texttt{scatter}) and from (\texttt{gather}) the grid. Our implementation exploits the fact that JVPs and VJPs of \texttt{scatter} and \texttt{gather} are again \texttt{scatter} and \texttt{gather} operations, involving (co)tangents and, in some cases, a derivative kernel, see Appendix~\ref{sec:custom_VJP_JVP}.
When the full Jacobian matrix of a mapping is required in \textsc{Jax}, 
it is built iteratively by applying the JVP (VJP) operation to Euclidean base vectors in forward (reverse) mode.

In the context of differentiable cosmological simulations, a crucial aspect concerns the memory requirements of the derivative computation. Forward-mode autodiff results in relatively low memory overhead, as it only requires propagating the tangent variables through the simulation, in addition to the `primal' variables (i.e.\ positions and momenta). Reverse-mode autodiff, in contrast, requires storing the entire computational graph in the forward pass, including intermediate states of the simulation, potentially leading to extremely high memory costs and in particular to an $O(N_{\mathrm{steps}})$ scaling in terms of the number of time steps $N_{\mathrm{steps}}$. Although it is possible to reduce the memory requirements by using a checkpointing scheme and recomputing forward-pass quantities in the backward pass (e.g.\ \cite{10.1145/347837.347846}), even the storage of a relatively small number of position-momentum states along the time integration is often prohibitive, particularly given the limited memory of GPUs. In order to circumvent this, we leverage the \textit{adjoint method} for differentiating through the $N$-body time integration with an $O(1)$ memory footprint in terms of $N_{\mathrm{steps}}$, as explained in the next section. 

\subsection{Adjoint method}
\label{sec:adjoint}
$N$-body simulations discretise the Vlasov--Poisson system \eqref{eq:Vlasov-Poisson} in space by tracing a finite number of $N$ tracers and in time by performing a finite number of $N_{\mathrm{steps}}$ time steps (see Sec.~\ref{sec:timestepping}). Hence, the $N$-body system at the $n$th step $(\vecb{X}^n, \vecb{V}^n)$ -- as well as any quantity derived from this state -- depends on the cosmological parameters $\theta$ and the white noise Gaussian random field $w$ that sets the specific realisation. Computing the gradient w.r.t.\ either is crucial in many applications. For outputs on the lightcone, particle positions and velocities are frozen at the moment a particle $i$ crosses the lightcone, i.e.\ observables depend on $(\vecb{X}^{n(i)}, \vecb{V}^{n(i)})$ (where a fractional value of $n(i)$ would correspond to interpolated values between time steps).

There are two different approaches for computing reverse-mode derivatives, known as `discretise-then-optimise' and `optimise-then-discretise' (e.g.\ \cite{kidger2021on}). 
The former is what happens by default when iterating with a \texttt{jax.lax.scan} (or \texttt{for}-loop) over the time steps: gradients are propagated backwards step by step through the operations executed by the integrator, i.e.\ every drift and kick. Since the acceleration in the kick depends non-linearly on the positions, however, the chain rule of differentiation requires that the positions computed in the forward pass be available also during each step of the backward pass. Hence, naively, the memory requirements scale as $O(N_{\mathrm{steps}})$, as explained above.

An interesting alternative is `optimise-then-discretise', also known as the adjoint method \cite{pontryagin2018mathematical, chen2018neural}. The key insight behind the adjoint method is that the gradients of any loss (or objective) function themselves satisfy a differential equation, the so-called adjoint equation, formulated in terms of adjoint (or cotangent) variables. These variables encode the sensitivity of the loss to the state at each time. Importantly, the initial conditions of the initial value problem for the adjoint variables are set at the \textit{final} time of the simulation, and the adjoint equation is solved backwards in time. Since the primary variables $(\vecb{X}, \vecb{V})$ also appear in the adjoint equation, as we will see below, solving the adjoint equation in the context of $N$-body simulations requires performing a second backwards-in-time simulation which, in addition to the phase-space variables, also tracks their adjoint variables.

In principle, the adjoint equation can be discretised arbitrarily and, in particular, independently of the discretisation used in the forward pass. In \codename, we choose the same integrator for both the forward and the backward pass: in this way, given that we use a simple non-adaptive, time-reversible (albeit not necessarily symplectic) stepping scheme, particles propagate backwards on the exact same trajectory without discretisation errors (however with round-off errors). 

In most generality, the adjoint equation can be derived by defining a constrained optimisation problem where the dynamics imposed by the ODE are enforced using Lagrange multipliers and one requires stationarity of the resulting Lagrangian w.r.t.\ the state variables and the Lagrange multipliers, see Ref.~\cite{Li2022DifferentiableMethod} for such a derivation in the context of $N$-body simulations. However, since we want to propagate the dynamics on the exact same DKD trajectory as in the forward pass, a more straightforward approach is to iteratively compute the pullback of the drift and kick operations (which is what standard reverse-mode autodiff would compute), which shows that the resulting iteration scheme defines an `adjoint simulation' that propagates the adjoint variables backwards in time. From this perspective, the only difference between the adjoint method and standard autodiff concerns the primal variables $(\vecb{X}, \vecb{V})$, required for the adjoint simulation, see below: standard autodiff (i.e.\ letting \textsc{Jax} automatically take care of the gradient computation) stores these values during the forward-in-time simulation such that they are available during the backward pass. While this is convenient in many applications, it is prohibitive in the present case, where $\vecb{X}$ and $\vecb{V}$ are high-dimensional. The crux of the adjoint method is to realise that instead of storing the primal variables during the simulation, we can also supply their values on the fly during the backward pass by tracing their trajectories backwards in time alongside the adjoint fields.

In order to clarify how this works in practice, let us consider the notion of the pullback in more detail. For a function $f: A \to B$, the pullback defines a linear mapping from the cotangent space $T^*\!B$ to the cotangent space $T^*\!A$. In the Euclidean case, one can identify $T^*\!A \cong A$ and $T^*\!B \cong B$ and compute the pullback as a VJP between $\overcirc{b} \in B$ and the Jacobian $\partial f / \partial a$ as $\overcirc{b}^\top \, (\partial f / \partial a)$, which then lives in $A$ via the identification $T^*\!A \cong A$.
In our setting, this corresponds to computing how a scalar-valued loss function $L$ depending on a state propagates gradients backward through the steps of the DKD integrator.

Let us define the cotangents $(\overcirc{\vecb{X}}^{n}, \overcirc{\vecb{V}}^{n})$, which represent the sensitivity of the loss function w.r.t.\ the positions and velocities after the $n$th time step (for instance the final one if $n = n_{\mathrm{steps}}$). Consider the drift. For propagating the gradients, we need to ask: what is the sensitivity of the loss to the state given by the positions and velocity \textit{before} that drift? This question is answered by the pullback of the cotangents through the drift. Defining a dummy state for the pre-drift velocity $\vecb{V}^{n-\nicefrac{1}{2}} = \vecb{V}^{n}$ (which, recall, is not affected by the drift and hence usually only takes integer superscripts for the time step), the second half of the drift computes
\begin{subequations}
\begin{align}
    \left(\vecb{X}^{n}, \vecb{V}^{n}\right)^\top = \left(\vecb{X}^{n - \nicefrac{1}{2}} + \tau_2 \vecb{V}^{n-\nicefrac{1}{2}}, \vecb{V}^{n-\nicefrac{1}{2}}\right)^\top.
\end{align}
\end{subequations}
Thus, by computing the Jacobian blocks 
\begin{equation}
\mathbf{J}_{\mathrm{Drift}}^{n-\nicefrac{1}{2}\to n} = 
\begin{pmatrix}
\frac{\partial \vecb{X}^{n}}{\partial \vecb{X}^{n-\nicefrac{1}{2}}} &
\frac{\partial \vecb{X}^{n}}{\partial \vecb{V}^{n-\nicefrac{1}{2}}} \\
\frac{\partial \vecb{V}^{n}}{\partial \vecb{X}^{n-\nicefrac{1}{2}}} &
\frac{\partial \vecb{V}^{n}}{\partial \vecb{V}^{n-\nicefrac{1}{2}}}
\end{pmatrix}
= 
\begin{pmatrix}
\mathbf{I} & \tau_2 \mathbf{I} \\
\mathbf{0} & \mathbf{I}
\end{pmatrix},
\end{equation}
one finds that the adjoints propagate backwards according to
\begin{equation}
    \left(\overcirc{\vecb{X}}^{n - \nicefrac{1}{2}},  \overcirc{\vecb{V}}^{n - \nicefrac{1}{2}}\right) = \left(\overcirc{\vecb{X}}^{n},  \overcirc{\vecb{V}}^{n}\right) \mathbf{J}_{\mathrm{Drift}}^{n-\nicefrac{1}{2}\to n} = \left(\overcirc{\vecb{X}}^{n}, \overcirc{\vecb{V}}^{n} + \tau_2 \overcirc{\vecb{X}}^{n}\right).
\label{eq:adjoint_drift}
\end{equation}
Notably, this is also a drift, but with the roles of $\overcirc{\vecb{X}}$ and $\overcirc{\vecb{V}}$ flipped in comparison to $\vecb{X}$ and $\vecb{V}$ in the forward pass. Thus, backpropagating the sensitivity of a loss function through a drift simply requires us to perform a drift with the adjoint variables. The adjoint position $\overcirc{\vecb{X}}$ does not participate in the drift; therefore, we will continue writing $\overcirc{\vecb{X}}^{n}$ rather than $\overcirc{\vecb{X}}^{n-\nicefrac{1}{2}}$ in what follows (just as we do not use $\vecb{V}^{n-\frac{1}{2}}$).
Since the drift is a linear operation, the adjoint drift is independent of the primal variables $(\vecb{X}, \vecb{V})$. However, this is not the case for the kick
\begin{equation}
        \left(\vecb{X}^{n-\nicefrac{1}{2}, \mathrm{K}}, \vecb{V}^{n}\right)^\top = \left(\vecb{X}^{n - \nicefrac{1}{2}}, \alpha \vecb{V}^{n-1} + \beta \vecb{A}(\vecb{X}^{n-\nicefrac{1}{2}}\right)^\top,
\end{equation}
for which one finds the Jacobian
\begin{equation}
\mathbf{J}_{\mathrm{Kick}}^{n-1 \to n} =
\begin{pmatrix}
\frac{\partial \vecb{X}^{n-\nicefrac{1}{2},\mathrm{K}}}{\partial \vecb{X}^{n-\nicefrac{1}{2}}} & \frac{\partial \vecb{X}^{n-\nicefrac{1}{2},\mathrm{K}}}{\partial \vecb{V}^{n-1}} \\
\frac{\partial \vecb{V}^{n}}{\partial \vecb{X}^{n-\nicefrac{1}{2}}} & \frac{\partial \vecb{V}^{n}}{\partial \vecb{V}^{n-1}}
\end{pmatrix}
=
\begin{pmatrix}
\mathbf{I} & \mathbf{0} \\
\beta \bnabla_{\vecb{X}} \vecb{A}(\vecb{X}^{n-\nicefrac{1}{2}}) & \alpha \mathbf{I}
\end{pmatrix},
\end{equation}
where $\bnabla_{\vecb{X}} \vecb{A}(\vecb{X}^{n+\nicefrac{1}{2}})$ is the Jacobian of the acceleration field. Also, we denote the (unchanged) positions after the kick as $\vecb{X}^{n-\nicefrac{1}{2},\mathrm{K}} = \vecb{X}^{n-\nicefrac{1}{2}}$. Although the kick does not affect the positions, we make this distinction explicit here, as we will see now that the adjoint variable $\overcirc{\vecb{V}}$, which has already been updated by the adjoint drift to take the value $\overcirc{\vecb{V}}^{n-\nicefrac{1}{2}}$, will again be updated by the adjoint kick due to the scaling $\alpha$ in the second row of the Jacobian diagonal, to a `post-kick' value $\overcirc{\vecb{V}}^{n-\nicefrac{1}{2},\mathrm{K}}$. Specifically, we obtain for the adjoint kick,
\begin{equation}
\begin{aligned}
\left(\overcirc{\vecb{X}}^{n - 1},  \overcirc{\vecb{V}}^{n - \nicefrac{1}{2}, \mathrm{K}}\right) &= \left(\overcirc{\vecb{X}}^{n},  \overcirc{\vecb{V}}^{n-\nicefrac{1}{2}}\right) \mathbf{J}_{\mathrm{Kick}}^{n-1\to n}  \\
    &= \left(\overcirc{\vecb{X}}^{n} + \beta \overcirc{\vecb{V}}^{n-\nicefrac{1}{2}} \cdot \bnabla_{\vecb{X}} \vecb{A}(\vecb{X}^{n-\nicefrac{1}{2}}), \alpha \overcirc{\vecb{V}}^{n-\nicefrac{1}{2}}\right).
\end{aligned}
\end{equation}
Some comments are in order. 
First, note that we named the resulting adjoint positions $\overcirc{\vecb{X}}^{n-1}$ although the drift $n - \nicefrac{1}{2} \to n - 1$ is yet to come, as we have already seen above that the adjoint drift does not affect $\overcirc{\vecb{X}}$. 
Second, the adjoint velocity $\overcirc{\vecb{V}}^{n-\nicefrac{1}{2}}$ computed in Eq.~\eqref{eq:adjoint_drift} is rescaled by $\alpha$ in the adjoint kick, giving rise to an intermediate value $\overcirc{\vecb{V}}^{n-\nicefrac{1}{2},\mathrm{K}}$, which further propagates through the remaining half drift. For many choices of integrators, such as \textsc{FastPM} and the standard symplectic leapfrog integrator written in terms of canonical variables, one has $\alpha \equiv 1$, and this step is therefore not necessary (see e.g.\ \cite[Eqs.~49]{Li2022DifferentiableMethod}).
Third, note that the update for $\overcirc{\vecb{X}}^n$ involves a VJP between the adjoint velocity $\overcirc{\vecb{V}}^{n-\nicefrac{1}{2}}$ and the Jacobian of the accelerations $\bnabla_{\vecb{X}} \vecb{A}(\vecb{X}^{n-\nicefrac{1}{2}})$. This is the reason why the primal variables must also participate in the backward-in-time simulation, requiring us to propagate the 4-tupel $(\vecb{X}, \vecb{V}, \overcirc{\vecb{X}}, \overcirc{\vecb{V}})$ to the initial time of the simulation. Rather than explicitly building the Jacobian matrix of the accelerations, we use \lstinline!jax.vjp! to obtain both $\vecb{A}(\vecb{X}^{n-\nicefrac{1}{2}})$ and $\bnabla_{\vecb{X}} \vecb{A}(\vecb{X}^{n-\nicefrac{1}{2}})$, where the former and latter are used for kicking the primal velocities and adjoint positions, respectively.

The remaining drift half is similar to the first one, for which reason we omit it here. In summary, this computation shows that `discretise-then-optimise' reverse-mode autodiff though the leapfrog integrator yields a backward-in-time leapfrog scheme for the adjoint variables. From this point of view, the key insight motivating the adjoint method is that although $\vecb{X}$ is required for computing the adjoint kick, it is not necessary to store checkpoints of the primal variables $(\vecb{X}, \vecb{V})$ during the forward pass: rather, one can let $(\vecb{X}, \vecb{V})$ participate in the backward pass together with the adjoint variables $(\overcirc{\vecb{X}}, \overcirc{\vecb{V}})$ and propagate them from their final to their initial states. Once there, the values of $\overcirc{\vecb{X}}^0$ and $\overcirc{\vecb{V}}^0$ represent the loss gradient w.r.t.\ the initial conditions. This point defines the programming boundary where \codename returns the custom VJPs evolved through the $N$-body simulation, and standard autodiff takes over again, until the white noise field $w$ and/or cosmological parameters $\theta$ are reached in the computing graph.

We summarise the trajectories of the primal and adjoint variables during the backward-in-time simulation in Table~\ref{tab:adjoint_dkd}, where the former immediately follow from flipping the arrow of time in Eqs.~\eqref{eq:dkd}. For completeness, we also include the adjoints w.r.t.\ the temporal parameters $\tau_1, \tau_2, \alpha, \beta$ -- also indicated with a star -- which are similarly derived from VJPs. Note that unlike the adjoint variables $(\overcirc{\vecb{X}}, \overcirc{\vecb{V}})$, which are overwritten in every step, the (scalar) adjoints for the temporal variables are stored for all steps.

\begin{table}[ht]
\centering
\renewcommand{\arraystretch}{1.6}
\setlength{\tabcolsep}{4pt}
\begin{tabular}{@{}ll@{}}
\toprule
\textbf{Step} & \textbf{Equation} \\
\midrule
Primal drift & $\vecb{X}^{n-\nicefrac{1}{2}} = \vecb{X}^{n} - \tau_2 \vecb{V}^{n}$ \\
Adjoint drift & $\overcirc{\vecb{V}}^{n-\nicefrac{1}{2}} = \overcirc{\vecb{V}}^{n} + \tau_2 \overcirc{\vecb{X}}^{n}$ \\
Adjoint for $\tau_2$ & $\overcirc{\tau}_2^{n-\nicefrac{1}{2}} = \overcirc{\vecb{X}}^{n} \cdot \vecb{V}^{n}$ \\
\addlinespace
Primal kick & $\vecb{V}^{n-1} = \alpha^{-1} \left(\vecb{V}^{n} - \beta \vecb{A}(\vecb{X}^{n-\nicefrac{1}{2}})\right)$ \\
Adjoint kick & $\overcirc{\vecb{X}}^{n-1} = \overcirc{\vecb{X}}^{n} + \beta \overcirc{\vecb{V}}^{n-\nicefrac{1}{2}} \cdot \bnabla_{\vecb{X}}\vecb{A}(\vecb{X}^{n-\nicefrac{1}{2}})$ \\
Rescale adjoint velocity & $\overcirc{\vecb{V}}^{n-\nicefrac{1}{2}, \text{K}} = \alpha \overcirc{\vecb{V}}^{n-\nicefrac{1}{2}}$ \\
Adjoint for $\alpha$ & $\overcirc{\alpha}^{n-\nicefrac{1}{2}} = \overcirc{\vecb{V}}^{n-\nicefrac{1}{2}} \cdot \vecb{V}^{n}$ \\
Adjoint for $\beta$ & $\overcirc{\beta}^{n-\nicefrac{1}{2}} = \overcirc{\vecb{V}}^{n-\nicefrac{1}{2}} \cdot \vecb{A}(\vecb{X}^{n-\nicefrac{1}{2}})$ \\
\addlinespace
Primal drift & $\vecb{X}^{n-1} = \vecb{X}^{n-\nicefrac{1}{2}} - \tau_1 \vecb{V}^{n-1}$ \\
Adjoint drift & $\overcirc{\vecb{V}}^{n-1} = \overcirc{\vecb{V}}^{n-\nicefrac{1}{2}, \text{K}} + \tau_1 \overcirc{\vecb{X}}^{n-1}$ \\
Adjoint for $\tau_1$ & $\overcirc{\tau}_1^{n-\nicefrac{1}{2}} = \overcirc{\vecb{X}}^{n-1} \cdot \vecb{V}^{n-1}$ \\
\bottomrule
\end{tabular}
\caption{Backward-in-time dynamics for the primal and adjoint (indicated with a star) variables for the DKD leapfrog integrator in Eq.~\eqref{eq:dkd}.}
\label{tab:adjoint_dkd}
\end{table}

\subsection{Discreteness suppression}
\label{sec:discreteness}
The role of the particles in $N$-body simulations is twofold: first, they act as tracers, which follow the characteristics and allow studying the evolution of gravitational collapse. In addition, the particles themselves source the gravitational potential, which governs the dynamics. 
However, approximating the continuous potential via the discrete $N$-body particles causes the particles to systematically deviate from the underlying continuous dynamics -- most notably at early times (e.g.\ \cite{Joyce:2005, Marcos:2006, Garrison:2016}).

In \codename, we implemented various discreteness reduction schemes, which remedy different imprints of the discretisation in the force computation:
\begin{itemize}
    \item higher-order mass assignment kernels (CIC, TSC, PCS, see Fig.~\ref{fig:kernels_fourier}) \cite{HockneyEastwood, 2004JCoPh.197..253C}
    \item de-aliasing accelerations by means of interlaced grids \cite{Chen:1974, HockneyEastwood, sefusatti2016accurate}
    \item deconvolution of the mass assignment kernel in Fourier space
    \item sheet-based particle resampling via spectral (or trilinear) interpolation of the displacement field \cite{Abel:2012, Shandarin:2012, Hahn:2013, stucker2020simulating}
\end{itemize}

For a more in-depth explanation of these techniques, we refer the interested reader to the Appendix of Ref.~\cite{List2024StartingBang}. There, we show that, in combination with LPT-informed integrators (see Sec.~\ref{sec:timestepping}), these techniques enable initialising cosmological simulations at time $a = 0$ -- without requiring LPT. 
The sheet-based resampling technique is particularly useful for accessing the fluid regime at early times. In this approach, the number of particles in the density computation is increased by introducing artificial particles, whose locations are determined by interpolating the displacement of the original particles in Lagrangian space.
An alternative to higher-order mass assignment kernels is given by NUFFT (see Sec.~\ref{sec:force} above), which uses a kernel with even wider support in order to guarantee spectral accuracy.

\subsection{Miscellaneous features}
For completeness, let us mention some other features currently implemented in \codename. For comparability with full $N$-body simulations, we implemented the random number generator of the popular \textsc{N-GenIC} initial condition generating tool \cite{springel2005simulations}. Creating the same realisation of a simulated universe as produced by \textsc{N-GenIC} is as easy as 
\lstinline!white_noise = dj.get_ngenic_noise(seed=...)!. Moreover, snapshots can be stored in \textsc{Gadget} \cite{Springel:2005} and \textsc{Swift} \cite{schaller2024swift} snapshot format: thus, one can e.g.\ perform the first few steps with \codename and then switch to a TreePM code at later times. 
For analysing snapshots, \codename includes a fast and differentiable (cross-)power spectrum routine, as well as a highly performant and differentiable bispectrum estimator adapted from \textsc{BFast}.\footnote{https://github.com/tsfloss/BFast}.
Furthermore, most components of \codename are implemented for $1$, $2$, and $3$ dimensions, enabling rapid experimentation in lower dimensions, e.g.\ for benchmarking perturbation theory against $N$-body simulations, testing numerical accuracy and convergence, developing or debugging machine learning components with reduced computational cost, and for educational purposes.

\subsection{Usage example}
To showcase \codename's ease of use, we include a simple exemplary $N$-body workflow:
\begin{lstlisting}
    from discodj import DiscoDJ
    dj = DiscoDJ(dim=3, res=256, boxsize=500.0, precision="single",
                 cosmo="Planck18EEBAOSN")
    dj = dj.with_timetables()
    dj = dj.with_linear_ps(transfer_function="DiscoEB")
    dj = dj.with_ics()
    dj = dj.with_lpt(n_order=2)
    X, P, a = dj.run_nbody(a_ini=0.02, a_end=1.0, n_steps=10, res_pm=512, 
                           stepper="bullfrog")
    delta = dj.get_delta_from_pos(X)
\end{lstlisting}
\codename is written in an object-oriented way, albeit complying with \textsc{Jax}'s functional programming paradigm that mandates pure functions. Therefore, methods that would change an internal state of a \texttt{DiscoDJ} object instead return a new object (which, however, does not imply that all the object's attributes need to be copied in memory). Note that most of the steps in the code snippet above can be further customised by passing additional parameters (or more flexible parameter values, such as a dictionary for the cosmology). 

Computing a partial derivative $\partial f/\partial x$ through \codename requires wrapping the forward model inside a function \texttt{f}, which takes the quantity $x$ as in input parameter. Then, one can call \lstinline!jax.grad(f)(x_eval)! (and similar functions) to evaluate $\partial f/\partial x$ at the value \lstinline!x_eval!.
For completeness, let us mention that the \codename class is registered as a \texttt{PyTree}, allowing users to differentiate w.r.t.\ \codename object, where the cosmological parameters and white noise are declared as the dynamic values with respect to which gradients are computed.

\section{Validation and performance}
\label{sec:validation}

\begin{table}[h]
\centering
\renewcommand{\arraystretch}{1.0}
\setlength{\tabcolsep}{4pt}
\begin{tabular}{lc}
\toprule
Parameter & Value \\
\midrule
$\Omega_m$ & $0.3158$ \\
$\Omega_b$ & $0.0494$ \\
$h$        & $0.67321$ \\
$n_s$      & $0.9661$ \\
$\sigma_8$ & $0.8102$ \\
\bottomrule
\end{tabular}
\caption{Fiducial flat $\Lambda$CDM cosmology adopted for the experiments in this work.}
\label{tab:cosmology}
\end{table}

\begin{table}[h]
\centering
\resizebox{\textwidth}{!}{%
\begin{tabular}{@{}lcl@{}}
\toprule
\textbf{Parameter} & \textbf{Default value} & \textbf{Explanation} \\
\midrule
\texttt{stepper}              & \lstinline!"bullfrog"!  & Time integrator (\lstinline!"bullfrog"!, \lstinline!"fastpm"!, \lstinline!"symplectic"!) \\
\texttt{n\_steps}             & \lstinline!100!         & Number of time steps \\
\texttt{method}               & \lstinline!"pm"!        & Force computation method (\lstinline!"pm"!, \lstinline!"nufftpm"!) \\
\texttt{res\_pm}              & \lstinline!2 * dj.res!    & Grid resolution per dim. $\sqrt[3]{N_g}$ \\
\texttt{time\_var}            & \lstinline!"D"!         & Time variable (\lstinline!"D"!, \lstinline!"a"!, \lstinline!"log\_a"!, \lstinline!"superconft"!) \\
\texttt{antialias}            & \lstinline!0!             & Anti-aliasing order via grid interlacing (\lstinline!0!, \lstinline!1!, \lstinline!2!, \lstinline!3!) \\
\texttt{grad\_kernel\_order}  & \lstinline!4!             & Gradient order (\lstinline!0!: $\mathrm{i}\vecb{k}$, \lstinline!2/4/6!: finite difference kernel) \\
\texttt{laplace\_kernel\_order} & \lstinline!0!            & Laplacian order (\lstinline!0!: $-k^2$, \lstinline!2/4/6!: finite difference kernel) \\
\texttt{n\_resample}          & \lstinline!1!             & Sheet-based resampling: copies of each particle per dim. \\
\texttt{worder}               & \lstinline!2!             & Order of the MAK (\lstinline!2!: CIC, \lstinline!3!: TSC, \lstinline!4!: PCS) \\
\texttt{deconvolve}           & \lstinline!False!         & Deconvolve the MAK in the force computation \\
\texttt{a\_ini}                & \lstinline!1 / (1 + 50)!  & Initial scale factor \\
\texttt{nlpt\_order\_ics}      & \lstinline!2!            & LPT order of the initial conditions \\
\bottomrule
\end{tabular}
}
\caption{Default parameter settings for the \lstinline!run_nbody! method in our numerical experiments. The acronym MAK stands for mass assignment kernel. The default value for the PM grid size per dimension is $N_g^{\nicefrac{1}{3}} = $ \lstinline!2 * dj.res! $= 2 N^{\nicefrac{1}{3}}$. In the dedicated numerical studies, these parameters are systematically varied. Also, the default particle resolution is $N = 512^3$, and corner modes are zeroed. However, since these settings need to be specified when initialising the \codename object and setting up the initial conditions, rather than in the \lstinline!run_nbody! method, they are not listed above. Instead of using LPT-based initial conditions, it is also possible to pass the parameter \lstinline!ic_method = "bullfrog"!, in which case an single discreteness-reduced time step will be performed from $a = 0$ to the desired initial time. The \lstinline!worder! parameter only affects local mass assignment; for NUFFT, there are separate parameters for specifying the settings (not listed above).}
\label{tab:default_parameters}
\end{table}

In this section, we validate the efficacy of \codename and demonstrate its computational efficiency. First, we will compare the results with our custom VJP and JVP implementations to the \textsc{Jax} default gradients to confirm the correctness of our implementation. In the reverse-mode case (relying on the VJP), we will further validate our adjoint method implementation against standard autodiff. We proceed by studying convergence in time and resolution through summary statistics. Then, we analyse the effect of different force computation settings on the accuracy. Finally, we measure the runtime of the PM simulations for different resolutions and settings.

In all our experiments, we use a flat $\Lambda$CDM cosmology with the parameters listed in Table~\ref{tab:cosmology}, and our fiducial white noise field has \textsc{N-GenIC} seed \texttt{54321}. All gradients are evaluated at these points in parameter space. The default numerical settings for our PM simulations in \codename are listed in Table~\ref{tab:default_parameters}.

For benchmarking the performance of \codename, we use the popular TreePM code \textsc{Gadget-4} \cite{Springel2020}. The \textsc{Gadget-4} reference runs were initialised with \textsc{Monofonic} \cite{Michaux:2021} at $z = 70$ with 2LPT and used $N = 1024^3$ particles and a PM grid size of $N_g = 2048^3$.

\subsection{Correctness of the adjoint method and custom derivatives}
\begin{figure}
    \centering
    \includegraphics[width=1\linewidth]{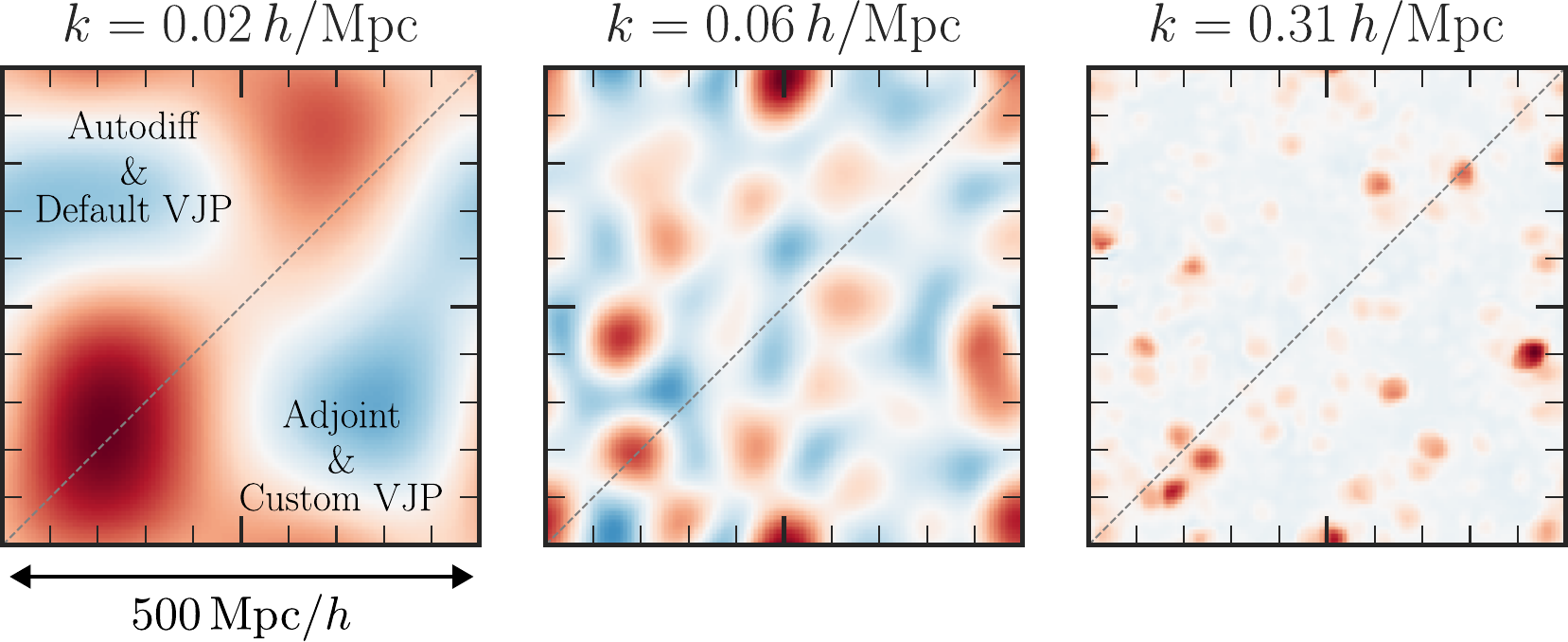}
    \caption{Thin slice of the gradient of the non-linear power spectrum w.r.t.\ the white noise field $\partial P/\partial w$ through a \codename simulation with 10 time steps using the \textsc{BullFrog} stepper, evaluated for three different $k$-bins. The results in the upper left corners have been computed using standard (discretise-then-optimise) reverse-mode autodiff and the default VJP for the \texttt{gather} and \texttt{scatter} operations for the density computation in the simulation, whereas the lower right corners show the results with the adjoint (optimise-then-discretise) method and custom VJP operations. The transition is smooth, confirming the correctness of our implementation (see Fig.~\ref{fig:dPk_dw_hist} for quantitative results). The colour scale is normalised individually for each $k$-bin.
    }
    \label{fig:dPk_dw}
\end{figure}

\begin{figure}
    \centering
    \includegraphics[width=0.66\linewidth]{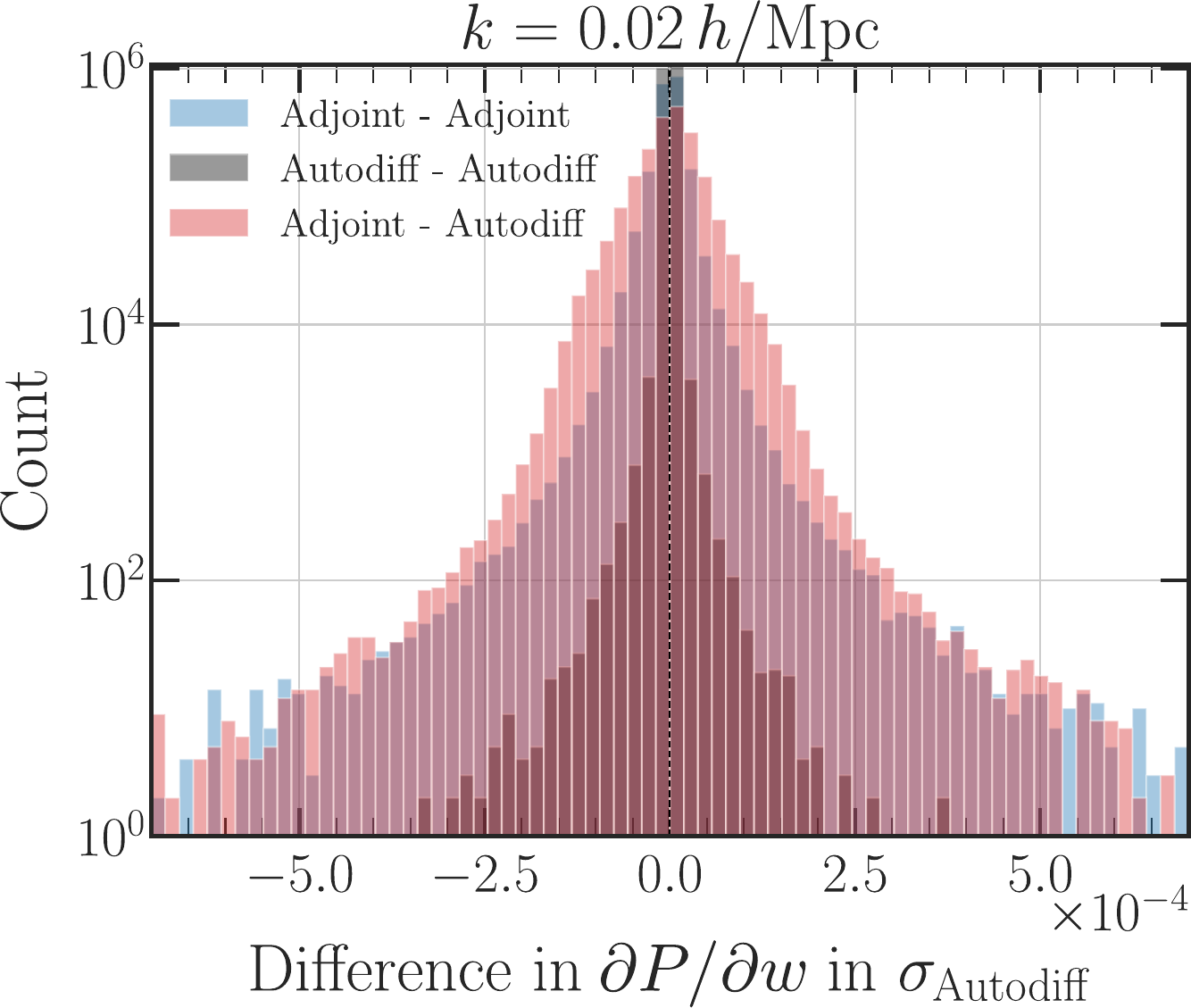}
    \caption{Histogram of differences in the power spectrum gradient $\partial P/\partial w$ for a single $k$-bin, normalised by the standard deviation of the autodiff result, for the same scenario as in Fig.~\ref{fig:dPk_dw}.
    We compare two independent adjoint runs (blue), two autodiff runs (grey), and a cross-comparison between adjoint and autodiff (red). 
    The vertical dashed line marks zero difference. The backwards integration leads to somewhat noisier gradients with the adjoint method than with standard autodiff; still, the deviations are small in comparison to the magnitude of $\partial P/\partial w$ itself.}
    \label{fig:dPk_dw_hist}
\end{figure}

\subsubsection*{Reverse mode w.r.t.\ initial phases}
First, we demonstrate the correctness of our custom VJP and adjoint method implementation. To this aim, we compute the gradient of the non-linear power spectrum $P(k)$ w.r.t.\ the initial white noise field $w$ (in real space) through a 10-step simulation with $N = 128^3$ particles (at PM grid resolution $N_g = 256^3$) at single precision, for two settings: 1) using the default VJPs of the scatter and gather operations computed by \textsc{Jax} and standard reverse-mode autodiff through the simulation and 2) using our custom VJP operations (see Appendix~\ref{sec:custom_VJP_JVP}) and computing gradients with the adjoint method (see Sec.~\ref{sec:adjoint}). As expected, the values of $\partial P/\partial w$ for small $k$ show sensitivity to large patches in Lagrangian space, and vice versa; see Fig.~\ref{fig:dPk_dw}.

We note that on the \textsc{Nvidia} A100 ($40 \, \mathrm{GB}$) on which we perform this test, computing the gradients using $N = N_g = 256^3$ without the adjoint method and default VJPs causes an out-of-memory error. With custom VJPs, the gradients for $N_{\mathrm{steps}} = 10$ time steps fit in memory; however, due to the $O(N_{\mathrm{steps}})$ scaling of the VJP with standard autodiff, increasing the number of time steps eventually causes an out-of-memory error, too. With the adjoint method and custom VJPs, arbitrary numbers of steps are possible. 

To quantify the agreement, we also plot a histogram of the pairwise differences between the adjoint vs standard autodiff gradients in Fig.~\ref{fig:dPk_dw_hist}, for the $k$-bin centred around $k = 0.02 \, h / \mathrm{Mpc}$, in units of the standard deviation of $\partial P/\partial w$ as computed with standard autodiff. In addition, we also show the pairwise differences between two independent calls of the gradient computation with either method. Since computations are performed in parallel on GPUs, the results of operations are generally non-deterministic.\footnote{It is possible to set \lstinline!os.environ["XLA_FLAGS"] = "--xla_gpu_deterministic_ops=true"! to enforce determinism in \textsc{Jax}; however, this may affect performance by preventing certain XLA optimisations.} Unsurprisingly, the pairwise differences between two autodiff runs are the smallest, as no additional errors are introduced by the backward simulation (see also \cite{Li2022DifferentiableMethod}). The distribution of Adjoint $-$ Adjoint is somewhat thinner around $0$ than Adjoint $-$ Autodiff. Compared to the magnitude of $\partial P/\partial w$, the magnitudes of all errors are modest. Note that all experiments presented here use single precision. If very high-accuracy gradients are required, \codename can also be used in double-precision mode by passing \lstinline!precision = "double"!. For completeness, let us also mention that computing the final density on which the power spectrum is based with a higher-order kernel such as PCS further reduces the scatter between multiple autodiff computations, however not between multiple adjoint realisations.
For a more detailed study on the reproducibility of derivatives in \textsc{Jax} in the context of $N$-body simulations, we refer the reader to Ref.~\cite{Li2022DifferentiableMethod}.

\subsubsection*{Forward mode w.r.t.\ cosmological parameters}
\begin{figure}
    \centering
    \includegraphics[width=0.66\linewidth]{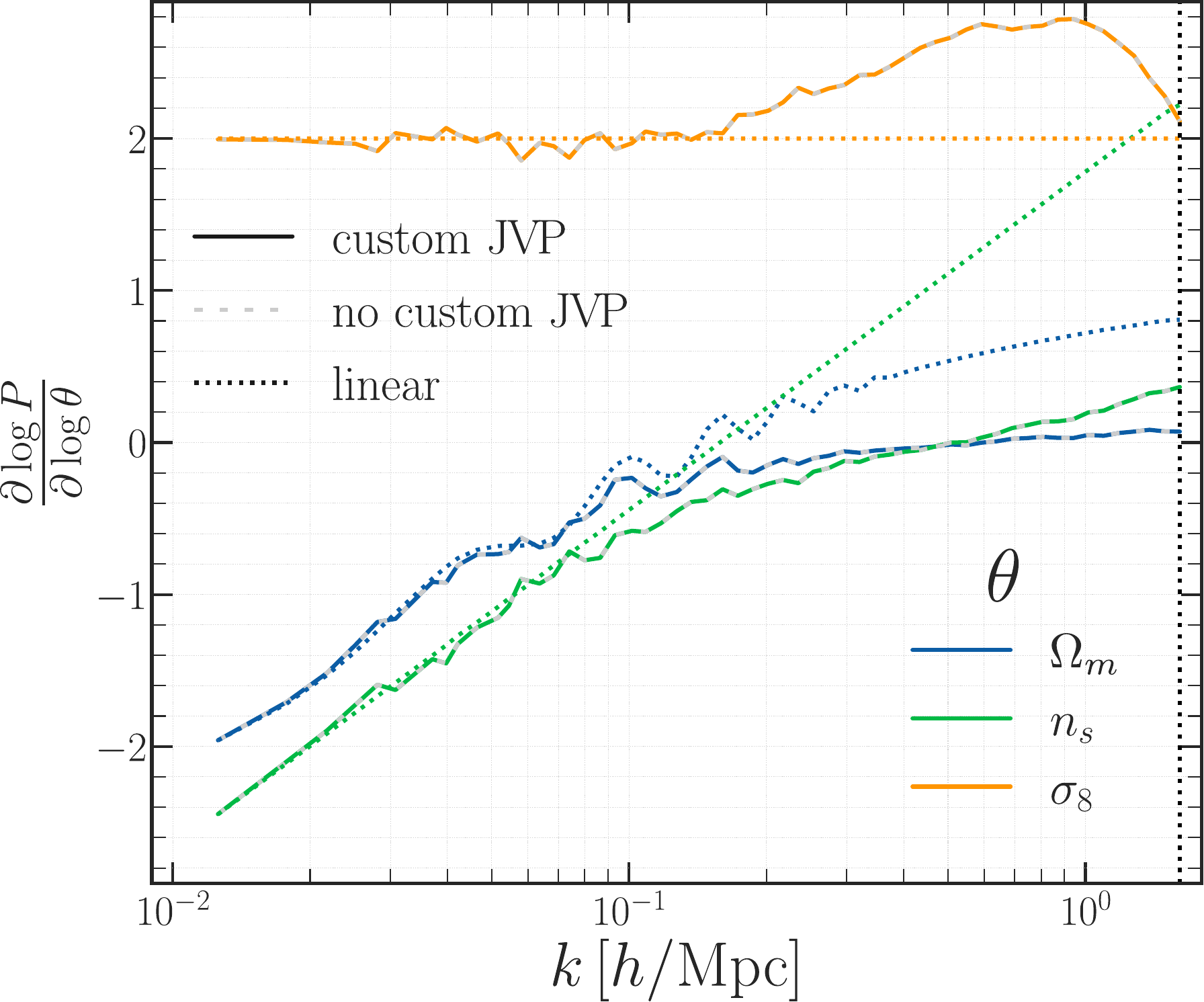}
    \caption{Logarithmic derivative $\partial \log P(k) / \partial \log \theta$  of the power spectrum $P(k)$ w.r.t.\ different cosmological parameters $\theta$ at $z = 0$. We compare the non-linear power spectrum (computed with the Einstein--Boltzmann + $N$-body modules) and the linear power spectrum (computed with the Einstein--Boltzmann module), obtained using forward-mode autodiff. In the non-linear case, our custom JVP implementation (solid lines) agrees with the default \textsc{Jax} implementation (grey dashed lines). The dotted vertical line marks the particle Nyquist mode $k_{\mathrm{Nyquist}}$.    
    }
    \label{fig:dPk_dtheta}
\end{figure}

Having validated the agreement between the gradients w.r.t.\ the white noise field, we now consider the derivatives w.r.t.\ cosmological parameters. Specifically, we compute the power spectrum gradient w.r.t.\ $\Omega_m$, $\sigma_8$, and $n_s$. Notably, we restrict the cosmology to be flat; therefore, gradients w.r.t.\ $\Omega_m$ point in the direction of decreasing $\Omega_\Lambda$ such that $\Omega_m + \Omega_\Lambda = 1$. Moreover, we keep $\Omega_b = 0.0494$ fixed. For stability reasons, the Einstein--Boltzmann module of \codename currently uses double precision. However, users have the option of switching to single precision for the $N$-body module. This mixed-mode precision allows for highly accurate linear power spectrum computations (which are usually not memory-limited), while reducing the memory footprint of the $N$-body part by a factor of two. Gradients propagated through the combined Einstein--Boltzmann + $N$-body pipeline are in single precision in this case.

We validate the differentiation through Einstein--Boltzmann + $N$-body by computing the gradient of the power spectrum w.r.t.\ cosmological parameters, while keeping the white noise field fixed. Specifically, we consider $\theta \in \{\Omega_m, \sigma_8, n_s\}$ for this experiment. The simulation setup is the same as for the derivatives w.r.t.\ the white noise field.
Since the resulting Jacobian matrix is tall (i.e.\ few cosmological parameters are mapped to the power spectrum in many bins), we use forward-mode differentiation. Similarly to the previous experiment where we verified the correctness of our custom VJP implementation and the adjoint method, we now compute the derivatives 1) using the default JVPs of the scatter and gather operations, and 2) using our custom JVPs. 

Figure \ref{fig:dPk_dtheta} shows the results. For comparison, we also plot the gradients of the \textit{linear} power spectrum for the same white noise realisation. These have been computed completely analogously to their non-linear counterparts; however, propagating the tangents only through the Einstein--Boltzmann module and not the $N$-body one. 
In all cases, the agreement between our custom JVP implementation and the \textsc{Jax} results is excellent.

For the matter density $\Omega_m$ (blue), we find that $\partial \log P(k) / \partial \log \Omega_m$ is negative on large scales. This behaviour arises from the fact that $\sigma_8$  is kept fixed; moreover, increasing $\Omega_m$ at fixed $z = 0$ leads to a younger universe, reducing the growth factor and hence suppressing the late-time power spectrum. On smaller scales, the effect of enhanced gravitational clustering begins to compensate, leading to a shallower derivative that approaches zero in the deeply non-linear regime. Comparing the linear and the non-linear case, the derivative rises more slowly in the latter case, indicative of the fact that halo formation causes the power spectrum to become more sensitive to internal halo properties and less so to the cosmological background.

The spectral index $n_s$ (green) tilts the shape of the initial power spectrum. Consequently, the derivative $\partial \log P(k) / \partial \log n_s$ is negative and positive on large and scales, respectively. Specifically, since $P_{\mathrm{lin}}(k) \propto (k/k_{\mathrm{pivot}})^{n_s - 1}$, the logarithmic derivative of the linear power spectrum scales linearly with $\log k$. The non-linear evolution leads to a flattening for $k \gtrsim 0.1 \, h / \mathrm{Mpc}$ and a shift of the zero-crossing point to higher $k$ compared to the linear theory.

The logarithmic derivative w.r.t.\ $\sigma_8$ (orange) approaches the theoretical expectation of $\partial \log P_\mathrm{lin}(k) / \partial \log \sigma_8 \equiv 2$ on large (linear) scales, consistent with the fact that the linear spectrum scales quadratically with $\sigma_8$. On smaller scales, the non-linear spectrum exhibits an enhanced sensitivity to $\sigma_8$, reaching values above 2 due to non-linear mode coupling, which amplifies structure formation beyond the linear prediction.

\subsection{Convergence in terms of time stepping}
\label{sec:time_convergence}
\begin{figure}
    \centering 
    \includegraphics[width=.75\linewidth]{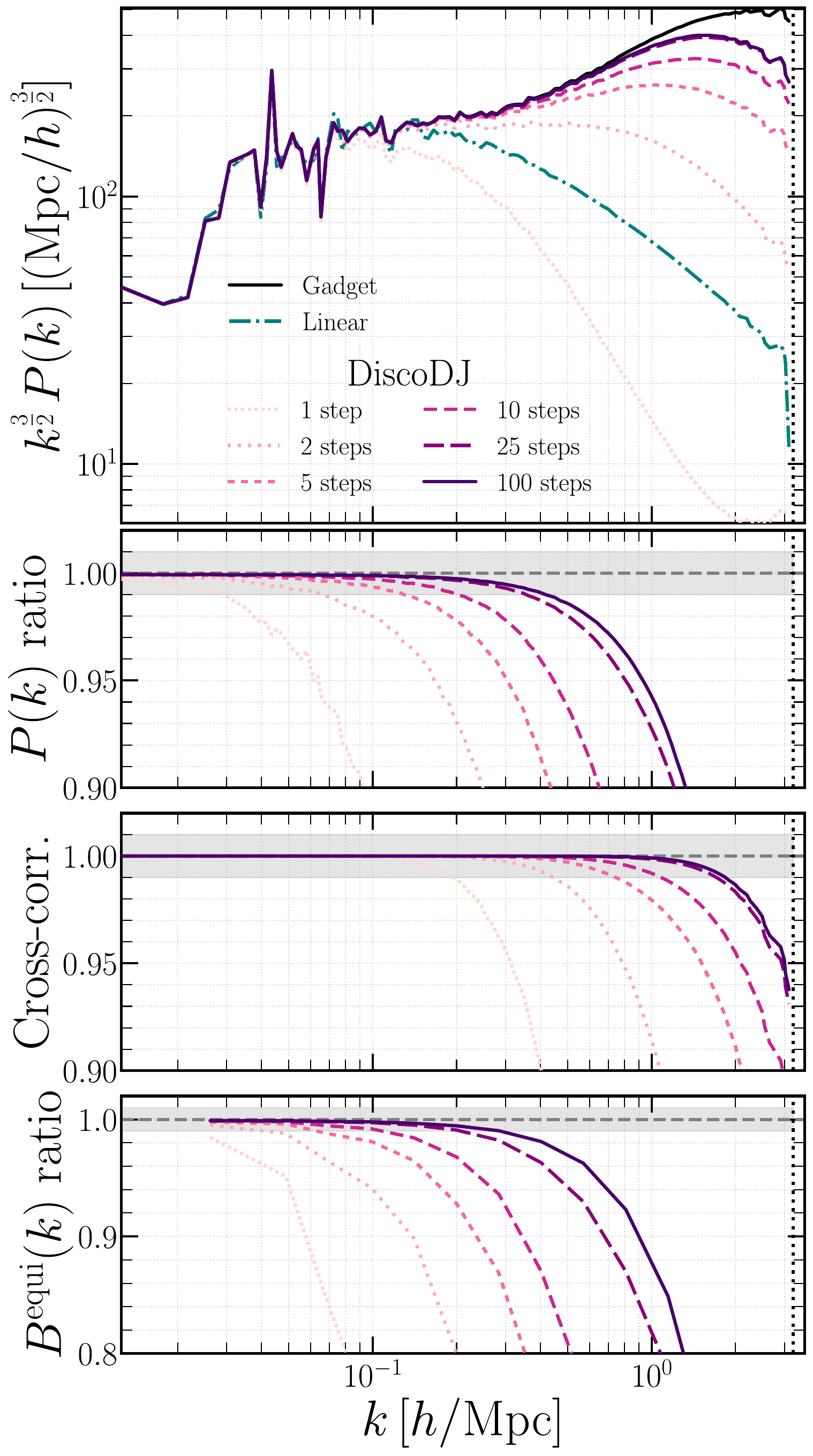}
    \caption{Convergence of the power spectrum, cross-correlation, and equilateral bispectrum at $z = 0$ as a function of the number of time steps for a box size of $L = 500 \, \mathrm{Mpc} / h$ and $N = 512^3$ particles. The reference simulation was performed using \textsc{Gadget-4} with 1024$^3$ particles. The teal dash-dotted line shows the linear power spectrum.}
    \label{fig:convergence_in_time}
\end{figure}
We now study the convergence in time as the number of time steps increases. We use the 2LPT-informed \textsc{BullFrog} integrator, recently introduced in Ref.~\cite{Rampf_2025}, which is the default in \codename. Focusing on the time integration aspect, the convergence study in that reference was performed relative to a time-converged PM simulation at the same resolution. In contrast, we now benchmark the performance of \codename against \textsc{Gadget-4} \cite{Springel2020}. We consider a box size of $L = 500 \, \mathrm{Mpc} / h$; results for a larger and a smaller box are presented in Appendix~\ref{sec:boxsize}.
For our \codename runs, we use $N = 512^3$ particles, $N_g = 1024^3$ grid cells for the PM-based force calculation, and the default parameters in Table~\ref{tab:default_parameters}.

Figure~\ref{fig:convergence_in_time} shows the power spectrum, cross-correlation (or normalised cross-power spectrum), and equilateral bispectrum at $z = 0$ in comparison with the \textsc{Gadget-4} reference for different numbers of time steps. The time-converged simulation with 100 steps achieves sub-per-cent accuracy in terms of $P(k)$ up to $k \lesssim 0.4 \, h / \mathrm{Mpc}$. (This can be pushed to $k = 1 \, h / \mathrm{Mpc}$ by using an $L = 100 \, \mathrm{Mpc} / h$ box, see Appendix~\ref{sec:boxsize}, or of course also by increasing the number of particles.) Performance with 25 steps is only slightly worse in terms of $P(k)$; however, the difference between 25 and 100 steps is slightly more pronounced for the equilateral bispectrum, which for the time-converged simulation is sub-per-cent accurate up to $k = 0.3 \, h / \mathrm{Mpc}$. The cross-correlation exceeds $99\%$ even at $k = 1 \, h / \mathrm{Mpc}$ already with 10 steps. The remaining residuals between \codename and \textsc{Gadget-4} are due to the lack of spatial and force resolution, and cannot be significantly reduced further by increasing the number of time steps. 

For reference, we list the number of \textsc{BullFrog} steps required to reach a certain error in terms of the power spectrum ratio at different scales (with $N = 512^3$ particles and default settings): \\[-0.3cm]

\renewcommand{\arraystretch}{1.0}
\setlength{\tabcolsep}{4pt}
\begin{center}
\begin{tabular}{lccc}
\toprule
Accuracy & $k=0.1\,h/\mathrm{Mpc}$ & $k=0.3\,h/\mathrm{Mpc}$ & $k=0.5\,h/\mathrm{Mpc}$ \\
\midrule
$<5\%$  & 2 steps & 5 steps  & 12 steps \\
$<1\%$  & 4 steps & 19 steps & -- \\
\bottomrule
\end{tabular}
\end{center}

The dash indicates that the target was not reached within 100 steps for this box size. Note that for a given setup, it might be possible to satisfy the error bounds with fewer steps, e.g.\ by adjusting the initial redshift (which is fixed to $z_{\mathrm{ini}} = 50$ here), customising the time step spacing, or using different force computation settings (e.g.\ with NUFFT) -- the above table is intended rather to provide some rough guidance for users. For a given scale on which the time-converged \codename solution satisfies the tolerance, we observe only a weak dependence of the minimum number of steps on the box size. For instance, with $L = 100 \, \mathrm{Mpc} / h$ instead of $L = 500 \, \mathrm{Mpc}/h$, the number of steps required for per-cent accuracy of the power spectrum at $k = 0.3 \, h / \mathrm{Mpc}$ reduces slightly from 19 to 17. In that case, per-cent accuracy at $k = 0.5 \, h / \mathrm{Mpc}$ is achieved with 46 steps, and $\sim 100$ steps yield per-cent accuracy up to $k = 1 \, h / \mathrm{Mpc}$ (see Fig.~\ref{fig:convergence_in_time_small} in the appendix).

\begin{figure}
    \centering
    \includegraphics[width=1\linewidth]{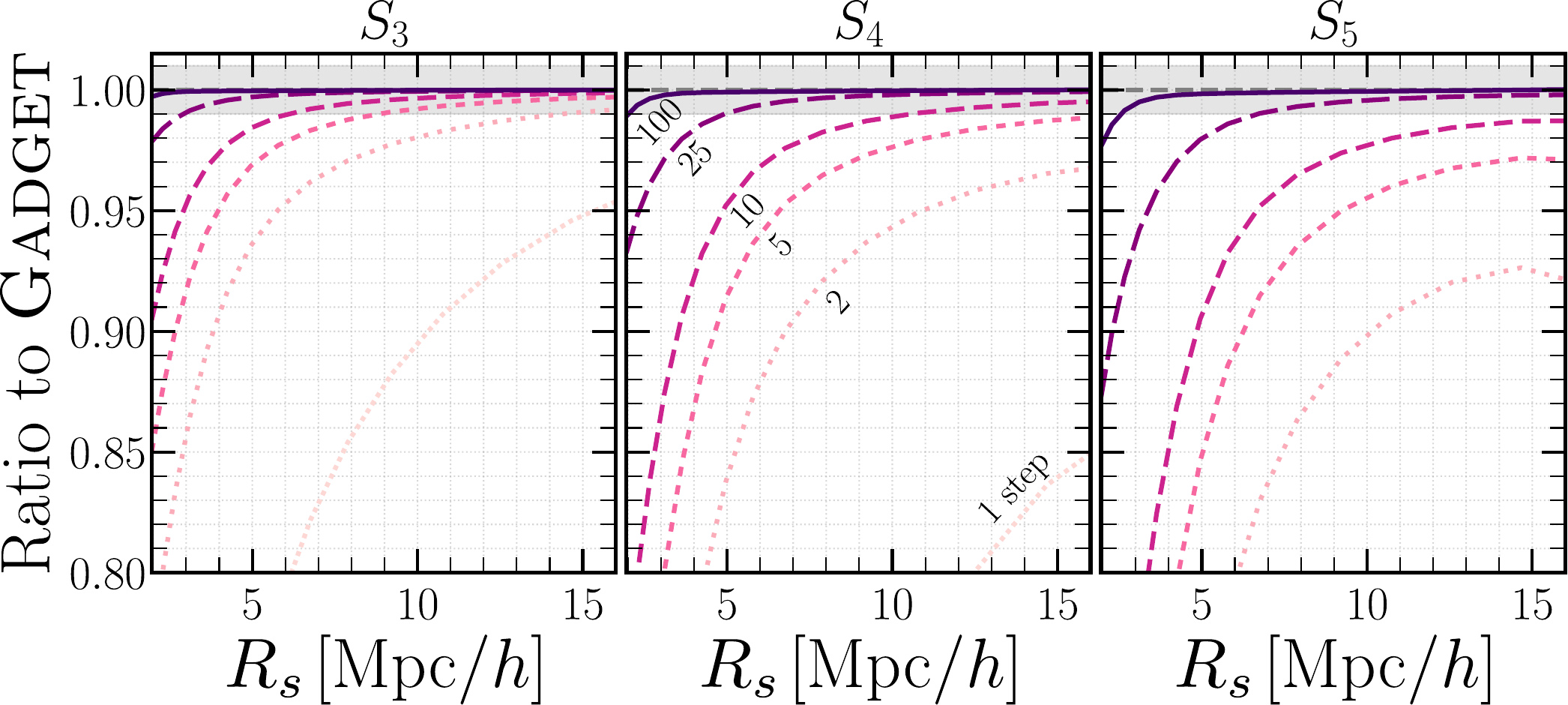}
    \caption{Reduced cumulants $S_3$, $S_4$, and $S_5$ of the $z = 0$ density field as a function of the Gaussian smoothing scale $R_s$ for different numbers of time steps (as indicated next to the curves in the $S_4$ panel), compared to the \textsc{Gadget-4} reference.}
    \label{fig:reduced_cumulants}
\end{figure}

Computing $n$-point correlation functions for $n > 3$ (or their Fourier versions, i.e.\ spectra beyond the bispectrum) quickly becomes infeasible. Instead, to measure the convergence of higher-order terms as the number of time steps increases, it is informative to study the cumulants of the smoothed $z = 0$ density field (e.g.\ \cite{1994A&A...291..697B}). Recall that for a mean-free field such as the density contrast $\delta$, the first cumulant vanishes by construction, while the next few are given by
\begin{align}
\kappa_2 = \langle \delta^2 \rangle, \quad
\kappa_3 = \langle \delta^3 \rangle, \quad
\kappa_4 = \langle \delta^4 \rangle - 3 \langle \delta^2 \rangle^2, \quad
\kappa_5 = \langle \delta^5 \rangle - 10 \langle \delta^3 \rangle \langle \delta^2 \rangle.
\end{align}
Thus, $\kappa_2$ is simply the variance, i.e.\ the integral of the power spectrum over all modes, whose convergence we analysed above. Similarly, the skewness $\kappa_3$ is the integral of the bispectrum over all triangle configurations, etc. For $n \geq 3$, these connected moments isolate the genuinely non-Gaussian contributions, removing the disconnected parts that are already determined by lower-order statistics.
It is customary to define the \emph{reduced cumulants} as
\begin{equation}
S_n := \frac{\kappa_n}{\kappa_2^{\,n-1}}, \qquad n \geq 3,
\end{equation}
which suppresses the trivial scaling with the overall fluctuation amplitude.

To determine the agreement between the \codename density field and the \textsc{Gadget-4} reference as a function of scale, we evaluate the reduced cumulants on Gaussian-smoothed density fields,
\begin{equation}
\delta_{R_s}(\vecb{x}) = \int W_{R_s}(|\vecb{x}-\vecb{y}|) \, \delta(\vecb{y}) \, \mathrm{d}^3y,  
\qquad 
W_{R_s}(r) = \frac{1}{(2\pi R_s^2)^{\nicefrac{3}{2}}} \exp\!\left(-\frac{r^2}{2 R_s^2}\right),
\end{equation}
where $R_s$ is the smoothing radius. In Fourier space, this corresponds to multiplying by $\exp(-k^2 R_s^2/2)$, which suppresses modes with $k \gtrsim 1/R_s$.

The results, shown in Fig.~\ref{fig:reduced_cumulants}, demonstrate that the reduced cumulants $S_3, S_4$, and $S_5$ measured from \codename approach the \textsc{Gadget-4} benchmark as the number of time steps increases. At large smoothing scales, the ratios are already close to unity even for few steps, while convergence is slower at smaller scales. Unsurprisingly, correctly capturing correlations of higher order becomes gradually more difficult, with respective errors for the 100-step simulation of $0.3\%$, $1.1\%$, and $2.3\%$ for $S_3$, $S_4$, and $S_5$ for the smallest considered smoothing scale $R_s = 1.95 \, \mathrm{Mpc} / h$.

\subsection{Convergence in terms of resolution}
\begin{figure}
    \centering
    \includegraphics[width=1\linewidth]{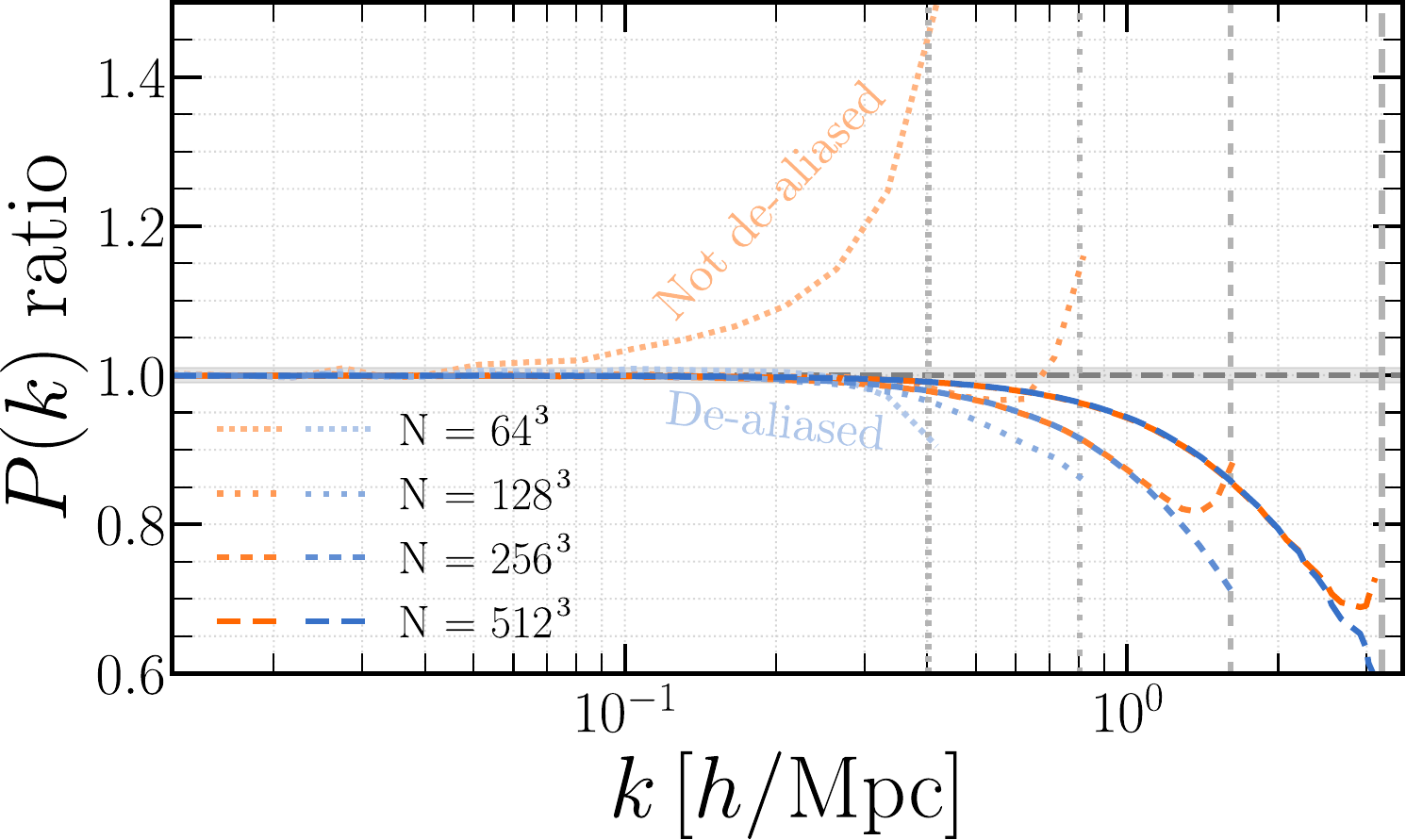}
    \caption{Power spectrum ratio w.r.t.\ the \textsc{Gadget-4} reference as a function of particle resolution $N$ -- with (blue) and without (orange) de-aliasing via grid interlacing (with two grids) during the simulation and the final density computation. The grey vertical lines indicate the particle Nyquist mode for each resolution.
    }
    \label{fig:convergence_in_res}
\end{figure}
Next, we study the effect of particle resolution in \codename. To this end, we compare the $z = 0$ power spectrum ratios resulting from \codename with our \textsc{Gadget-4} reference, varying the particle resolution $N$ in \codename. As will be seen in the following, aliasing severely affects the results at low spatial resolution; therefore, in addition to our standard simulation, we also perform a run with de-aliased force computation via an interlaced PM grid, see Sec.~\ref{sec:discreteness}. For the de-aliased simulations, we also de-alias the final $z = 0$ density from which the power spectrum is determined. The \textsc{Gadget-4} density at all considered resolutions is obtained using de-aliased CIC. In our experiments, we noticed that at low resolutions, the power spectrum becomes slightly noisy when setting $N_g = N$ in the final density computation; therefore, we use $\sqrt[3]N_g = \sqrt[3]N + 2$ here. We leave a detailed theoretical investigation of this phenomenon for future work.
Apart from the de-aliasing, we take the default settings for the $N$-body run in \codename (in particular a PM grid size of $N_g = 2^3 N$ during the simulation), and use $N_{\mathrm{steps}} = 100$ steps, thus considering the time-converged limit. Moreover, here we only consider the CIC mass assignment kernel.

Figure~\ref{fig:convergence_in_res} shows the results. As expected, aliasing affects the power spectrum in the $k$-range close to the Nyquist mode, deteriorating in magnitude as the particle resolution decreases. De-aliasing with a single additional grid (blue) is effective in suppressing the undesirable aliasing imprint. For $N \gtrsim 256^3$, aliasing only affects $k \gtrsim 1 \, h / \mathrm{Mpc}$ at this box size, making de-aliasing dispensable for many applications. As an alternative to interlaced grids, higher-order mass assignment kernels also reduce aliasing.

\subsection{Force computation}
\label{sec:force_results}
\begin{figure}
    \centering
    \includegraphics[width=1\linewidth]{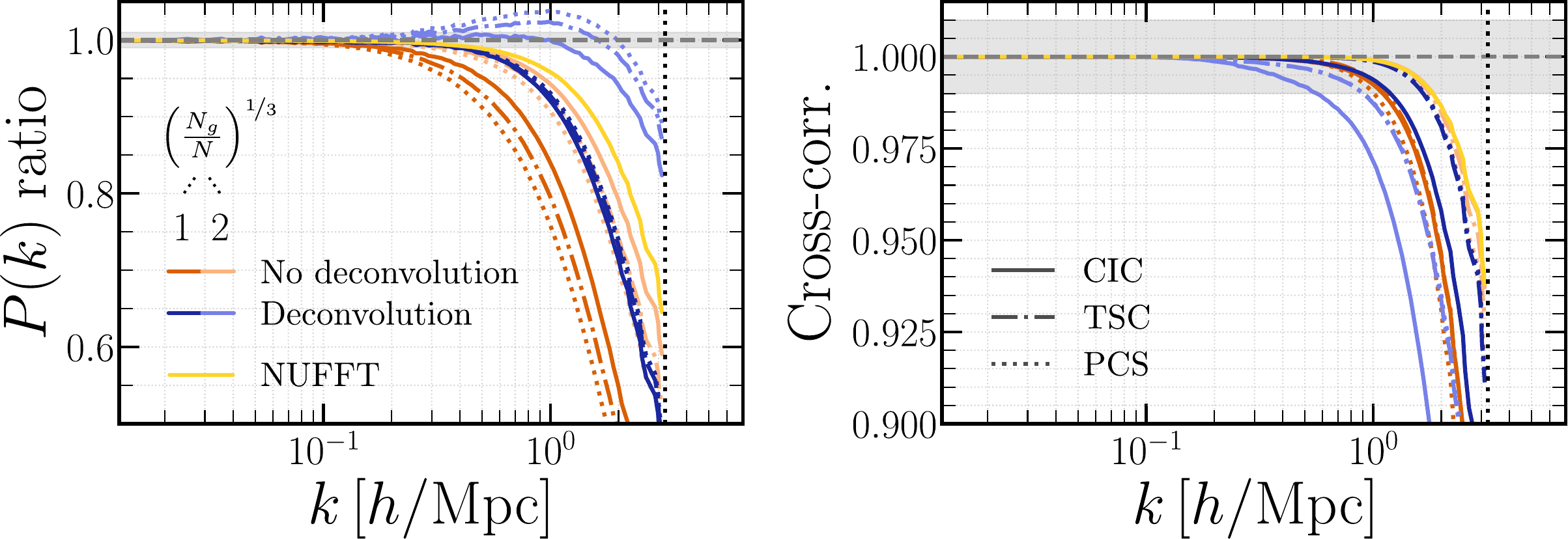}
    \caption{Power spectrum ratio and cross-correlation w.r.t.\ the \textsc{Gadget-4} reference at $z = 0$ for different force computation settings. 
    Blue lines show the results when deconvolving the density field in the force calculation of the $N$-body simulation with the mass assignment kernel squared (accounting for interpolation onto the grid and back to the particles), whereas no deconvolution is performed for the orange lines. Dark and light hues correspond to $N_g = N$ and $N_g = 2^3 N$, respectively. Different line styles indicate the order of the mass assignment kernel. For comparison, we also show in yellow a single line for the results with NUFFT-based force calculation, which performs best. For consistency, the final density is always computed with de-aliased and deconvolved CIC (however, this deconvolution kernel is the same for \codename and \textsc{Gadget-4} and therefore cancels out in the considered statistics).
    }
    \label{fig:mass_assignment_study}
\end{figure}
We now investigate the impact of different mass assignment kernels in the force computation of the $N$-body simulation, as well as the effect of deconvolving the density field with the corresponding kernel squared (accounting for the interpolation to and from the mesh). Figure~\ref{fig:mass_assignment_study} compares the power spectrum ratio and the cross-correlation w.r.t.\ the \textsc{Gadget-4} reference for various combinations of kernel order, PM mesh resolution, and deconvolution settings. For localised kernels such as CIC, TSC, and PCS, two settings stand out as performing best in our default scenario:
\begin{enumerate}[label=(\roman*),topsep=1pt,itemsep=-4ex,partopsep=1ex,parsep=1ex]
\item using a mesh resolution of $N_g = 2^3 N$ \textit{without} deconvolution (light orange curves),
\label{case:1}
\item using $N_g = N$ \textit{with} deconvolution (dark blue curves). 
\label{case:2}
\end{enumerate}
Option \ref{case:1} avoids small-scale noise amplification below the particle Nyquist mode -- which otherwise adversely affects the cross-correlation when deconvolving at high mesh resolution -- while the high mesh resolution mitigates the drop in power induced by the interpolation that acts as a low-pass filter. For this setting, the best $P(k)$ ratio is achieved with CIC, which introduces the least smoothing, and the cross-correlation is very similar for all kernel orders.

Interestingly, option \ref{case:2} achieves a very similar accuracy with lower mesh cost. The $P(k)$ ratios in this case are comparable for CIC, TSC, and PCS -- unsurprisingly, as the low-pass filtering effect is neutralised by the deconvolution -- and going to higher order improves the cross-correlation (the lines for TSC and PCS overlap). In practice, the optimal force computation settings depend on the box size and resolution, see Appendix~\ref{sec:boxsize}.

Using no deconvolution at $N_g = N$ (dark orange) causes an early power drop and is therefore not recommendable. Although the deconvolved $N_g = 2^3 N$ case (light blue) with CIC has a $P(k)$ ratio that remains in the 1\% error band to small scales $k \sim 1 \, h / \mathrm{Mpc}$, the cross-correlation is worst for this setting, indicating that the phases of these small-scale modes are misaligned w.r.t.\ the truth.

In Ref.~\cite[App.~A]{buisman2025differentiable}, we found that further increasing the grid resolution to e.g.\ $N_g = 4^3 N$ for relative small (e.g.\ $100 \, \mathrm{Mpc} / h$) box sizes can further alleviate the power drop on small scales and improve the halo mass function; however, for large box sizes, the increased discreteness reflected by many empty PM grid cells can actually make the results worse.

The NUFFT-based force calculation (yellow) yields superior agreement with the reference: both the power spectrum and the cross-spectrum drop later than with option \ref{case:1}. For completeness, we recall that some settings for the NUFFT-based computation differ from the localised case, as described in Sec.~\ref{sec:force}; e.g.\ the 
gradient kernel is taken to be $\mathrm{i}\vecb{k}$ in that case.
We therefore recommend localised MAK with option~\ref{case:1} or option~\ref{case:2}, depending on available memory and desired mesh size, or NUFFT as an attractive alternative.

\subsection{Runtime}
\begin{figure}
    \centering
    \includegraphics[width=0.75\linewidth]{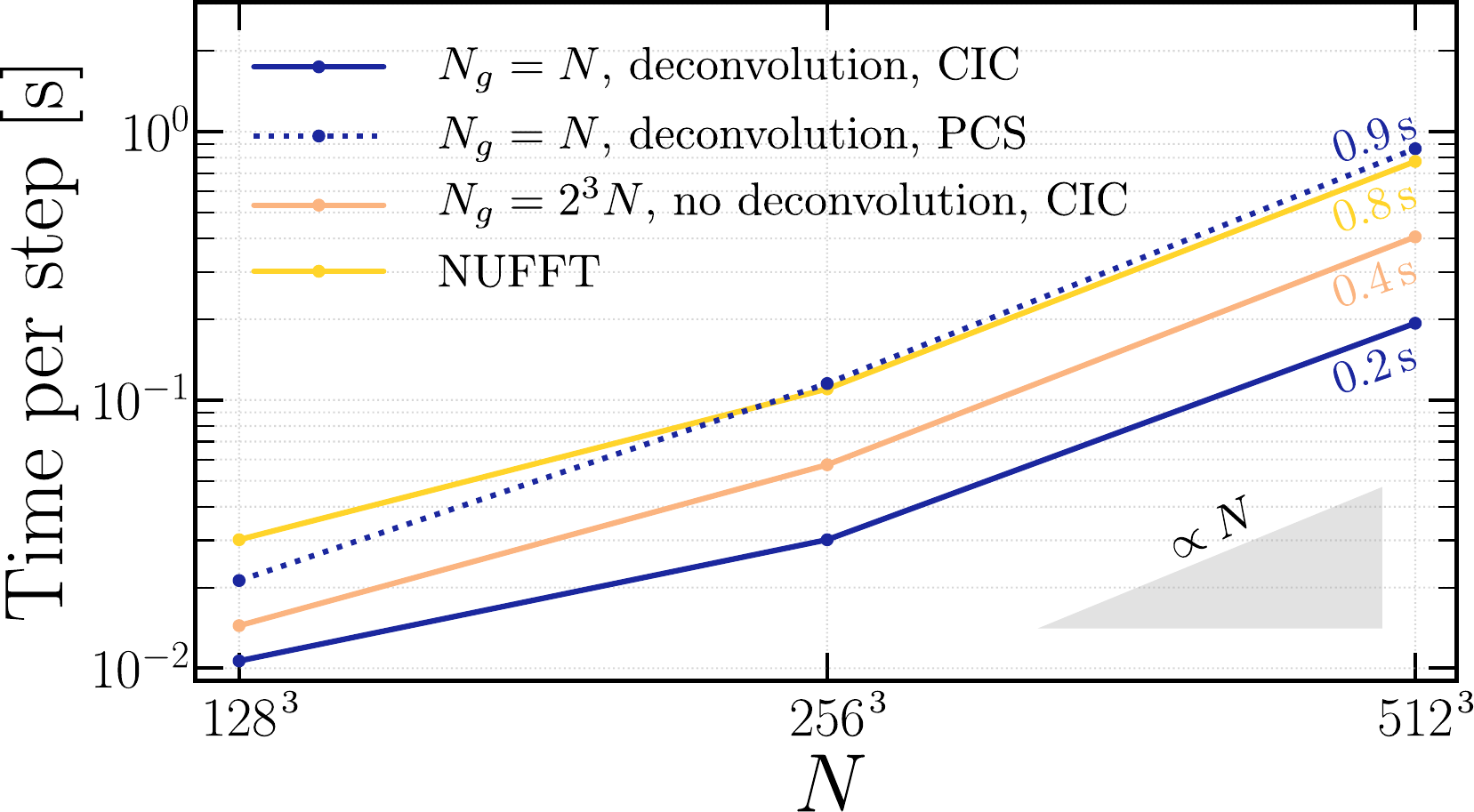}
    \caption{Average wall time per time step in a 100-step simulation on an \textsc{Nvidia} A100 ($40 \, \mathrm{GB}$). Line colours and styles correspond to different force computation settings and have the same meaning as in Fig.~\ref{fig:mass_assignment_study}. The slope of the inset triangle corresponds to a linear $N$ vs runtime scaling.}
    \label{fig:runtime_scaling}
\end{figure}

A main intended use case of the \codename framework is inference of cosmological parameters and/or initial conditions, either via explicit (usually gradient-aided) inference or via modern (usually deep learning-aided) implicit inference methods. In both cases, simulator speed is crucial, as many posterior evaluations are necessary for sampling, and deep learning models require large amounts of training data.

We benchmark the runtime of the PM $N$-body simulations in our default scenario by performing $N_{\mathrm{steps}} = 100$ time steps and computing the average wall time per time step, using an \textsc{Nvidia} A100 GPU. Specifically, we consider the two recommended settings with local mass assignment based on Sec.~\ref{sec:force_results} \ref{case:1} ($N_g = 2^3 N$ without deconvolution) and \ref{case:2} ($N_g = N$ with deconvolution) with CIC, additionally the PCS version of \ref{case:2}, and NUFFT-based force computation (with a support of $5^3$ cells for the interpolation kernel).
Figure~\ref{fig:runtime_scaling} shows the results for $N \in \{128^3, 256^3, 512^3\}$ particles. With $N = 512^3$ particles -- which is close to the memory limit for a $40 \, \mathrm{GB}$ GPU with $N_g = 2^3 N$ -- even the high-accuracy simulations with PCS and NUFFT require less than a second per step. With CIC, options \ref{case:1} and \ref{case:2} take $0.4$ and $0.2 \, \mathrm{s} / \mathrm{step}$, respectively. For $N = 256^3$ particles -- which is, for example, the resolution of the \textsc{Camels} simulation suite \cite{2023ApJS..265...54V} that is popular for machine learning applications -- roughly $0.1 \, \mathrm{s} / \mathrm{step}$ or less are required with all considered settings, and only $0.03 \, \mathrm{s} / \mathrm{step}$ for option \ref{case:2} with CIC. When going from $N = 256^3$ to $N = 512^3$ the runtime scales roughly linearly.

For completeness, we remark that when de-aliasing via interlaced grids is used in the simulation, a runtime increase roughly proportional to the number of employed grids can be expected (e.g.\ a doubling for two grids, which is typically sufficient in practice, see Fig.~\ref{fig:convergence_in_res}), as the force computation is by far the most computationally expensive ingredient of the simulations. 
We leave further improvements in terms of runtime, such as through custom CUDA kernels, for future work.

\section{Application: field-level cosmological inference}
\label{sec:inference}
In this section, we demonstrate the use of \codename{} in field-level cosmological inference. This approach to cosmological inference circumvents the data compression, through summary statistics, that is used in conventional cosmological analysis. Instead, field-level inference aims to forward model the entire data \emph{field}, e.g. the three-dimensional distribution of galaxies. This introduces the initial density fluctuations as latent parameters that are to be inferred alongside the cosmological parameters of interest, resulting in an exceptionally high-dimensional inference problem (typically upwards of $10^6$ parameters). Using standard random-walk MCMC methods in this many dimensions would be infeasible, due to a low acceptance rate of proposed steps, and we have to instead resort to more sophisticated methods. HMC is an MCMC variant that employs gradients of the posterior w.r.t.\ the parameters of interest to make informed proposals, significantly improving the acceptance rate in high-dimensional inference problems \cite{neal2011mcmc}. As described in Sec.~\ref{sec:autodiff}, the automatic differentiability of \codename{} gives access to such gradients, thus enabling field-level cosmological inference applications.

\begin{figure}
    \centering
    \includegraphics[width=1.0\linewidth]{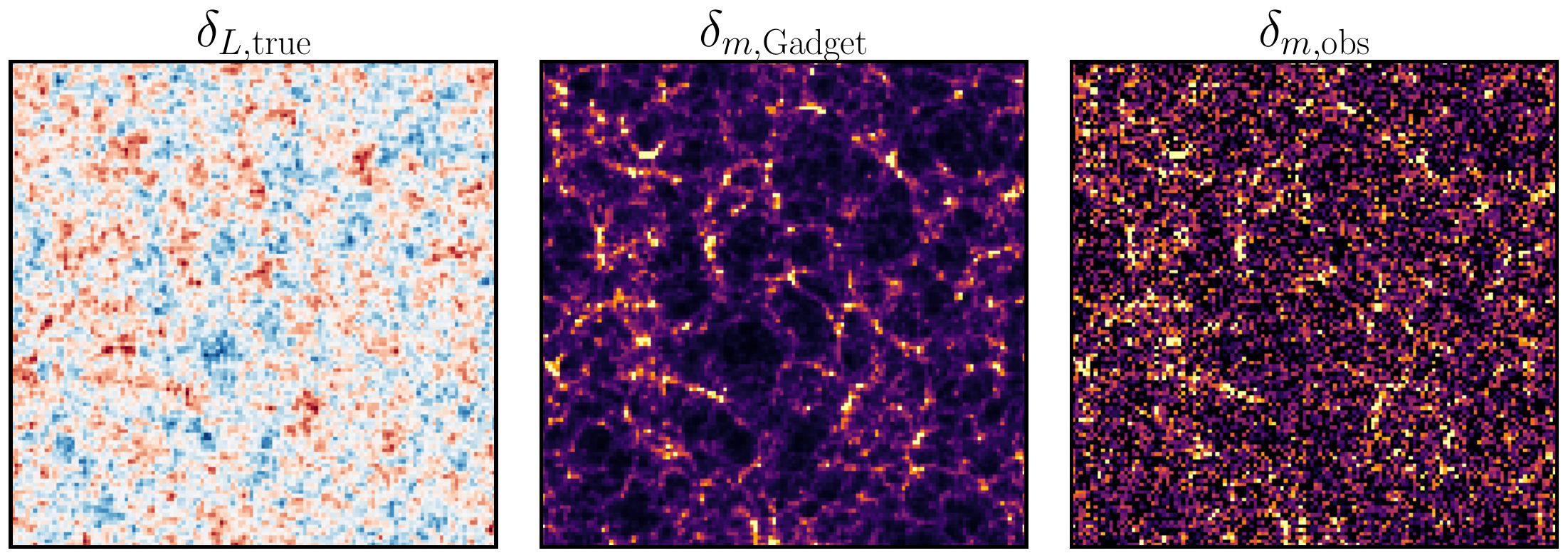}
    \includegraphics[width=1.0\linewidth]{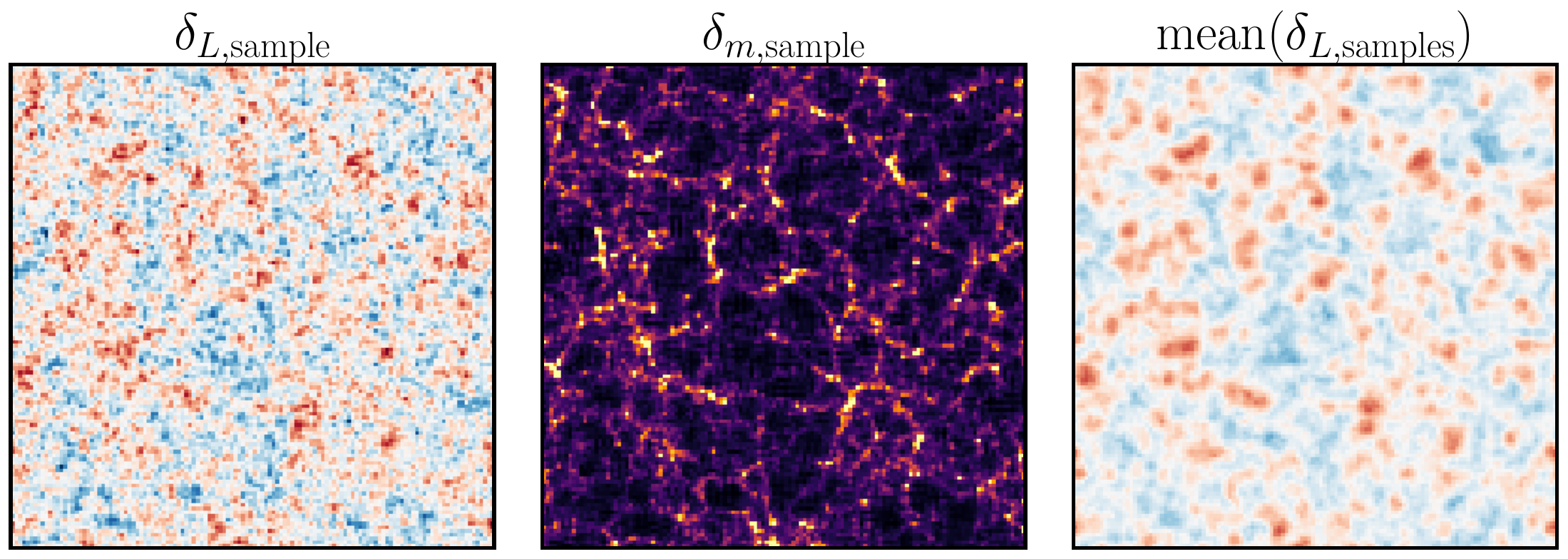}
    \caption{Top row: slices of the fields used to create the mock observation. Top left: the input (`true') linear density field for the \textsc{Gadget} simulation. Top middle: the corresponding output matter field of the \textsc{Gadget} simulation. Top right: the mock observation, consisting of the \textsc{Gadget} matter field and Gaussian noise. Bottom row: slices of samples generated during inference. Bottom left: a sample of the linear density field. Bottom middle: the corresponding output matter field of the \codename{} forward model. Bottom right: the mean over $60$ independent initial condition samples. All slices are averaged over $31.25 \, \mathrm{Mpc}/h$ along one dimension.}
    \label{fig:inference_fields}
\end{figure}

As a simple demonstration of this, we will show that \codename{} can be used as an accurate forward model for extracting cosmological information from the matter field. To this end, we take the \textsc{Gadget} matter field with size $L = 500 \, \mathrm{Mpc}/h$ at redshift $z=0$ that was introduced earlier in the paper, and create a noisy mock observation as
\begin{equation}
    \delta_{m,\mathrm{obs}} = \delta_{m,\mathrm{Gadget}} + \epsilon_{\mathrm{noise}},
\end{equation}
where the noise is taken to be Gaussian with a flat spectrum $P_{\mathrm{noise}} = 954 \, (\mathrm{Mpc}/h)^3$. The top row of Fig.~\ref{fig:inference_fields} shows slices of the fields used to create the mock observation.

The goal of field-level inference is then to compute the posterior of cosmological parameters $\theta$ and white-noise initial conditions $w(\vecb{k})$, given the observed field. We will use the following Fourier-space posterior that extracts information from all scales larger than a cutoff $k_{\mathrm{max}}$:
\begin{equation}
    \log \mathcal{P}(\theta, w(\vecb{k}) | \delta_{m,\mathrm{obs}}) = - \!\! \sum_{|\vecb{k}|<k_{\mathrm{max}}} \! \frac{|\delta_{m,\mathrm{DJ}}(\theta, w(\vecb{k})) - \delta_{m, \mathrm{obs}}(\vecb{k})|^2}{2 P_{\mathrm{noise}}} - \sum_{|\vecb{k}|<\Lambda} \! \frac{|w(\vecb{k})|^2}{2}
    + \log \mathcal{P}(\theta).    
\end{equation}
The value of $\Lambda$, which is often referred to as the UV-cutoff of the initial conditions, is taken to be the radius of the Nyquist sphere, i.e. $|\vecb{k}| < k_{\mathrm{Nyq}}$. In practice, this UV-cutoff is enforced by performing the \codename simulations without corner modes (see also Appendix~\ref{sec:corner_modes}).
Contrary to the perturbative EFT approach to field-level inference (e.g. \cite{Schmidt2020AnStructure}), here our aim is to perform a UV-complete simulation of the modes up to the likelihood cutoff $k_{\mathrm{max}}$, which we therefore take to be sufficiently lower than the UV-cutoff, as $k_{\mathrm{max}} = \Lambda/2$. 

Our forward model simulates the matter field using a PM simulation with $N = 128^3$ particles, and $16$ \textsc{BullFrog} steps, resulting in a $2.5\%$ accurate power spectrum out to the smallest scales of interest, $k_{\mathrm{max}} = 0.4 \, h / \mathrm{Mpc}$, which is sufficiently accurate given the impact of the Gaussian noise on these scales. Besides the initial white-noise modes, we infer the amplitude of the linear power spectrum as parameterised by $\sigma_8$, keeping all other cosmological parameters fixed. For this setup, a single evaluation of the posterior gradient takes less than $300 \, \mathrm{ms}$ on an \textsc{Nvidia} A100 GPU. We perform inference using the standard HMC implementation of the \textsc{Blackjax}\footnote{https://github.com/blackjax-devs/blackjax} package, with a mass matrix tuned using the provided warm-up algorithm. 

\begin{figure}
    \centering
    \includegraphics[width=0.45\linewidth]{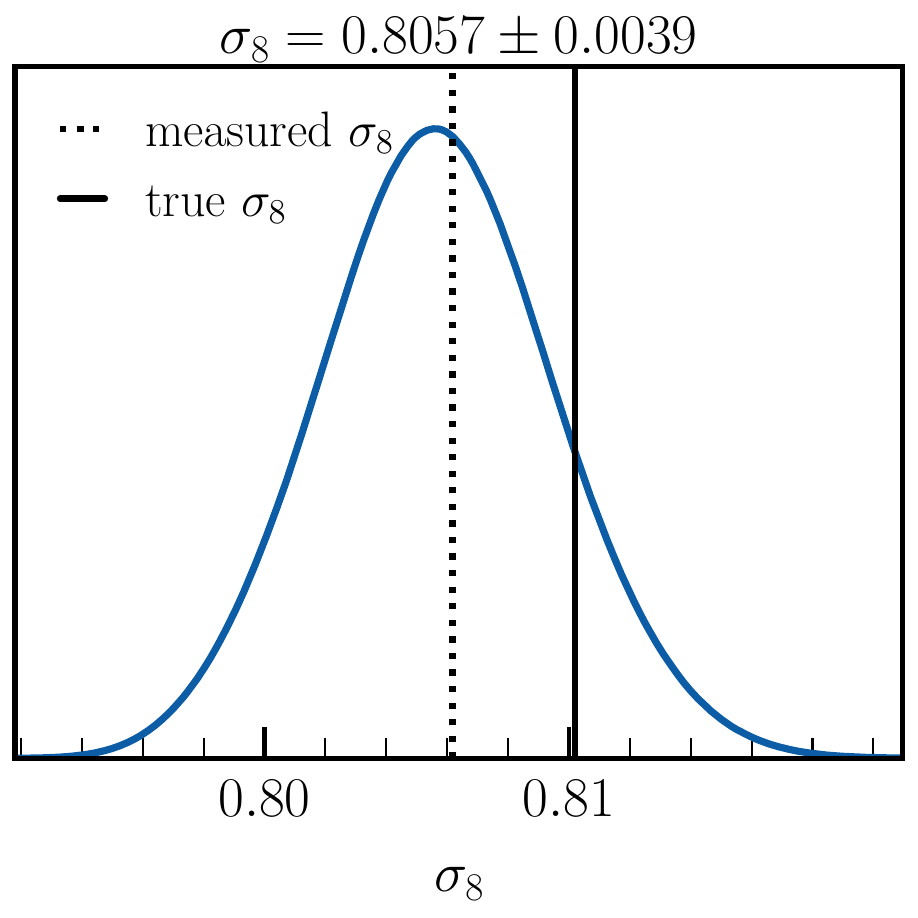}
    \caption{Marginal posterior of $\sigma_8$ inferred from the mock observation. The black solid line indicates the true input value of $\sigma_8 = 0.8102$, whereas the dotted line marks the \emph{effective} (i.e.\ cosmic variance affected) $\sigma_8=0.8062$ of this realisation, as inferred from the power spectrum measured from the true linear density field.}
    \label{fig:inference_sigma8}
\end{figure}

Summarising the results, the bottom row in Fig.~\ref{fig:inference_fields} shows slices of an inferred realisation of the initial conditions and the corresponding $z=0$ matter density field as simulated using our forward model, as well as the mean over several inferred initial conditions, all showing clear correlation with the true fields in the top row. Figure~\ref{fig:inference_sigma8} shows the inferred marginal posterior of $\sigma_8$, agreeing well with the true input value, though centred slightly low. As usual in cosmological inference, such a shift can simply occur due to cosmic variance of the individual realisation, and indeed a measurement of $\sigma_8$ from the power spectrum of the true linear density field yields a lower \emph{effective} $\sigma_8$ for this realisation, in close agreement with the value inferred at the field-level. Although beyond the scope of the proof-of-concept application in this section, we stress that a rigorous assessment of the accuracy (bias) and precision (variance) of cosmological inference requires repeated application of the inference pipeline to many realisations (i.e.\ coverage tests). 

The above application assumed access to the three-dimensional matter field, which is unobservable in reality. Instead, observations of the three-dimensional distribution of baryonic matter, such as through spectroscopic galaxy surveys or line-intensity mapping, yield biased tracers of this underlying matter field. Such biased tracers can be parameterised on top of the matter field as simulated by \codename{}, using various bias models (see e.g.\ \cite{Desjacques:2016bnm}). Furthermore, observations are made on the lightcone and in redshift space, rather than real space, requiring additional modelling as well. The development of these models in \codename{}, and their application to field-level cosmological inference from biased tracers, will be studied in future publications.

\section{Conclusions}
\label{sec:conclusions}
In this article, we have presented the particle-mesh (PM) $N$-body module of \codename{} (\textbf{DI}fferentiable \textbf{S}imulations for \textbf{CO}smology -- \textbf{D}one with \textbf{J}\textsc{ax}). 
This module features fast time integration methods, such as the recently developed \textsc{BullFrog} integrator, which enable accurate predictions up to mildly non-linear scales $k \sim 0.1 - 0.3 \, h / \mathrm{Mpc}$ in a few seconds on a GPU. Various techniques for improving the PM force computation are implemented, such as a custom non-uniform FFT, de-aliasing via interlaced grids, particle resampling based on the matter sheet in phase space, higher-order mass assignment kernels, different derivative kernels, and the choice whether or not a deconvolution of the mass assignment kernel is performed.
Although default settings are sufficient in many scenarios, these techniques prove useful for various applications, for example to achieve highly accurate field-level agreement with perturbative methods in the fluid regime (see \cite{List2024StartingBang}) or when higher-order differentiability is required. We expect that many of our findings on the effect of different numerical parameters carry over to other cosmological PM codes.

We envision \codename to be a powerful and versatile tool for cosmological field-level inference. Explicit (i.e.\ likelihood-based) inference using gradient-based optimisation and sampling methods is facilitated by the automatic differentiability of the code. Furthermore, the adjoint method used for backpropagation through the time steps keeps the memory footprint constant as the number of time steps increases. Our custom VJP implementation and the possibility to loop over chunks of particles in the interpolation functions further reduce the memory usage. In Sec.~\ref{sec:inference} we presented a first proof-of-concept example of explicit inference, demonstrating that \codename can be used to simultaneously constrain initial conditions and cosmological parameters from field quantities without resorting to summary statistics. 

For implicit (i.e.\ likelihood-free) inference based on machine learning models such as neural networks, \textsc{Jax}'s batching abilities (e.g.\ with \texttt{vmap} and \texttt{pmap}) enable the simultaneous execution of multiple simulations, making \codename well suited for the fast generation of large training datasets. If desired, the output can be further processed by downstream tasks such as halo finding, halo occupation distribution (which can even be performed differentiably \cite{2024MNRAS.529.2473H}), subhalo abundance matching, or the modelling of biased fields etc. 
In addition, solver-in-loop approaches that introduce learnable parameters as part of the simulation (e.g.\ to mimic the effect of baryonic physics \cite{2021PNAS..11820324D} or to improve the small-scale force accuracy \cite{2022mla..confE..60L}) can be seamlessly integrated into the pipeline. Recently, it has also been shown that halo properties of fast PM simulations, such as provided by \codename, can be improved by adjusting halo finder parameters \cite{2024MNRAS.531.4944W}.

In the near future, we will incorporate lightcone generation with redshift-space distortions and bias modelling, further bridging the gap between simulations and observations. A distributed multi-GPU version is currently under development and will be presented in an upcoming publication. \codename is released as an open source package\footnote{\url{https://github.com/cosmo-sims/DISCO-DJ}}, and we look forward to its further enhancement through community-driven contributions.


\acknowledgments
The authors thank Cornelius Rampf, Jens St\"{u}cker, and the participants of the ESI Workshop ``Putting the Cosmic Large-scale Structure on the Map: Theory Meets Numerics'' held in Vienna (Sept.~22 $-$ Sept.~26, 2025) for insightful discussions. We thank Lena Einramhof for preliminary work on the convergence of reduced cumulants in fast simulations in her BSc thesis. The computational results presented have been achieved using the Austrian Scientific Cluster (ASC) infrastructure. The authors declare no conflicts of interest or external support in the preparation of this manuscript. 

\bibliographystyle{JHEP}
\bibliography{main}

\appendix

\section{Additional checks}
\label{sec:additional_checks}
This section contains further numerical experiments with \codename concerning the effect of the box size, initialisation redshift and LPT order, corner modes, and derivative kernels in the force computation.

\subsection{Effect of the box size}
\label{sec:boxsize}
In this appendix, we study the convergence in time for a larger ($L = 1.5 \, \mathrm{Gpc}/h$) and a smaller ($L = 100 \, \mathrm{Mpc}/h$) box than our $L = 500 \, \mathrm{Mpc}/h$ baseline considered in the main body. All other numerical settings are the same as in Sec.~\ref{sec:time_convergence}.

\begin{figure}
    \centering
    \includegraphics[width=0.75\linewidth]{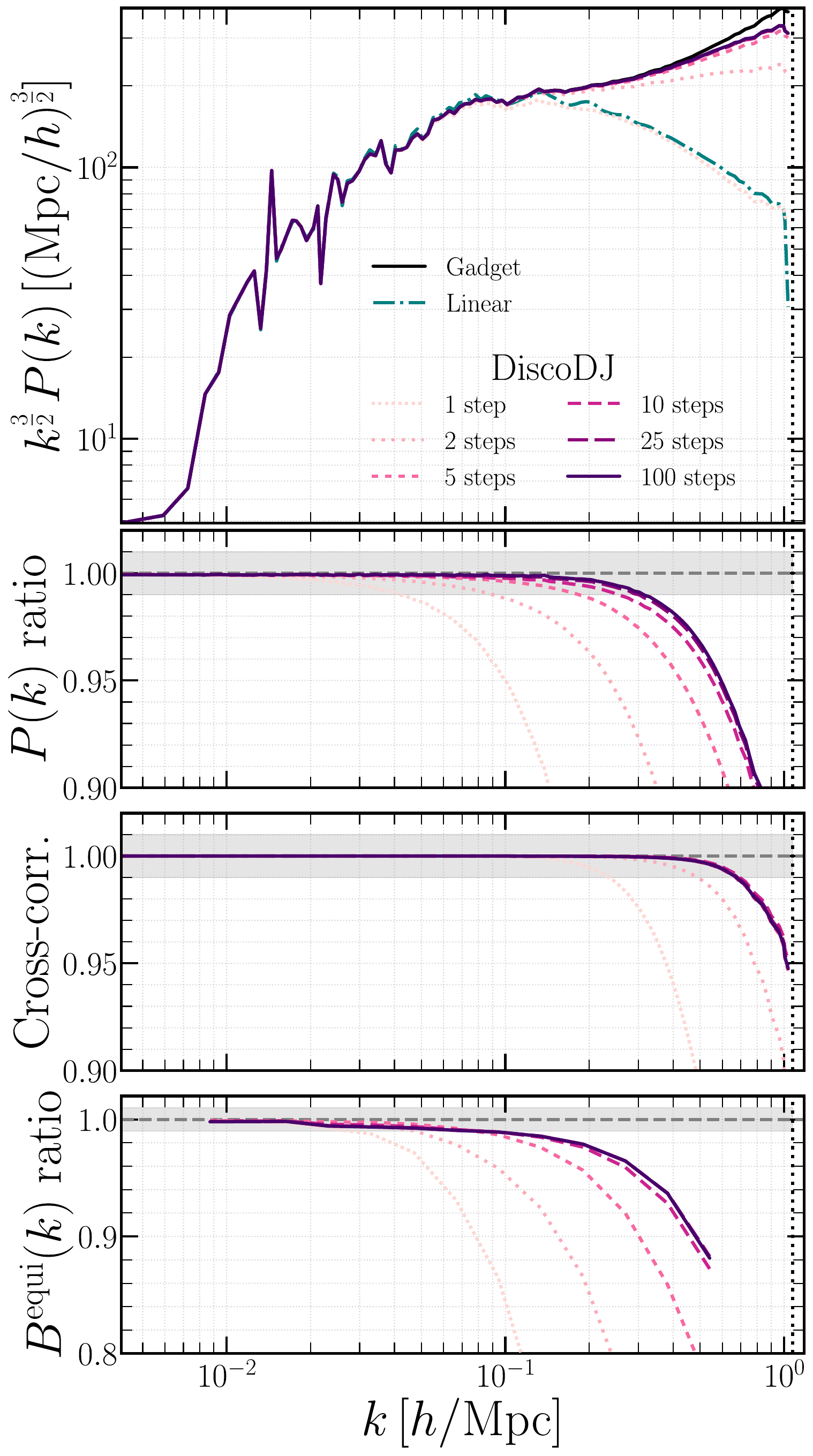}
    \caption{Same as Fig.~\ref{fig:convergence_in_time}, but for a \textbf{large} simulation box ($L = 1.5 \, \mathrm{Gpc}/h$). In this mildly non-linear case with $k_{\mathrm{Nyquist}} = 1.07 \, h / \mathrm{Mpc}$, using more than 10 time steps only yields small improvements.}
    \label{fig:convergence_in_time_big}
\end{figure}
\paragraph{Convergence in time}
The results for the $L = 1.5 \, \mathrm{Gpc} / h$ case are shown in Fig.~\ref{fig:convergence_in_time_big}. The time-converged power spectrum is sub-percent accurate up to $k = 0.3 \, h / \mathrm{Mpc}$ in this case -- not far below the $k = 0.4 \, h / \mathrm{Mpc}$ achievable with the $L = 500 \, \mathrm{Mpc}/h$ box -- although the equilateral bispectrum already deviates from the truth by $4 \%$ at this scale. As few as 10 \textsc{BullFrog} time steps are sufficient to get close to temporal convergence, and the improvement provided by 25 or 100 steps is small. The remaining irreducible error on smaller scales is due to the missing spatial and force resolution.

\begin{figure}
    \centering
    \includegraphics[width=0.75\linewidth]{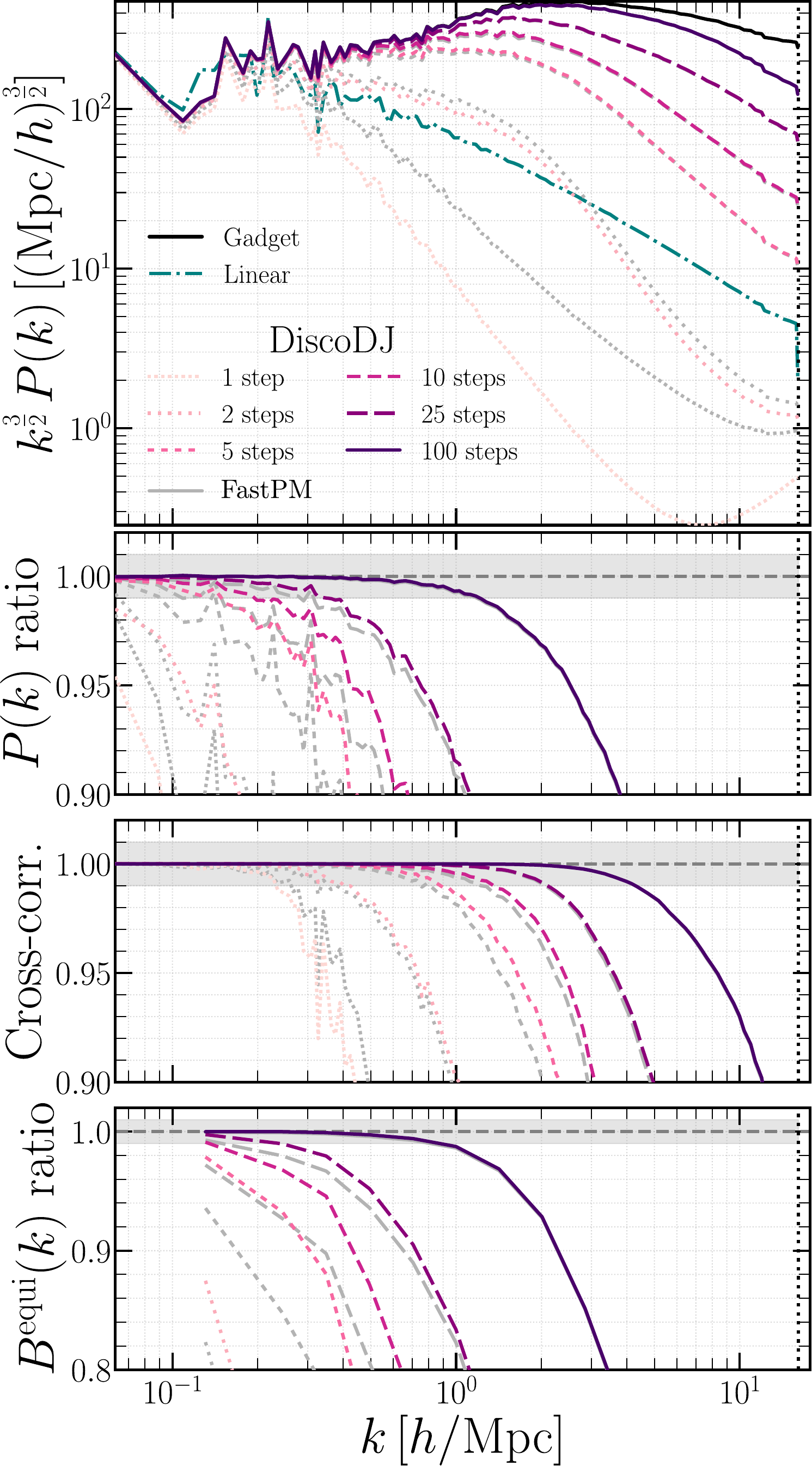}
    \caption{Same as Fig.~\ref{fig:convergence_in_time}, but for a \textbf{small} simulation box ($L = 100 \, \mathrm{Mpc}/h$). The grey lines show the results when using the \textsc{FastPM} integrator \cite{Feng:2016} instead of \textsc{BullFrog} \cite{Rampf_2025}. Percent-level accuracy in terms of all plotted statistics up to $k \lesssim 1 \, h / \mathrm{Mpc}$ can be achieved in this case, but $\sim 100$ steps are required.}
    \label{fig:convergence_in_time_small}
\end{figure}
The $L = 100 \, \mathrm{Mpc} / h$ case is plotted in Fig.~\ref{fig:convergence_in_time_small}. Now, a much larger number of time steps is required to accurately resolve these strongly non-linear scales (with a particle Nyquist wave number given by $k_{\mathrm{Nyquist}} = 16.1 \, h / \mathrm{Mpc}$ for our $N = 512^3$ particles). With 25 \textsc{BullFrog} steps, all considered statistics are suppressed compared to the \textsc{Gadget-4} reference simulation. With 100 steps, the errors in the power spectrum and bispectrum decrease to $\lesssim 1\%$ up to $k \lesssim 1 \, h / \mathrm{Mpc}$. 
The \textsc{BullFrog} integrator employed by default in \codename is not symplectic, but rather exactly matches the 2LPT trajectory (up to higher-order terms) in the pre-shell-crossing regime. While it is clear that this is beneficial on perturbative scales, it is interesting to study whether it is advantageous to switch to the only 1LPT consistent but symplectic \textsc{FastPM} integrator for this smaller box. Therefore, we also show the results with the DKD variant of \textsc{FastPM} (grey lines). For very few steps ($\leq 2$), \textsc{FastPM} indeed outperforms \textsc{BullFrog} (reminiscent of 1LPT often performing better than 2LPT when evaluated in the non-perturbative regime, where the 2LPT correction may be detrimental). However, starting from $\gtrsim 5$ time steps, \textsc{BullFrog} overtakes \textsc{FastPM} in terms of accuracy. With 100 time steps, the results with both steppers match closely, indicating that time convergence has been achieved. Thus, even in the strongly non-linear regime $k \gtrsim 1 \, h / \mathrm{Mpc}$, \textsc{BullFrog} is an effective choice.

\begin{figure}
    \centering
    \includegraphics[width=1\linewidth]{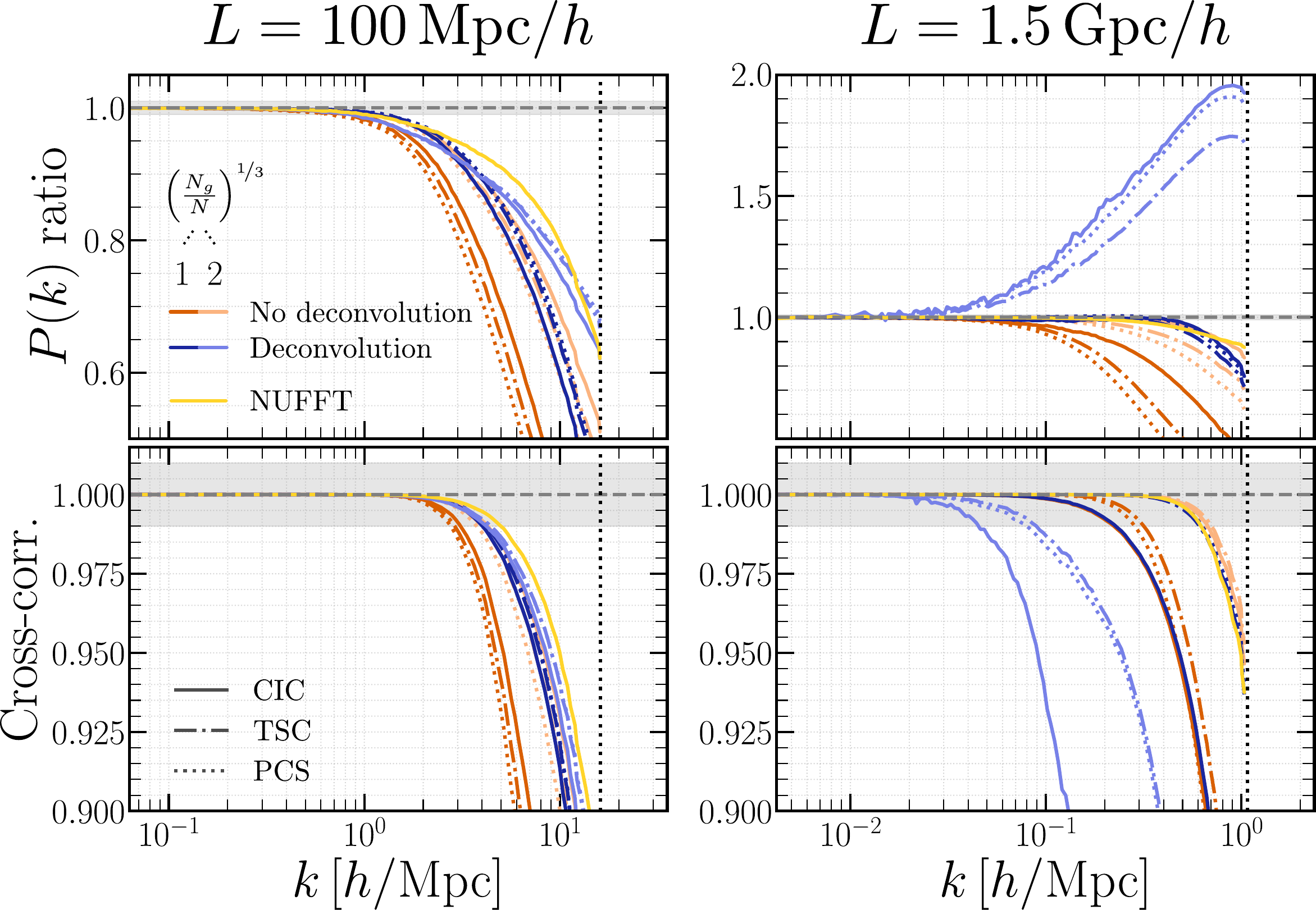}
    \caption{Same as Fig.~\ref{fig:mass_assignment_study}, but for a \textbf{small} ($L = 100 \, \mathrm{Mpc} / h$, left) and \textbf{large} ($L = 1.5 \, \mathrm{Gpc}/h$, right) simulation box instead of our default $L = 500 \, \mathrm{Mpc} / h$.}
    \label{fig:mass_assignment_study_small_large}
\end{figure}

\paragraph{Force computation}
It is also interesting to study how favourable settings for the PM grid size and deconvolution (see Sec.~\ref{sec:force_results}) vary when considering larger or smaller scales. Figure~\ref{fig:mass_assignment_study_small_large} shows the power spectrum ratio and cross-correlation with the \textsc{Gadget-4} reference for different settings -- similarly to Fig.~\ref{fig:mass_assignment_study} in the main body, but now for $L = 100 \, \mathrm{Mpc} / h$ and $1.5 \, \mathrm{Gpc} / h$. Our main findings are as follows.
\begin{itemize}
    \item For all box sizes, NUFFT yields competitive results. 
    \item Regardless of the box size, $N_g = N$ without deconvolution (dark orange) leads to an early drop in power. For the small box, the cross-correlation is also the worst for this choice.
    \item Option~\ref{case:1} recommended in the main body ($N_g = 2^3 N$, no deconvolution, light orange) leads to a $P(k)$ ratio and cross-correlation that are on par with NUFFT for the large box when using CIC. Higher-order kernels that suppress power more strongly are not beneficial. For the small box, this combination also gives good results, although not as good as NUFFT.
    \item Option~\ref{case:2} ($N_g = N$, deconvolution, dark blue) generally achieves a similar accuracy to option~\ref{case:1} at lower memory and runtime (with CIC). For the large box, however, CIC leads to worse cross-correlation, which can be remedied by using TSC instead.
    \item For the combination of $N_g = 2^3 N$ with deconvolution (light blue), we observe a large effect of the box size: for the small box, these settings yield excellent results in terms of the $P(k)$ ratio and cross-correlation -- not far behind NUFFT. In contrast, for the large box, the power spectrum overshoots significantly, and the cross-correlation drops very early.
    These findings suggest that when a deconvolution of the interpolation kernel is applied, a necessary condition for safely harnessing information beyond the particle Nyquist mode $k_{\mathrm{Nyquist}}$ through a finer PM grid $N_g > N$ is that the considered scales are non-linear enough for the beyond-$k_{\mathrm{Nyquist}}$-modes to be sufficiently sampled by the particles. We leave a detailed investigation for future work. 
\end{itemize}

\subsection{Effect of the initial redshift and LPT order}
In the main body of this work, all \codename simulations have been initialised with 2LPT at $z = 50$. For fast simulations with few time steps, it can be delicate to determine the optimal trade-off between starting sufficiently early such that truncation errors in the initial conditions are subdominant while still leaving enough steps for late times to capture the non-linear growth.

\begin{figure}
    \centering
    \includegraphics[width=1\linewidth]{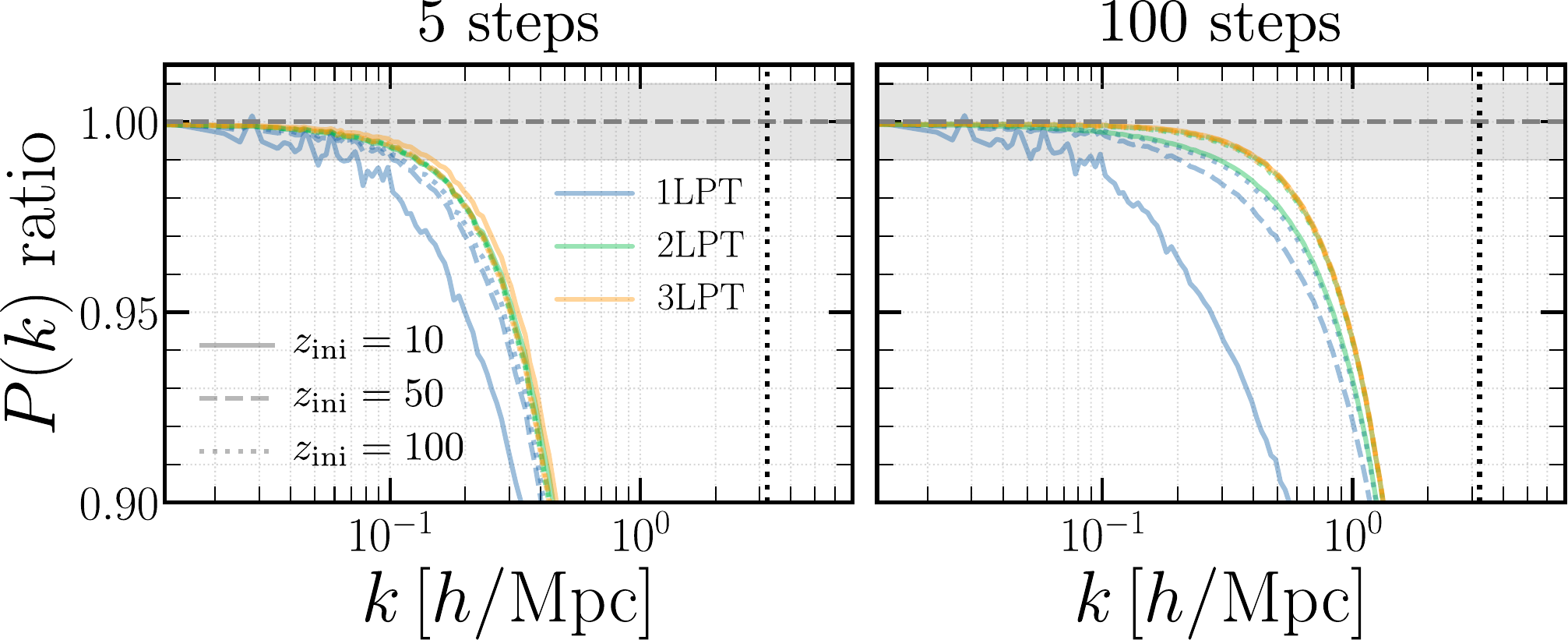}
    \caption{Effect of the initialisation redshift $z_{\mathrm{ini}}$ (different line styles) and LPT order (different line colours) on the $z = 0$ power spectrum compared to the \textsc{Gadget-4} reference, using 5 (left) and 100 (right) time steps between $z_{\mathrm{ini}}$ and $z = 0$. For 5 steps, the 2LPT lines for different $z_{\mathrm{ini}}$ and the 3LPT lines for $z_{\mathrm{ini}} = 50$ and $100$ all overlap, with 3LPT initial conditions at $z_{\mathrm{ini}} = 10$ performing best. For 100 steps, 3LPT for any considered $z_{\mathrm{ini}}$ and 2LPT for $z_{\mathrm{ini}} \geq 50$ yield very similar accuracy.}
    \label{fig:zstart_study}
\end{figure}

Figure \ref{fig:zstart_study} compares the impact of varying the LPT order and the starting redshift $z_{\mathrm{ini}}$ on the recovered power spectrum for very coarse ($N_{\mathrm{steps}} = 5$, left) and well-resolved ($N_{\mathrm{steps}} = 100$, right) time integration with \textsc{BullFrog}. 
In both cases, simulations initialised with 1LPT show the largest power deficits, and the situation worsens if the initial conditions are generated too late due to the growing importance of missing higher-order terms. At 5 steps, the trade-off between accurate initial conditions and time step placement becomes evident: starting very early (e.g.\ at $z_{\mathrm{ini}} = 100$) reduces truncation errors at low LPT order, but pushes the few available steps into the linear regime, leaving late-time non-linear growth poorly sampled. In contrast, starting late at $z_{\mathrm{ini}} = 10$ with 3LPT shifts the limited time steps to the non-linear epoch and provides the best overall match to the reference (solid orange line), consistent with the findings of Ref.~\cite{Michaux:2021}. With 100 steps, the time integration is sufficiently well resolved that the difference between $z_{\mathrm{ini}} = 50$ and $100$ with 2LPT and 3LPT largely disappears. In this context, recall that the \textsc{BullFrog} integrator automatically captures the 2LPT term exactly at each time step. In contrast, with standard, non-LPT-informed time integrators, it is often disadvantageous to start at early times, as the 2LPT term of the resulting trajectory generally only converges to the truth in the limit of infinitely many time steps -- requiring many steps at early times to properly capture the comparably simple pre-shell-crossing dynamics. In the case of late 2LPT initialisation at $z_{\mathrm{ini}} = 10$, the missing third-order terms in the initial conditions leave a noticeable imprint. Overall, this experiment highlights that particularly with few time steps, it can be advantageous to initialise late with at least 2LPT to concentrate integration effort at low redshift. Of course, when considering smaller scales, LPT ceases to be valid already at earlier times, so an earlier initialisation is required.

\subsection{Effect of corner modes}
\label{sec:corner_modes}
\begin{figure}
    \centering
    \includegraphics[width=0.66\linewidth]{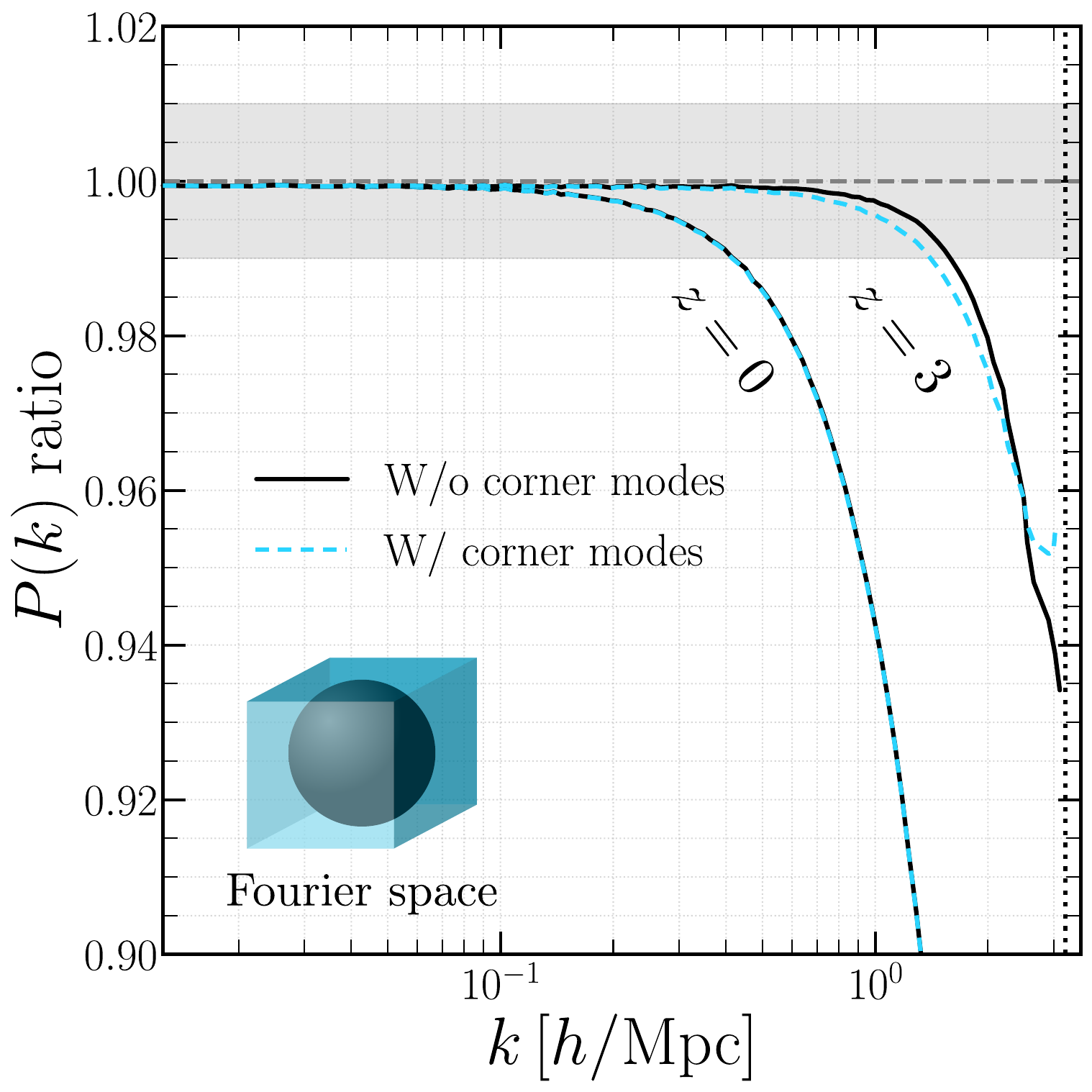}
    \caption{Effect of including the corner modes on the power spectrum at $z = 0$ and $z = 3$ in \codename. At $z = 3$, the power without corner modes drops slightly later than with corner modes, whereas the difference at $z = 0$ is negligible. The illustration in the lower left corner shows the support of the Fourier modes for the two cases, which are contained in a cube with corner modes and in a sphere without them.}
    \label{fig:corner_modes}
\end{figure}
The cubical geometry of cosmological simulation boxes leads to anisotropic coverage of the Fourier modes in the initial conditions. While the one-dimensional Nyquist frequency $k_{\mathrm{Nyquist}}$ defines the maximum resolvable mode along each axis, modes with magnitudes up to $\sqrt{3} k_{\mathrm{Nyquist}}$ can appear along the cube's space diagonal. As in many other codes, \codename allows users to choose whether these so-called `corner modes' with $k > k_{\mathrm{Nyquist}}$ -- which occupy $\approx \pi / 6$ and thus roughly half of the Fourier volume -- shall be initialised with noise according to the linear power spectrum or set to zero. For a dedicated numerical study on the impact of corner modes, we refer the reader to Ref.~\cite{Falck2017TheSimulations}. 

We again consider the same scenario as in Fig.~\ref{fig:convergence_in_time} with default numerical settings (see Table~\ref{tab:default_parameters}) and compute the power spectrum ratio between \codename and \textsc{Gadget-4}, with and without corner modes in \codename. Note that the reference simulation includes corner modes, but recall that there we used $N = 1024^3$, compared to $N = 512^3$ for \codename in this experiment. The results are shown in Fig.~\ref{fig:corner_modes}, at redshifts $z = 0$ and $z = 3$. At $z = 3$, \codename achieves sub-per-cent accuracy up to $k \approx 1 \, h / \mathrm{Mpc}$, and the difference between the two cases is already very small. By $z = 0$, the imprint of the corner modes has been erased.

\subsection{Effect of the derivative kernel}
\label{sec:effect_of_derivative_kernel}
\begin{figure}
    \centering
    \includegraphics[width=1\linewidth]{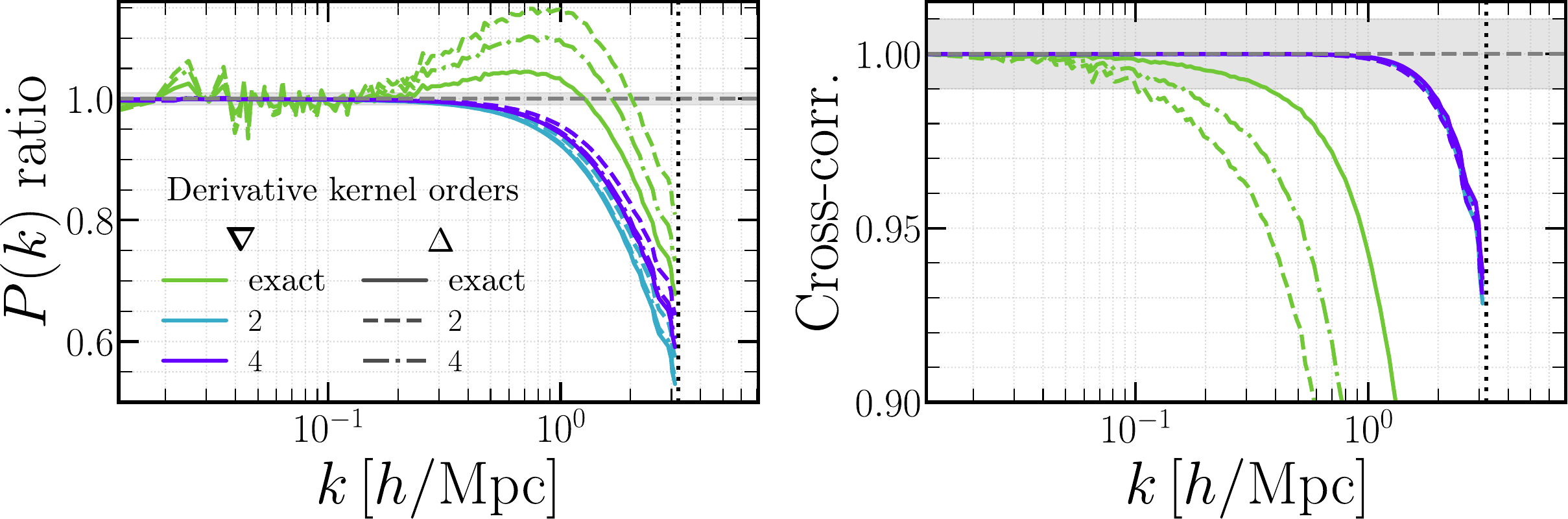}
    \caption{Power spectrum ratio (left) and cross-correlation (right) w.r.t.\ the \textsc{Gadget-4} reference at $z = 0$ for different derivative kernels. The line colours and line styles represent the gradient and Laplacian kernel orders, respectively. The 4$^{\mathrm{th}}$-order gradient kernel yields the best results. For this choice, the different Laplacian kernels perform comparably.
    }
    \label{fig:derivative_kernel_study}
\end{figure}

Finally, we investigate the effect of the gradient and Laplacian kernels on the power spectrum and cross-correlation. For this experiment, we again take $N_{\mathrm{steps}} = 100$ and $N = 512^3$, and perform simulations for each of the 3 $\times$ 3 combinations of $2^{\mathrm{nd}}$- and $4^{\mathrm{th}}$order finite difference kernels, and exact (i.e.\ spectral) kernel for both differential operators, see Sec.~\ref{sec:force}.

The results are shown in Fig.~\ref{fig:derivative_kernel_study}. When using the PM-only setup currently implemented in \codename without any additional discreteness suppression techniques, a finite difference gradient kernel is imperative to obtain accurate results: with the exact $\mathrm{i}\vecb{k}$ kernel, the power spectrum overshoots, and the cross-correlation is poor, regardless of the Laplacian implementation (however, see Fig.~4 in \cite{List2024StartingBang}, which shows that the exact gradient kernel improves the accuracy at early times when used together with particle resampling). The 4$^{\mathrm{th}}$-order gradient kernel is slightly superior compared to the 2$^{\mathrm{nd}}$-order version in terms of the power spectrum, and their cross-spectra are comparable, for which reason we take the former as our default. For this gradient choice, the 2$^{\mathrm{nd}}$-order Laplacian yields the best power-spectrum ratio; however, the exact Laplacian has a slightly higher cross-correlation, justifying the choice of the latter as our default.

\section{Exact vs approximate growth in $\Lambda$CDM}
\label{sec:growth}
\begin{figure}
    \centering
    \includegraphics[width=0.66\linewidth]{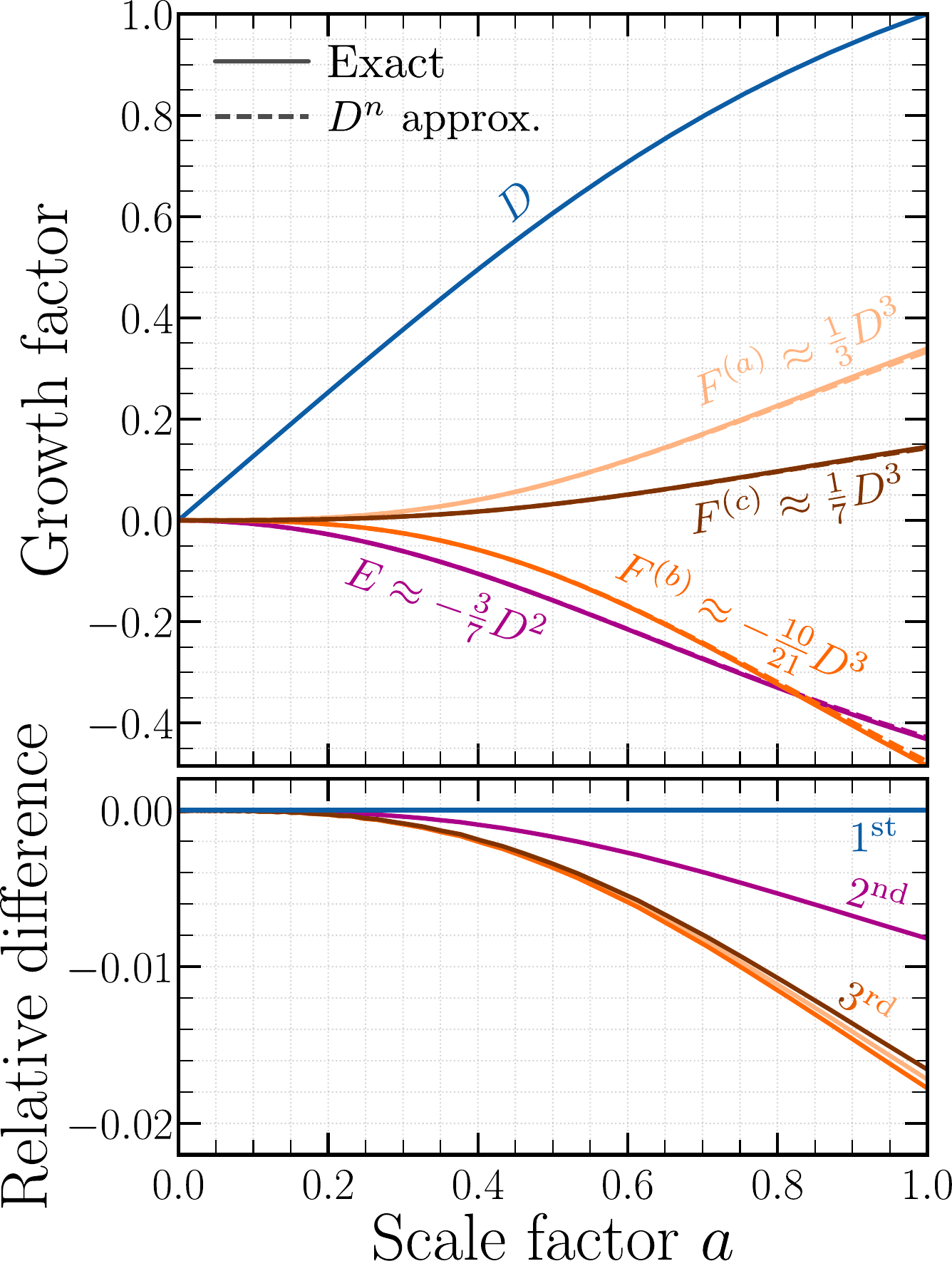}
    \caption{Comparison of exact $\Lambda$CDM growth (solid lines) and the $D^n$ approximation (dashed lines) as a function of scale factor $a$, for our benchmark cosmology. The first, second, and the three third-order growth functions are denoted as $D$, $E$, and $F^{(a)}$, $F^{(b)}$, $F^{(c)}$ respectively. The growth factors are normalised such that $D(a = 1) = 1$. The bottom panel shows the relative difference of the $D^n$ approximation towards exact $\Lambda$CDM. }
    \label{fig:growth}
\end{figure}
In Sec.~\ref{sec:lpt}, we presented our arbitrary-order LPT implementation and explained that, by default, we use the $D^n$ approximation of the higher-order growth factors in order to be able to lump together all spatial kernels of the same order. Here, we will show that for the initialisation of $N$-body simulations, the truncation error due to higher-order contributions (which are $O(\Omega_\Lambda)$ for $\Lambda$CDM cosmologies, see e.g.\ \cite{Rampf2022}) is negligible. 

As in the main body, let $D$ and $E$ be the first- and second-order growth factors, respectively, and let $F^{(a)}, F^{(b)}, F^{(c)}$ be the three third-order contributions, where $F^{(a)}$ corresponds to $(n_1, n_2, n_3) = (1, 1, 1)$ and $F^{(b)}$ to $(n_1, n_2, n_3) = (2, 1, 0)$ and permutations thereof in the longitudinal terms in Eq.~\eqref{eq:LPT_L}, and $F^{(c)}$ to the transversal term for $n = 3$ in Eq.~\eqref{eq:LPT_T}.

In addition, we define the (sign-aware) growth rates
\begin{equation}
\begin{aligned}
    \mathfrak{d}(a) & = \frac{\mathrm{d} \ln D}{\mathrm{d} \ln a}, 
    & \qquad \mathfrak{e}(a) & = \frac{\mathrm{d} \ln (-E)}{\mathrm{d} \ln a}, \\ 
    \mathfrak{f}^{(a)}(a) & = \frac{\mathrm{d} \ln F^{(a)}}{\mathrm{d} \ln a}, 
    & \qquad \mathfrak{f}^{(b)}(a) & = \frac{\mathrm{d} \ln (-F^{(b)})}{\mathrm{d} \ln a}, 
    & \qquad \mathfrak{f}^{(c)}(a) & = \frac{\mathrm{d} \ln F^{(c)}}{\mathrm{d} \ln a}.
\end{aligned}
\end{equation}

The signs in the numerators are chosen in such a way that the arguments of the logarithms are positive for flat $\Lambda$CDM cosmologies. In this case, these functions satisfy the following ordinary differential equations \cite{Bouchet:1995}:

\begin{subequations}
\begin{align}
\frac{\mathrm{d} \mathfrak{d}}{\mathrm{d} \ln a} &= 3 \gamma - \mathfrak{d}^2 - \mathfrak{d} \left(1 - \gamma\right), \\ 
\frac{\mathrm{d} \mathfrak{e}}{\mathrm{d} \ln a} &= 3 \gamma \left(1 - \frac{D^2}{E}\right) - \mathfrak{e}^2 - \mathfrak{e} \left(1 - \gamma\right), \\ 
\frac{\mathrm{d} \mathfrak{f}^{(a)}}{\mathrm{d} \ln a} &= 3 \gamma \left(1 + 2 \frac{D^3}{F^{(a)}}\right) - (\mathfrak{f}^{(a)})^2 - \mathfrak{f}^{(a)} \left(1 - \gamma\right), \\
\frac{\mathrm{d} \mathfrak{f}^{(b)}}{\mathrm{d} \ln a} &= 3\gamma \left(1 + \left(2 \frac{D E - D^3}{F^{(b)}}\right)\right) - (\mathfrak{f}^{(b)})^2 - \mathfrak{f}^{(b)} \left(1 - \gamma\right), \\
\mathfrak{f}^{(c)} &= \frac{D E}{F^{(c)}} \left(\mathfrak{d} - \mathfrak{e} \right),
\label{eq:growth_odes}
\end{align}
where 
\begin{equation}
\gamma := \frac{\Omega_m H_0^2}{2 a^3 H^2(a)}.
\end{equation}
\end{subequations}
Since the higher-order growth rates themselves are derivatives of the growth factors that appear in the source terms on the right-hand side, this is a second-order system, which we solve with a simple Runge--Kutta integrator with uniform steps in $\ln a$.

Figure~\ref{fig:growth} compares the $D^n$ approximation with the exact $\Lambda$CDM growth functions defined according to Eqs.~\eqref{eq:growth_odes}, for the default $\Lambda$CDM cosmology considered throughout this work. Even at $z = 0$, the error at second order remains less than $1\%$, while the third-order terms deviate by roughly $2\%$. At typical $N$-body initialisation redshifts ($z \gtrsim$ 10 even for fast approximate PM simulations), the deviation is on the order $\lesssim 10^{-4}$.

\section{Custom VJP and JVP for particle-mesh interpolation operators}
\label{sec:custom_VJP_JVP}
\def\myspace{0.3cm}
This section presents a formal derivation of the vector-Jacobian products (VJPs) and Jacobian-vector products (JVPs) for particle-to-mesh (\texttt{scatter}) and mesh-to-particle (\texttt{gather}) interpolation operators in \codename, which we leverage in our custom implementation of these operations. VJPs are required for reverse-mode differentiation, whereas JVPs are used for forward-mode differentiation. While \textsc{Jax} supports automatic transposition of custom JVPs (acting on tangents) to obtain VJP rules (acting on cotangents)\footnote{\url{https://docs.jax.dev/en/latest/notebooks/Custom_derivative_rules_for_Python_code.html}}, we found that the resulting VJPs incur significant memory overhead due to the internal graph construction. For this reason, we implement both the JVP and VJP rules explicitly, ensuring efficient memory usage and control over the adjoint computations. In particular, when vectorising the operations over chunks of particles, the same chunk size is used for the \texttt{scatter} and \texttt{gather} operations arising in the derivative operations (see below).

In what follows, we will derive the (co)tangent mappings for each input, and formalise the logic used in our custom VJP and JVP implementations. Note that while we mostly use the notion of ``adjoint'' variables in the presentation of the adjoint method in Sec.~\ref{sec:adjoint}, we now refer to adjoints as cotangents to emphasise the analogy with the tangents for JVPs.

Let:
\begin{itemize}
  \item $\mathcal{M} \subset \mathbb{R}^3$ be a regular grid of $N_g$ points,
  \item $\vecb{X} = \{\vecb{X}_i\}_{i=1}^{N} \in  \mathbb{R}^{N\times3}$ be an array of $N$ particle positions,
  \item $w = \{w_i\}_{i=1}^{N} \in \mathbb{R}^N$ be corresponding particle weights,
  \item $\rho = \{\rho_j\}_{j=1}^{N_g} \in \mathbb{R}^{N_g}$ be the scalar field defined on the grid,
  \item $W: \mathbb{R}^3 \to \mathbb{R}$ be a differentiable interpolation kernel with compact support.
\end{itemize}

We define the shorthand $W_j(\vecb{X}_i) := W(\vecb{r}_j - \vecb{X}_i)$ and $\bnabla W_j(\vecb{X}_i) := \bnabla W(\vecb{r}_j - \vecb{X}_i)$.

\paragraph{\bf Scatter: particles $\to$ mesh}
The \texttt{scatter} operation maps particle data to the grid:
\begin{equation}
    \rho_j = \sum_{i=1}^{N} w_i \, W_j(\vecb{X}_i),
\end{equation}
where $j \in \{1, \ldots, N_g\}$ indexes the grid points.

\paragraph{\bf Gather: mesh $\to$ particles}
The \texttt{gather} operation interpolates grid values back to particle positions:
\begin{equation}
    v_i = \sum_{j=1}^{N_g} \rho_j \, W_j(\vecb{X}_i),
\end{equation}
where $i \in \{1, \ldots, N\}$.
Due to the compact kernel of $W$, only nearby particles / grid cells have to be taken into account in the sums for the \texttt{scatter} / \texttt{gather}.
Since all spaces are Euclidean, tangent and cotangent spaces are identified with the base spaces.

\subsection{VJP and JVP derivations}
Let $\overcirc{\rho}$ and $\overcirc{v}$ denote the cotangent of a scalar loss $L$ w.r.t.\ the operator outputs $\rho$ and $v$, respectively. VJPs return cotangent contributions to the inputs, which we denote as $\overcirc{\rho}_{\mathrm{in}}$, $\overcirc{\vecb{X}}$, and $\overcirc{w}$.
For the JVPs, we denote input tangents as $\tilde{\rho}_{\mathrm{in}}, \tilde{\vecb{X}}, \tilde{w}$, and output tangents as $\tilde{\rho}$ and $\tilde{v}$, which we split up into contributions from the different inputs, i.e.\ $\tilde{\rho}_{\rho_{\mathrm{in}}}$, $\tilde{\rho}_{\vecb{X}}$, etc.

\subsubsection*{Scatter (input tangents: $(\tilde{\rho}_{\mathrm{in}}, \tilde{\vecb{X}}, \tilde{w})$, output cotangent: $\overcirc{\rho} \in T^*_\rho \cong \mathbb{R}^{N_g}$)}

\paragraph{\normalfont\textit{W.r.t.\ input grid}\\}
The scatter adds to any pre-existing value $\rho_{\mathrm{in}}$ on the mesh, and therefore
\begin{equation}
    \frac{\partial \rho}{\partial \rho_{\mathrm{in}}} = \mathbf{I}.
\end{equation}
Thus, tangents and cotangents propagate unchanged, i.e.\
\begin{equation}
    \overcirc{\rho}_{\mathrm{in}} = \overcirc{\rho} \qquad \text{and} \qquad \tilde{\rho}_{\rho_{\mathrm{in}}} = \tilde{\rho}_{\mathrm{in}}.
\end{equation}

\paragraph{\normalfont\textit{W.r.t.\ positions}\\}
Computing the response of $\rho$ w.r.t.\ particle positions yields
\begin{equation}
    \frac{\partial \rho_j}{\partial \vecb{X}_i} = -w_i \, \bnabla W_j(\vecb{X}_i).
\end{equation}
Hence,
\begin{equation}
    \overcirc{\vecb{X}}_i = -w_i \sum_{j=1}^{N_g} \overcirc{\rho}_j \bnabla W_j(\vecb{X}_i) \qquad \text{and} \qquad
    \tilde{\rho}_{\vecb{X}, j} = -\sum_{i=1}^{N} w_i \, \tilde{\vecb{X}}_i \cdot \bnabla W_j(\vecb{X}_i).
\end{equation}
The VJP is implemented as a gather of the cotangent mesh $\overcirc{\rho}_j$ with the gradient kernel $\bnabla W$, with a $w_i$ weighting for the $i$th particle, whereas the JVP is a scatter with gradient kernel $\bnabla W$ and dot-product with the tangent positions $\tilde{\vecb{X}}_i$.

\paragraph{\normalfont\textit{W.r.t.\ weights}\\}
The scatter operation is linear in the weights $w$. Therefore, one finds
\begin{equation}
    \frac{\partial \rho_j}{\partial w_i} = W_j(\vecb{X}_i)
\end{equation}
and hence
\begin{equation}
    \overcirc{w}_i = \sum_{j=1}^{N_g} \overcirc{\rho}_j \, W_j(\vecb{X}_i) \qquad \text{and} \qquad
    \tilde{\rho}_{w, j} = \sum_{i=1}^{N} \tilde{w}_i \, W_j(\vecb{X}_i),
\end{equation}
i.e.\ the VJP is a scalar gather of the cotangent values $\overcirc{\rho}_j$, and the JVP is a regular scatter with $\tilde{w}_i$.

\subsubsection*{Gather (input tangents: $(\tilde{\rho}, \tilde{\vecb{X}})$, output cotangent: $\overcirc{v} \in T^*_{v} \cong \mathbb{R}^{N}$)}

\paragraph{\normalfont\textit{W.r.t.\ mesh}\\}
Computing the derivative of a gathered value $v_i$ w.r.t.\ a mesh value $\rho_j$ yields
\begin{align}
    \frac{\partial v_i}{\partial \rho_j} &= W_j(\vecb{X}_i) \\
    \Rightarrow \quad
    \overcirc{\rho}_j = \sum_{i=1}^{N} \overcirc{v}_i \, W_j(\vecb{X}_i) \qquad &\text{and} \qquad
    \tilde{v}_{\rho, i} = \sum_{j=1}^{N_g} \tilde{\rho}_j \, W_j(\vecb{X}_i).
\end{align}
The VJP is a scatter of the cotangent values $\overcirc{v}_i$ using $W$, whereas the JVP is a standard gather with $\tilde{\rho}_j$. 

\paragraph{\normalfont\textit{W.r.t.\ positions}\\}
Note that the value $v_i$ for particle $i$ in the gather depends on the particle position $\vecb{X}_i$ via the kernel, but not on any other particle positions. We compute
\begin{equation}
    \frac{\partial v_i}{\partial \vecb{X}_i} = -\sum_{j=1}^{N_g} \rho_j \, \bnabla W_j(\vecb{X}_i),
\end{equation}
which yields
\begin{equation}
    \overcirc{\vecb{X}}_i = -\overcirc{v}_i \sum_{j=1}^{N_g} \rho_j \bnabla W_j(\vecb{X}_i) \qquad \text{and} \qquad
    \tilde{v}_{\vecb{X}, i} = -\sum_{j=1}^{N_g} \rho_j \, \tilde{\vecb{X}}_i \cdot \bnabla W_j(\vecb{X}_i).
\end{equation}
Thus, the VJP is a gather using $\bnabla W$ with weights $\overcirc{v}_i$, and the JVP is a gather of $\bnabla W$ dotted with $\tilde{\vecb{X}}$.

For completeness, we also state the full JVPs obtained by collecting the different contributions:
\begin{subequations}
\begin{align}
    \tilde{\rho}_j &= \tilde{\rho}_{\mathrm{in}, j} + \sum_{i=1}^{N} \left[ \tilde{w}_i \, W_j(\vecb{X}_i) - w_i \, \tilde{\vecb{X}}_i \cdot \bnabla W_j(\vecb{X}_i) \right], \\
    \tilde{v}_i &= \sum_{j=1}^{N_g} \left[ \tilde{\rho}_j \, W_j(\vecb{X}_i) - \rho_j \, \tilde{\vecb{X}}_i \cdot \bnabla W_j(\vecb{X}_i) \right].
\end{align}
\end{subequations}

Since all arising operations (except the identity case) are again scatter and gather operations, we implement a core scatter \& gather function in \codename, which is called with a flag indicating whether a forward or JVP/VJP pass is currently being performed. A summary of the operations is provided in the following table:

\begin{center}
\renewcommand{\arraystretch}{1.2}
\begin{tabular}{cccccc}
\toprule
\rotatebox[origin=c]{90}{} & Operator & Input & Type & Kernel & Implementation details \\
\midrule
\multirow{5}{*}{\rotatebox[origin=c]{90}{VJP}} 
  & Scatter & Mesh      & Identity & ---           & $\overcirc{\rho}_{\mathrm{in}} = \overcirc{\rho}$ \\
  & Scatter & Positions & Gather   & $\bnabla W$   & weighted by $w_i$ \\
  & Scatter & Weights   & Gather   & $W$           & scalar gather \\
  & Gather  & Mesh      & Scatter  & $W$           & weighted by $\overcirc{v}_i$ \\
  & Gather  & Positions & Gather   & $\bnabla W$   & weighted by $\overcirc{v}_i$ \\
\midrule
\multirow{5}{*}{\rotatebox[origin=c]{90}{JVP}} 
  & Scatter & Mesh      & Identity & ---           & $\tilde{\rho}_{\rho_{\mathrm{in}}} = \tilde{\rho}_{\mathrm{in}}$ \\
  & Scatter & Positions & Scatter  & $\bnabla W$   & weighted by $w_i$, dotted with $\tilde{\vecb{X}}_i$ \\
  & Scatter & Weights   & Scatter  & $W$           & weighted by $\tilde{w}_i$ \\
  & Gather  & Mesh      & Gather   & $W$           & weighted by $\tilde{\rho}_j$ \\
  & Gather  & Positions & Gather   & $\bnabla W$   & weighted by $\rho_j$, dotted with $\tilde{\vecb{X}}_i$ \\
\bottomrule
\end{tabular}
\end{center}

\subsection{Scatter and gather are adjoint to each other}

From a theoretical point of view, it is interesting to observe that the \texttt{scatter} and \texttt{gather} operators are adjoint under the $\ell^2$ pairing of the mesh and particle spaces, i.e.
\begin{equation}
    \langle \texttt{scatter}(w), \rho \rangle_{\text{mesh}} 
    = \langle w, \texttt{gather}(\rho) \rangle_{\text{particles}},
\end{equation}
where we define the discrete pairings as
\[
    \langle a, b \rangle_{\text{mesh}} := \sum_{j=1}^{N_g} a_j \, b_j,
    \qquad
    \langle a, b \rangle_{\text{particles}} := \sum_{i=1}^{N} a_i \, b_i.
\]

This follows immediately by computing:
\begin{equation}
\begin{aligned}
    \langle \texttt{scatter}(w), \rho \rangle_{\text{mesh}} 
    &= \sum_{j=1}^{N_g} \left( \sum_{i=1}^{N} w_i \, W_j(\vecb{X}_i) \right) \rho_j
    = \sum_{i=1}^{N} w_i \left(\sum_{j=1}^{N_g} \rho_j \, W_j(\vecb{X}_i) \right) \\
    &= \sum_{i=1}^{N} w_i \ \texttt{gather}(\rho)_i
    = \langle w, \texttt{gather}(\rho) \rangle_{\text{particles}}.
\end{aligned}
\end{equation}

This confirms that \texttt{scatter} and \texttt{gather} are adjoint operators in the $\ell^2$ sense:
\[
    \texttt{scatter} = \texttt{gather}^*,
    \qquad
    \texttt{gather} = \texttt{scatter}^*.
\]
Since the VJP of a linear map is its adjoint, and both \texttt{scatter} and \texttt{gather} are linear (in terms of $w$ and $\rho$, not $\vecb{X}$!) and mutually adjoint under the $\ell^2$ pairing, the VJP of one is naturally given by the other, justifying our unified implementation.

\end{document}